\documentclass[twocolumn,superscriptaddress, reqno, nobibnotes, aps, prb,showpacs,footinbib]{revtex4-2}
\usepackage{amsmath}
\usepackage{mathtools}
\usepackage{graphicx}
\usepackage{amsfonts}
\usepackage{amssymb}
\usepackage{bm}
\usepackage{dsfont}
\usepackage{array}
\usepackage{tikz}
\usepackage{comment}
\usepackage{subcaption}
\usepackage{enumitem}
\usetikzlibrary{math}
\usetikzlibrary{matrix}
\usepackage{pgfplots}
\usepackage{makecell}
\usepackage{csquotes}
\usepackage{multirow}
\usepackage{boldline}
\usepackage{ragged2e}
\pgfplotsset{compat=1.16}
\usepackage{epsfig,float,afterpage}
\usepackage[colorlinks,linkcolor=blue,citecolor=blue,urlcolor=blue]{hyperref}
\usepackage{pdfcomment}
\usepackage{soul,xcolor}
\allowdisplaybreaks
\newcommand{\Rom}[1]
{\MakeUppercase{\romannumeral #1}}

\newcommand{\beq}{\begin{equation}}
	\newcommand{\eeq}{\end{equation}}
\newcommand{\bes}{\begin{subequations}}
	\newcommand{\ees}{\end{subequations}}
\newcommand{\bea}{\begin{eqnarray}}
	\newcommand{\eea}{\end{eqnarray}}
\newcommand{\ba}{\begin{array}}
	\newcommand{\ea}{\end{array}}
\newcommand{\beqn}{\begin{eqnarray*}}
	\newcommand{\eeqn}{\end{eqnarray*}}

\begin{document}
	\title{Leading and beyond leading-order spectral form factor in chaotic quantum many-body systems across all Dyson symmetry classes}
	\author{Vijay Kumar}
	\affiliation{Raman Research Institute, Bangalore 560080, India}
	\author{Toma$\check{\rm z}$ Prosen}
	\affiliation{Physics Department, Faculty of Mathematics and Physics, University of Ljubljana, Jadranska 19, SI-1000 Ljubljana, Slovenia}
    \affiliation{Institute of Mathematics, Physics and Mechanics, Jadranska 19, SI-1000 Ljubljana, Slovenia}
    
	\author{Dibyendu Roy}
	\affiliation{Raman Research Institute, Bangalore 560080, India}
	
	\begin{abstract}
		We show the emergence of random matrix theory (RMT) spectral correlations in the chaotic phase of generic periodically kicked interacting quantum many-body systems by analytically calculating spectral form factor (SFF), $K(t)$, up to two leading orders in time, $t$. We explicitly consider the presence or absence of time reversal ($\mathcal{T}$) symmetry to investigate all three Dyson's symmetry classes. Our derivation only assumes random phase approximation to enable ensemble average. For $\mathcal{T}$-invariant systems with $\mathcal{T}^2=1$, we show that beyond the Thouless time $t^*$, the SFF takes the form $K(t)\simeq 2t-2t^2/\mathcal{N}$ up to second order in time, where $\mathcal{N}$ is the Hilbert space dimension. This is identical to the result from circular orthogonal ensemble of RMT. In the absence of $\mathcal{T}$-symmetry, we show that $K(t)\simeq t$ beyond $t^*$,  and there is no universal term in the second order, unlike the $\mathcal{T}^2=1$ case, in agreement with the result of circular unitary ensemble. For $\mathcal{T}$-invariant systems with $\mathcal{T}^2=-1$, we show that $K(t)\simeq 2t+2t^2/\mathcal{N}$ up to two orders in time beyond $t^*$, in agreement with the result of circular symplectic ensemble. In all three cases, the system-size, $L$, scaling of $t^*$ is determined by eigenvalues of a doubly stochastic matrix $\mathcal{M}$. For strongly interacting fermionic chains, $\mathcal{M}$ is $SU(2)$ invariant in all three cases, leading to $t^*\propto L^2$ in the presence of $U(1)$ symmetry. In the absence of $U(1)$ symmetry, we find $t^*\propto L^0$, due to gapped non-degenerate second-largest eigenvalue of $\mathcal{M}$ or $t^*\propto \ln(L)$ due to gapped second-largest eigenvalue with degeneracy $\propto L^\zeta$. Our calculation of SFF is plausible in higher space dimensions as well, where similar system-size scalings of $t^*$ can be obtained.
	\end{abstract}
	\maketitle
	
	\section{Introduction}\label{Intro}
	The quantum origin of chaos is still very actively studied even after fifty years of intensive research. An epoch in the field started with observing universal statistical properties of the spectra of quantum systems whose classical analogs are chaotic. These spectral properties are excellently described by random matrix theory (RMT) \cite{BohigasPRL1984,McDonaldPRL1979,Casati1980,Berry1977,Berry1981}, which was introduced in nuclear physics by Eugene Wigner to predict the distribution of energy level spacings of heavy nuclei \cite{Mehta2004,Wigner1955,Wigner1959,Dyson1962,Dyson_1970}. The success of RMT in capturing the universal spectral properties has been so remarkable that the presence of RMT behavior in a physical system is investigated to diagnose quantum chaos. Further advancing our understanding of quantum chaos requires an explanation for the emergence of RMT behavior in physical systems.
	
	A widely recognized and meaningful approach in this context involves analytically identifying statistical measures for random matrices and quantum systems with chaotic classical analogs. The results for the two cases are expected to match, but the steps leading to such a match would explain why such quantum systems behave identically to random matrices. The spectral form factor (SFF), a measure of correlation between energy levels, is extensively studied to diagnose quantum chaos. For time reversal ($\mathcal{T}$) invariant systems with $\mathcal{T}^2=1$ representing Gaussian orthogonal ensembles, the SFF takes the following form in RMT \cite{Mehta2004}:
	\begin{align}
		K_{\rm GOE}(t)=2t-t\ln\left(1+\frac{2t}{t_\text{H}}\right)=2t-\frac{2t^2}{t_\text{H}}+\frac{2t^3}{t_\text{H}^2}-...,
	\end{align}
	where $t$ is the time and $t_\text{H}$ is the Heisenberg time, which is related to the Hilbert space dimension $\mathcal{N}$. Berry \cite{Berry1985}, calculated the SFF up to leading order in time using the semiclassical periodic orbit theory, which was later extended up to second order in time by Sieber and Richter \cite{Sieber2001,Sieber2002}. Finally, the complete derivation, up to all orders in time, was performed by M{\"u}ller {\it{et al.}} \cite{MullerPRL2004,MullerPRE2005}. Additionally, a rigorous explanation has been possible only for quantum graphs \cite{PhysRevLett.112.144102,Pluhar_2015}. These studies provided valuable information on the reasons for the effectiveness of RMT. However, the RMT spectral statistics has also been observed to emerge in quantum many-body systems without classical analogs and is used to diagnose chaos in such systems as well. Therefore, a proper explanation of the emergence of RMT statistics in quantum many-body systems without classical analogs requires approaches that extend beyond the semiclassical periodic orbit theory.
	
	In recent years, the SFF has been calculated in different studies to understand quantum chaos in interacting many-body systems and its connection to RMT \cite{Dubertrand2016,KosPRX2018,ChanPRX2018,ChanPRL2018,BertiniPRL2018,Bertini2019PRL,BertiniPRX2019,FriedmanPRL2019,RoyPRE2020,Moudgalya21,BertiniCMP,PhysRevX.11.021051, RoyPRE2022, Richter2022,Winer22,Swingle22,Liao2022,Joshi2022,Dag23,Kumar2024, Vikram2024}. Some of these studies showed the emergence of RMT SFF in quantum circuits with random unitary gates within the limits of large local Hilbert space dimensions \cite{Dubertrand2016,ChanPRL2018,FriedmanPRL2019,Moudgalya21,Richter2022}. While quantum circuits are powerful theoretical tools, the requirement of a large local Hilbert space dimension leaves the question unanswered for physical systems with finite local Hilbert space dimensions like interacting fermions or qubits. Other studies involve effective field theory description \cite{Swingle22}, but a general mechanism explaining the emergence of RMT behavior for generic many-body quantum systems is still under question. Kos {\it{et al.}}\cite{KosPRX2018} made a significant advancement in this direction by deriving the SFF up to two leading orders in time. They considered a one-dimensional (1D) system of qubits with long-range interactions and a periodically kicked magnetic field in a transverse direction. Following the random phase approximation (RPA), they could express the leading-order contribution of SFF in terms of a partition function of a classical 1D Ising model (on a ring of circumference $t$), which can be calculated via a $2\times 2$ transfer matrix. It was plausible as the transverse field does not lead to any coupling among qubits in the driving Hamiltonian. This calculation was further generalized for the generic driving Hamiltonian in \cite{RoyPRE2020} to derive the SFF only in leading order in time. The higher-order terms of SFF for generic many-body quantum systems remain yet to be found. Furthermore, the models studied in \cite{KosPRX2018, RoyPRE2020,RoyPRE2022, Kumar2024} have $\mathcal{T}$ invariance. From a technical point of view, it is not clear how the formalism can be extended to systems in the absence of $\mathcal{T}$-symmetry or half-integer spin systems with $\mathcal{T}$ invariance where $\mathcal{T}$ satisfies $\mathcal{T}^2=-1$. In particular, while the number of diagrams that contribute in the first and second order in $t$ for $\mathcal{T}$-invariant systems with $\mathcal{T}^2=1$ and in the absence of $\mathcal{T}$-symmetry is finite, the number of such diagrams is exponentially large in $t$ for $\mathcal{T}$-invariant systems with $\mathcal{T}^2=-1$. Thus, one must find an alternative scheme to evaluate the SFF for $\mathcal{T}$-invariant systems with $\mathcal{T}^2=-1$. We have achieved this goal in this work, which is one of the main highlights of this paper. 
	
	This work calculates the SFF up to two leading orders in $t$ for generic periodically kicked interacting many-body quantum systems with or without $\mathcal{T}$-symmetry, which allows us to investigate all three Dyson's circular ensembles \cite{Dyson1962,Dyson_1970}. A recent study computed the exact SFF of non-interacting fermions with Dyson statistics \cite{ikeda2024exactspectralformfactors}. Our approach involves the RPA to perform ensemble average, which has been checked to work very well for such systems in the presence of random onsite potentials and long-range interactions \cite{RoyPRE2020,RoyPRE2022,Kumar2024}. We show the emergence of universal RMT SFF for Dyson's symmetry classes, namely, circular orthogonal ensemble (COE) for $\mathcal{T}^2=1$, circular unitary ensemble (CUE) in the absence of $\mathcal{T}$-symmetry, and circular symplectic ensemble (CSE) for $\mathcal{T}^2=-1$. The RMT predictions for the SFF of these ensembles are
	\begin{align}
		K_{\rm COE}(t)&=2t-t\ln\left(1+\frac{2t}{\mathcal{N}}\right),
		\label{COE_RMT_SFF}\\
		K_{\rm CUE}(t)&=t,
		\label{CUE_RMT_SFF}\\
		K_{\rm CSE}(t)&=2t-t\ln \big|1-\frac{2t}{\mathcal{N}}\big|,
		\label{CSE_RMT_SFF}
	\end{align}
	for $0<t<\mathcal{N}$ \cite{Dyson1962, Dyson_1970, Haake2001, ikeda2024exactspectralformfactors}. The SFF $K_{\text{CSE}}(t)$ differs from the CSE SFF in Refs. \cite{Dyson1962,Haake2001} by a factor of 1/4 because CSE SFF in these references is defined with an extra factor of 1/4. However, we exclude this factor in our derivation for a unified description. Our explicit derivation of SFF is crucial because it explains how (e.g., the mechanism, nonuniversal behavior) and when (timescales) many-body quantum systems acquire a universal RMT form. For $\mathcal{T}$-invariant systems with $\mathcal{T}^2=1$, the leading order in $t$ contribution to the SFF can be interpreted as a return probability $P_t(\underline{n})$ to an initial state $|\underline{n}\rangle$ after $t$ time steps. The diagrams contributing to the second order in $t$ of the SFF resemble the Sieber-Richter pairs of semiclassical periodic orbit theory \cite{Sieber2001,Sieber2002}. Most importantly, our results for the SFF not only give the universal SFF at longer time but also the nonuniversal part of SFF at short time, which goes beyond the RMT predictions. We particularly notice that the nonuniversal part of SFF at shorter times mainly comes from leading order in the $t$ contributions of the SFF. We also determine the system-size scaling of the Thouless timescales $t^*$ beyond which the SFF takes the universal RMT form in chains of spinless or spinful interacting fermions in the presence or absence of a $U(1)$ symmetry. We observe $P_t(\underline{n})\sim 1$ when $t\ll t^*$ and $P_t(\underline{n})\sim 1/\mathcal{N}$ when $t\geq t^*$. While 1D long-range models require more control to probe our predictions, checking them in higher dimensions experimentally would be easier. Our calculation of SFF is also applicable to higher space dimensions. We specifically notice the validity of the RPA for a shorter range of interactions with increasing coordination numbers that are easily feasible in higher dimensions. This work is highly significant as it explains the emergence of RMT behavior in generic many-body quantum systems. 	
	
	The rest of the paper is organized as follows. In Sec. \ref{Models_and_Observables}, we present the basic models, the RPA used to perform ensemble average, and the key technical developments as four theorems in computing $K(t)$. We summarize the main findings in Sec. \ref{Outline_of_the_Results}. In Secs. \ref{COE_Rules}-\ref{Exact_cancellation_of_reduced_diagrams}, we prove all the main technical steps. In Sec. \ref{leading_order_SFF}, we derive the leading-order SFF and the system-size scaling of $t^*$ for each case of $\mathcal{T}$-symmetry. In Sec. \ref{Second_order_correction}, we explain the second order in time terms of the RMT SFF. In Sec. \ref{Higher_dimensions}, we extend our $K(t)$ calculation to higher spatial dimensions with an example showing identical system-size scaling of $t^*$ as in lower dimension. Finally, we present our conclusion in Sec. \ref{Conclusion}. Further technical details are presented in five appendices.
	
	\section{Models and observables of interest}
	\label{Models_and_Observables}
	The main goal of this paper is to provide an explanation for the emergence of RMT behavior in generic strongly interacting quantum systems. Thus, we study a general class of periodically kicked systems, described by the Hamiltonian
	\begin{align}
		\hat{H}(t)&=\hat{H}_0+\tau_p\hat{H}_1\sum_{n\in\mathds{Z}}\delta(t-n\tau_p),
		\label{Ht}
	\end{align}
	where $\mathds{Z}$ is the set of integers, $\delta(t)$ is the Dirac delta function, $\hat{H}_0$ and $\hat{H}_1$ are two non-commuting Hermitian operators, and $\tau_p$ is the period of kicking. We set $\tau_p=1$ for the rest of the discussion. We take $\hat{H}_0$ to be $\mathcal{T}$-invariant, and $\hat{H}_1$ with or without $\mathcal{T}$-symmetry. In Tab.~\ref{T_const}, we summarize the constraints imposed by these symmetries on relevant quantities related to $\hat{H}_0$ and $\hat{H}_1$.
	\begin{table*}[t]
		\centering
		\begin{tabular}{|c|c|c|c|c|}
			\hline
			&\makecell{\textbf{Basis states}\\ \textbf{under time reversal}}& \textbf{Phases} $\boldsymbol{\theta_{\underline{n}}}$& $\boldsymbol{\hat{H}_1}$ \textbf{matrix}& $\boldsymbol{\hat{V}}(\equiv e^{-i\hat{H}_1})$ \textbf{matrix} \\
			\hline
			$\boldsymbol{\mathcal{T}^2=1}$& $|\mathcal{T}\underline{n}\rangle=|\underline{n}\rangle$ & non-degenerate & real symmetric& symmetric unitary\\
			\hline
			\makecell{Absence of\\ $\boldsymbol{\mathcal{T}}$-symmetry}& $|\mathcal{T}\underline{n}\rangle=|\underline{n}\rangle$& non-degenerate & complex Hermitian& non-symmetric unitary\\
			\hline
			$\boldsymbol{\mathcal{T}^2=-1}$& $\langle\underline{n}|\mathcal{T}\underline{n}\rangle=0$& \makecell{doubly degenerate,\\ $\theta_{\mathcal{T}\underline{n}}=\theta_{\underline{n}}$} & \makecell{complex Hermitian with \\ $\langle\mathcal{T}\underline{n}|\hat{H}_1|\mathcal{T}\underline{n}'\rangle=\langle\underline{n}|\hat{H}_1|\underline{n}'\rangle^*$,\\ $\langle\underline{n}|\hat{H}_1|\mathcal{T}\underline{n}'\rangle=-\langle\mathcal{T}\underline{n}|\hat{H}_1|\underline{n}'\rangle^*$,\\ $\langle\mathcal{T}\underline{n}|\hat{H}_1|\underline{n}'\rangle=-\langle\underline{n}|\hat{H}_1|\mathcal{T}\underline{n}'\rangle^*$}& \makecell{non-symmetric unitary with\\ $\langle\mathcal{T}\underline{n}|\hat{V}|\mathcal{T}\underline{n}'\rangle=\langle\underline{n}'|\hat{V}|\underline{n}\rangle$,\\ $\langle\underline{n}|\hat{V}|\mathcal{T}\underline{n}'\rangle=-\langle\underline{n}'|\hat{V}|\mathcal{T}\underline{n}\rangle$,\\ $\langle\mathcal{T}\underline{n}|\hat{V}|\underline{n}'\rangle=-\langle\mathcal{T}\underline{n}'|\hat{V}|\underline{n}\rangle$}\\
			\hline
		\end{tabular}
		\caption{\justifying\small Constraints imposed on relevant quantities related to $\hat{H}_0$ and $\hat{H}_1$ by the presence or absence of time reversal $(\mathcal{T})$ symmetry.}
		\label{T_const}
	\end{table*}
	To study the emergence of RMT behavior in such systems we study the SFF, which is defined as the Fourier transform of a connected two-point correlation function of spectral density \cite{KosPRX2018,Haake2001}. For periodically driven systems, the spectral density is defined as
	\begin{align}
		\rho(\phi)&=\frac{2\pi}{\mathcal{N}}\sum_{n=1}^{\mathcal{N}}\delta(\phi-\phi_n),
	\end{align}
	where $\phi_n$ for $n=1,...,\mathcal{N}$ are eigenphases of the Floquet operator, $\hat{U}$. The connected two-point correlation function of $\rho(\phi)$ is defined as
	\begin{align}
		C(\theta)=\frac{1}{2\pi}\int_0^{2\pi}d\phi\: \rho(\phi+\theta/2)\rho(\phi-\theta/2)-1.
	\end{align}
	We obtain the SFF by taking its Fourier transform:
	\begin{align}
		K(t)&=\frac{\mathcal{N}^2}{2\pi}\int_0^{2\pi}d\theta\: C(\theta)e^{-i\theta t}=[\text{tr}\hat{U}^t][\text{tr}\hat{U}^{-t}]-\mathcal{N}^2\delta_{t,0}.
	\end{align}
	Since $K(t)$ is not self averaging, we perform additional ensemble averaging denoted by $\langle ...\rangle$ and explained later in this section. Thus, the SFF takes the final form:
	\begin{align}
		K(t)&=\langle[\text{tr}\hat{U}^t][\text{tr}\hat{U}^{-t}]\rangle -\mathcal{N}^2\delta_{t,0}.
		\label{SFF}
	\end{align}
	For the Hamiltonian in Eq.~(\ref{Ht}), the operator $\hat{U}$ can be expressed in terms of operators $\hat{H}_0$ and $\hat{H}_1$ as
	\begin{align}
		\hat{U}&=\mathsf{T}e^{-i\int_0^1 dt \hat{H}(t)}=\hat{V}\hat{W},
	\end{align}
	where $\mathsf{T}$ represents time ordering, $\hat{V}=e^{-i\hat{H}_1}$, and $\hat{W}=e^{-i\hat{H}_0}$. We choose the eigenstates of $\hat{H}_0$, denoted by $|\underline{n}\rangle$, as the basis states to compute the SFF. Thus, the operators $\hat{H}_0$ and $\hat{W}$ act as
	\begin{align}
		\hat{H}_0|\underline{n}\rangle&=E_{\underline{n}}|\underline{n}\rangle,\\
		\hat{W}|\underline{n}\rangle&=e^{-i\theta_{\underline{n}}}|\underline{n}\rangle,
	\end{align}
	where the eigenvalues of $\hat{H}_0$ and the eigenphases of $\hat{W}$ are related by $\theta_{\underline{n}}=E_{\underline{n}}$ mod $2\pi$. Inserting identities $\sum_{\underline{n}_\tau}|\underline{n}_\tau\rangle\langle\underline{n}_\tau|=\mathds{1}_\mathcal{N}$ in $\text{tr}\hat{U}^t$ and $\sum_{\underline{n}'_\tau}|\underline{n}'_\tau\rangle\langle\underline{n}'_\tau|=\mathds{1}_\mathcal{N}$ in $\text{tr}\hat{U}^{-t}$ for $\tau=1,...,t$, we obtain
	\begin{align}
		K(t)&=\sum_{\underline{n}_1,...,\underline{n}_t}\sum_{\underline{n}_1',...,\underline{n}_t'}\langle e^{-i\sum_{\tau=1}^t\left(\theta_{\underline{n}_\tau}-\theta_{\underline{n}_\tau'}\right)}\notag\\
		&\times\prod_{\tau=1}^t V_{\underline{n}_\tau,\underline{n}_{\tau+1}}V^*_{\underline{n}_\tau',\underline{n}_{\tau+1}'}\rangle,
		\label{SFF_before_RPA}
	\end{align}
	where the trace in Eq.~(\ref{SFF}) forces periodic boundary condition (PBC) in time, $t+1\equiv 1$, and $V$ denotes the matrix representation of $\hat{V}$ in the computational basis with elements $V_{\underline{n},\underline{n}'}=\langle \underline{n}|\hat{V}|\underline{n}'\rangle$. We assume that the phases $\theta_{\underline{n}}$ are independent and uniformly distributed over the interval $[0,2\pi)$ apart from the case when $\mathcal{T}^2=-1$, where the Kramers' degeneracy leads to doubly degenerate phases, but still, uniformly distributed. The assumption of independent and uniformly distributed phases has been tested numerically to success in the presence of random onsite potentials and long-range interactions in $\hat{H}_0$ for 1D lattices \cite{KosPRX2018,RoyPRE2020,RoyPRE2022,Kumar2024}. We shall later discuss that it works better in higher spatial dimensions, even for relatively shorter-range interactions. Therefore, we get by averaging over the phases: 
	\begin{align}
		\langle e^{-i\sum_{\tau=1}^t\left(\theta_{\underline{n}_\tau}-\theta_{\underline{n}_\tau'}\right)}\rangle=\sum_{\pi}\prod_{\tau=1}^t\delta_{\theta_{\underline{n}'_\tau},\pi(\theta_{\underline{n}_{\tau}})},
		\label{RPA}
	\end{align}
	where $\pi$ is a permutation over $t$ phases $\{\theta_{\underline{n}_1},...,\theta_{\underline{n}_t}\}$. In general, one or more phases appear multiple times, thus, the collection of phases, $\{\theta_{\underline{n}_1},...,\theta_{\underline{n}_t}\}$, forms a multiset. If there are $l$ distinct phases of multiplicity $p_1,...,p_l$, satisfying $\sum_{i=1}^lp_i=t$, the permutations $\pi$ belong to the group $S_t/(S_{p_1}\times ...\times S_{p_l})$, where $S_x$ is the symmetric group of degree $x$ for $x=t,p_1,...,p_l$. For $\mathcal{T}$-invariant systems with $\mathcal{T}^2=1$ and systems without $\mathcal{T}$-symmetry, all the phases are non-degenerate, which implies that the permutation of phases is equivalent to the permutation of states $\underline{n}_\tau$. Thus,
	\begin{align}
		\pi(\theta_{\underline{n}_\tau})=\theta_{\underline{n}_{\pi(\tau)}}.
		\label{phase_state_perm_COE_CUE}
	\end{align}
	For $\mathcal{T}$-invariant systems with $\mathcal{T}^2=-1$, the phases are doubly degenerate, e.g., $\theta_{\underline{n}}=\theta_{\mathcal{T}\underline{n}}$. Thus,
	\begin{align}
		\pi(\theta_{\underline{n}_\tau})=\theta_{\underline{n}_{\pi(\tau)}},\theta_{\mathcal{T}\underline{n}_{\pi(\tau)}},
		\label{phase_state_perm_CSE}
	\end{align}
	where the subscript $\mathcal{T}\underline{n}_{\pi(\tau)}$ represents the time-reversed state $|\mathcal{T}\underline{n}_{\pi(\tau)}\rangle$ of a state $|\underline{n}_{\pi(\tau)}\rangle$. Substituting Eqs.~(\ref{RPA},\ref{phase_state_perm_COE_CUE}) in Eq. (\ref{SFF_before_RPA}) gives the SFF, $K_1(t)$ and $K_0(t)$, respectively, for COE and CUE as
	\begin{align}
		K_1(t)&=\sum_{\underline{n}_1,...,\underline{n}_t}\sum_{\pi}\prod_{\tau=1}^tV_{\underline{n}_\tau,\underline{n}_{\tau+1}}V^*_{\underline{n}_{\pi(\tau)},\underline{n}_{\pi(\tau+1)}},
		\label{SFF_COE_RPA}\\
		K_0(t)&=\sum_{\underline{n}_1,...,\underline{n}_t}\sum_{\pi}\prod_{\tau=1}^tV_{\underline{n}_\tau,\underline{n}_{\tau+1}}V^*_{\underline{n}_{\pi(\tau)},\underline{n}_{\pi(\tau+1)}}.
		\label{SFF_CUE_RPA}
	\end{align}
	We stress that while the Eqs.~(\ref{SFF_COE_RPA},\ref{SFF_CUE_RPA}) seem to have identical expressions on the right-hand side, the $V$ matrix is symmetric in Eq.~(\ref{SFF_COE_RPA}) and non-symmetric in Eq.~(\ref{SFF_CUE_RPA}). Substituting Eqs.~(\ref{RPA},\ref{phase_state_perm_CSE}) in Eq.~(\ref{SFF_before_RPA}) gives the SFF, $K_{-1}(t)$ for CSE:
	\begin{align}
		K_{-1}(t)&=\sum_{\underline{n}_1,...,\underline{n}_t}\sum_\pi\sum_{\vec{\sigma}}\prod_{\tau=1}^tV_{\underline{n}_\tau,\underline{n}_{\tau+1}}V^*_{\underline{n}^{(\sigma_\tau)}_{\pi(\tau)},\underline{n}^{(\sigma_{\tau+1})}_{\pi(\tau+1)}},
		\label{SFF_CSE_RPA}
	\end{align}
	where $\vec{\sigma}=(\sigma_1,...,\sigma_t)$ and $\sigma_\tau=0,1$ for $\tau=1,...,t$ such that $\underline{n}^{(0)}_{\pi(\tau)}=\underline{n}_{\pi(\tau)}$ and $\underline{n}^{(1)}_{\pi(\tau)}=\mathcal{T}\underline{n}_{\pi(\tau)}$. Since the subgroups of permutations $\pi$ depend on the multiplicity of different phases in the multiset $\{\theta_{\underline{n}_1},...,\theta_{\underline{n}_t}\}$, the summations $\sum_{\underline{n}_1,...,\underline{n}_t}$ and $\sum_{\pi}$ in Eqs.~(\ref{SFF_COE_RPA}), (\ref{SFF_CUE_RPA}), and (\ref{SFF_CSE_RPA}) can not be performed independently. To make further progress, it is necessary to decouple the two summations. We introduce the following procedure to achieve this.\\
	\textbf{Step 1}: We first take $\pi\in S_t$, irrespective of multiplicity of phases in the multiset $\{\theta_{\underline{n}_1},...,\theta_{\underline{n}_t}\}$. For a multiset containing $l$ distinct phases of multiplicity $p_1,..,p_l$, this leads to each distinct permutation being considered $\prod_{i=1}^lp_i!$ times. Therefore, all the extra appearances of each distinct permutation must be removed. The following step explains how this can be achieved.\\
	\textbf{Step 2}: We start by fixing the minimal case of repetition, where a phase appears twice in the multiset, $\theta_{\underline{n}_{\tau_1}}=\theta_{\underline{n}_{\tau_2}}$. Each distinct permutation of such a multiset appears twice. We remove the contributions of such extra appearances by explicitly calculating and then subtracting their contributions as below:
	\begin{align}
		K_1(t)&=\sum_{\pi\in S_t}\mathcal{X}_\pi-\sum_{\pi\in S_t/S_2}\mathcal{X}_\pi^{\{\underline{n},\underline{n}\}}+...,
		\label{SFF_RPA_X_pi}\\
		K_0(t)&=\sum_{\pi\in S_t}\mathcal{Y}_\pi-\sum_{\pi\in S_t/S_2}\mathcal{Y}_\pi^{\{\underline{n},\underline{n}\}}+...,
		\label{SFF_RPA_Y_pi}\\
		K_{-1}(t)&=\sum_{\pi\in S_t}\sum_{\vec{\sigma}}\mathcal{Z}_{\pi,\vec{\sigma}}-\sum_{\pi\in S_t/S_2}\sum_{\vec{\sigma}}\mathcal{Z}_{\pi,\vec{\sigma}}^{\{\underline{n},\underline{n}\}}\notag\\
		&-\sum_{\pi\in S_t/S_2}\sum_{\vec{\sigma}}\mathcal{Z}_{\pi,\vec{\sigma}}^{\{\underline{n},\mathcal{T}\underline{n}\}}+...,
		\label{SFF_RPA_Z_pi}
	\end{align}
	where 
	\begin{align}
		\mathcal{X}_\pi&=\sum_{\{\underline{n}\}}\prod_{\tau=1}^t V_{\underline{n}_\tau,\underline{n}_{\tau+1}}V^*_{\underline{n}_{\pi(\tau)},\underline{n}_{\pi(\tau+1)}},
		\label{X_pi}\\
		\mathcal{X}_\pi^{\{\underline{n},\underline{n}\}}&=\sum_{\{\underline{n}\}}\sum_{\tau_1<\tau_2}\delta_{\underline{n}_{\tau_1},\underline{n}_{\tau_2}}\prod_{\tau=1}^t V_{\underline{n}_\tau,\underline{n}_{\tau+1}}V^*_{\underline{n}_{\pi(\tau)},\underline{n}_{\pi(\tau+1)}},\\
		\mathcal{Y}_\pi&=\sum_{\{\underline{n}\}}\prod_{\tau=1}^t V_{\underline{n}_\tau,\underline{n}_{\tau+1}}V^*_{\underline{n}_{\pi(\tau)},\underline{n}_{\pi(\tau+1)}},
        \label{Y_pi}\\
		\mathcal{Y}_\pi^{\{\underline{n},\underline{n}\}}&=\sum_{\{\underline{n}\}}\sum_{\tau_1<\tau_2}\delta_{\underline{n}_{\tau_1},\underline{n}_{\tau_2}}\prod_{\tau=1}^t V_{\underline{n}_\tau,\underline{n}_{\tau+1}}V^*_{\underline{n}_{\pi(\tau)},\underline{n}_{\pi(\tau+1)}},\\
		\mathcal{Z}_{\pi,\vec{\sigma}}&=\sum_{\{\underline{n}\}}\prod_{\tau=1}^t V_{\underline{n}_\tau,\underline{n}_{\tau+1}}V^*_{\underline{n}^{(\sigma_\tau)}_{\pi(\tau)},\underline{n}^{(\sigma_{\tau+1})}_{\pi(\tau+1)}},
		\label{Z_pi_sigma}\\
		\mathcal{Z}_{\pi,\vec{\sigma}}^{\{\underline{n},\underline{n}\}}&=\sum_{\{\underline{n}\}}\sum_{\tau_1<\tau_2}\delta_{\underline{n}_{\tau_1},\underline{n}_{\tau_2}}\prod_{\tau=1}^t V_{\underline{n}_\tau,\underline{n}_{\tau+1}}V^*_{\underline{n}^{(\sigma_\tau)}_{\pi(\tau)},\underline{n}^{(\sigma_{\tau+1})}_{\pi(\tau+1)}},\\
		\mathcal{Z}_{\pi,\vec{\sigma}}^{\{\underline{n},\mathcal{T}\underline{n}\}}&=\sum_{\{\underline{n}\}}\sum_{\tau_1<\tau_2}\delta_{\underline{n}_{\tau_1},\mathcal{T}\underline{n}_{\tau_2}}\prod_{\tau=1}^t V_{\underline{n}_\tau,\underline{n}_{\tau+1}}V^*_{\underline{n}^{(\sigma_\tau)}_{\pi(\tau)},\underline{n}^{(\sigma_{\tau+1})}_{\pi(\tau+1)}},
	\end{align}
	where $\sum_{\{\underline{n}\}}=\sum_{\underline{n}_1,...,\underline{n}_t}$. The symbols $\mathcal{X}_\pi$ and $\mathcal{Y}_\pi$ denote, respectively, the total contribution to SFFs $K_1(t)$ and $ K_0(t)$ of a permutation $\pi\in S_t$ of all the multisets of states $\{\underline{n}_1,...,\underline{n}_t\}$. The symbols $\mathcal{X}_\pi^{\{\underline{n},\underline{n}\}}$ and $\mathcal{Y}_\pi^{\{\underline{n},\underline{n}\}}$ represent, respectively, the correction for a permutation $\pi\in S_t/S_2$ of all the multisets $\{\underline{n}_1,...,\underline{n}_t\}$, where at least one state appears two times. These terms fix the double counting of permutation $\pi$ of such multisets in $\mathcal{X}_\pi$ and $\mathcal{Y}_\pi$. However, the overcounting resulting from a higher number of repetitions is still present and requires further correction terms. In this paper, we restrict ourselves to double repetitions only. Similarly, the symbol $\mathcal{Z}_{\pi,\vec{\sigma}}$  denotes a contribution to the SFF, $K_{-1}(t)$, of a permutation $\pi\in S_t$ and configuration $\vec{\sigma}$ of all the multisets of states $\{\underline{n}_1,...,\underline{n}_t\}$. In this case, when a phase repeats two times in $\{\theta_{\underline{n}_1},...,\theta_{\underline{n}_t}\}$, due to the Kramers' degeneracy, either a state appears two times in $\{\underline{n}_1,...,\underline{n}_t\}$ or a state and its time-reversed state appear in $\{\underline{n}_1,...,\underline{n}_t\}$. Thus, there are two correction terms, denoted by $\mathcal{Z}_{\pi,\vec{\sigma}}^{\{\underline{n},\underline{n}\}}$ and $\mathcal{Z}_{\pi,\vec{\sigma}}^{\{\underline{n},\mathcal{T}\underline{n}\}}$.
	
	\begin{figure}[h]
		\centering
		\begin{subfigure}[t]{0.45\linewidth}
			\centering
			\begin{tikzpicture}
				\def\rad{1.5}
				\draw[blue,dashed] (0,0) circle(\rad);
				\draw[red,thick,->] (30:\rad) arc(30:90:\rad);\draw[red,thick] (90:\rad) arc(90:150:\rad)--(0:\rad)--(210:\rad);\draw[red,thick,->] (210:\rad) arc(210:270:\rad);\draw[red,thick] (270:\rad)  arc(270:330:\rad)--(180:\rad)--(30:\rad); 
				\draw (95:\rad) node[above left]{1};
				\draw (85:\rad) node[above right]{$t$};
				\draw (180:\rad) node[left]{$\tau_1$};
				\draw (0:\rad) node[right]{$\tau_2$};
			\end{tikzpicture}
			\caption{}
			\label{COE_CUE_diag_example}
		\end{subfigure}
		\hfill
		\begin{subfigure}[t]{0.45\linewidth}
			\centering
			\begin{tikzpicture}
				\def\radis{1}
				\def\radib{1.5}
				\draw[blue,dashed] (0,0) circle(\radib);
				\draw[dashed] (0,0) circle(\radis);
				\draw[red,thick,->] (30:\radib) arc(30:90:\radib);\draw[red,thick] (90:\radib) arc(90:150:\radib);
				\draw[red,thick,->] (210:\radib) arc(210:270:\radib);\draw[red,thick] (270:\radib) arc(270:330:\radib);
				\draw[red,thick] (150:\radib)--(0:\radis)--(210:\radib);
				\draw[red,thick] (330:\radib)--(180:\radis)--(30:\radib);
				\draw (95:\radib) node[above left]{1};
				\draw (85:\radib) node[above right]{$t$};
				\draw (180:\radib) node[left]{$\tau_1$};
				\draw (0:\radib) node[right]{$\tau_2$};
			\end{tikzpicture}
			\caption{}
			\label{CSE_diag_example}
		\end{subfigure}
		\caption{\justifying\small (a) A diagrammatic representation of a transposition for systems with $\mathcal{T}^2=1$ and without $\mathcal{T}$-symmetry, where states $\underline{n}_{\tau_1}$ and $\underline{n}_{\tau_2}$ are interchanged. The dashed blue circle represents initial configuration of states, $\{\underline{n}_1,...,\underline{n}_{\tau_1},...,\underline{n}_{\tau_2},...,\underline{n}_{t}\}$. The subscripts $(1,...,t)$ are in increasing order along the counterclockwise direction on the blue circle. The red curve represents the configuration of states after transposition $\{\underline{n}_1,...,\underline{n}_{\tau_2},...,\underline{n}_{\tau_1},...,\underline{n}_{t}\}$. (b) A diagrammatic representation of a transposition for $\mathcal{T}^2=-1$ case, where states $\underline{n}_{\tau_1}$ and $\underline{n}_{\tau_2}$ are interchanged and $\sigma_{\tau_1}=\sigma_{\tau_2}=1,\sigma_{\tau}=0$ for $\tau=1,\dots,t$ excluding $\tau_1,\tau_2$. The inner black dashed circle represent time reversed version of the states on the outer circle.}
		\label{perm_diag_example}
	\end{figure}
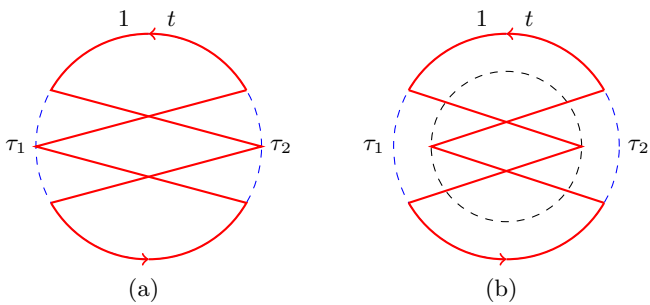 
	
	Calculating the SFF for each case of $\mathcal{T}$-symmetry requires a procedure to calculate $\mathcal{X}_\pi$, $\mathcal{X}_\pi^{\{\underline{n},\underline{n}\}}$, $\mathcal{Y}_\pi$, $\mathcal{Y}_\pi^{\{\underline{n},\underline{n}\}}$, $\mathcal{Z}_{\pi,\vec{\sigma}}$, $\mathcal{Z}_{\pi,\vec{\sigma}}^{\{\underline{n},\underline{n}\}}$, and $\mathcal{Z}_{\pi,\vec{\sigma}}^{\{\underline{n},\mathcal{T}\underline{n}\}}$ for different permutations $\pi$ and configurations $\vec{\sigma}$. Based on a diagrammatic representation of $\pi$ and $\vec{\sigma}$ (e.g., Fig. \ref{perm_diag_example}), we present rules to calculate the contribution of each permutation in Sec. \ref{COE_Rules}-\ref{CSE_rules}.
	
	We further find that, unlike the $\mathcal{T}^2=1$ and in the absence of $\mathcal{T}$-symmetry cases, the matrix $V$ has a richer structure for $\mathcal{T}^2=-1$ as given in Tab. \ref{T_const}. More specifically, the property $\langle\mathcal{T}\underline{n}|\hat{V}|\mathcal{T}\underline{n}'\rangle=\langle\underline{n}'|\hat{V}|\underline{n}\rangle$ immediately leads to the following theorem for such systems:
	\\\\
	\textbf{Theorem 1}: For a permutation $\pi$ and a configuration $\vec{\sigma}=(\sigma_1,...,\sigma_t)$, there exists another configuration $\vec{\sigma}'=(1-\sigma_t,...,1-\sigma_1)$, such that
	\begin{align}
		\mathcal{Z}_{\pi,\vec{\sigma}}=\mathcal{Z}_{\mathcal{R}\pi,\vec{\sigma}'},
	\end{align}
	where $\mathcal{R}$ is a reflection/anticyclic permutation, which reverses the order of all states in $\{\underline{n}_1,...,\underline{n}_t\}$. Therefore, collectively, permutation $\mathcal{R}\pi$ changes the order of states to $\{\underline{n}_{\pi(t)},...,\underline{n}_{\pi(1)}\}$.\\\\
	\textbf{Proof}: From Eq.~(\ref{Z_pi_sigma}), we write
	\begin{align}
		\mathcal{Z}_{\pi,\vec{\sigma}}&=\sum_{\underline{n}_1,...,\underline{n}_t}\prod_{\tau=1}^tV_{\underline{n}_\tau,\underline{n}_{\tau+1}}V^*_{\underline{n}_{\pi(\tau)}^{(\sigma_\tau)},\underline{n}_{\pi(\tau+1)}^{(\sigma_{\tau+1})}}\notag\\
		&=\sum_{\underline{n}_1,...,\underline{n}_t}\prod_{\tau=1}^tV_{\underline{n}_\tau,\underline{n}_{\tau+1}}V^*_{\mathcal{T}\underline{n}_{\pi(\tau+1)}^{(\sigma_{\tau+1})},\mathcal{T}\underline{n}_{\pi(\tau)}^{(\sigma_\tau)}}\notag\\
		&=\sum_{\underline{n}_1,...,\underline{n}_t}\prod_{\tau=1}^tV_{\underline{n}_\tau,\underline{n}_{\tau+1}}V^*_{\underline{n}_{\pi(\tau+1)}^{(1-\sigma_{\tau+1})},\underline{n}_{\pi(\tau)}^{(1-\sigma_\tau)}}\notag\\
		&=\mathcal{Z}_{\mathcal{R}\pi,\vec{\sigma}'}.
	\end{align}
	
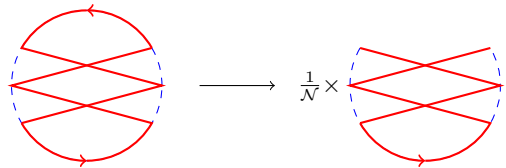
\begin{figure}[h]
		\centering
		\begin{tikzpicture}
			\def\rad{1}
			\draw[blue,dashed] (0,0) circle(\rad);
			\draw[red,thick,->] (30:\rad) arc(30:90:\rad);\draw[red,thick] (90:\rad) arc(90:150:\rad)--(0:\rad)--(210:\rad);\draw[red,thick,->] (210:\rad) arc(210:270:\rad);\draw[red,thick] (270:\rad)  arc(270:330:\rad)--(180:\rad)--(30:\rad); 
			\draw[->] ({\rad+0.5},0)--({\rad+1.5},0);
			\begin{scope}[shift={(4.5,0)}]
				\draw[blue,dashed] (150:\rad) arc(150:390:\rad);
				\draw[red,thick] (150:\rad)--(0:\rad)--(210:\rad);\draw[red,thick,->] (210:\rad) arc(210:270:\rad);\draw[red,thick] (270:\rad)  arc(270:330:\rad)--(180:\rad)--(30:\rad); 
				\draw (180:{\rad}) node[left]{$\frac{1}{\mathcal{N}}\times$};
			\end{scope}
		\end{tikzpicture}
		\caption{\justifying\small A reduced diagram obtained by inserting a factor $1/\mathcal{N}$ for a red arc in a diagram representing a transposition.}
		\label{red_diag_example}
	\end{figure}

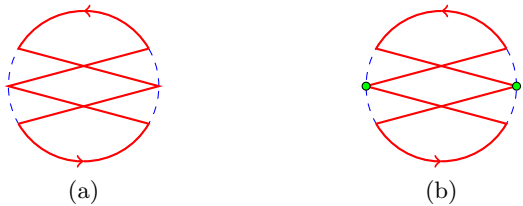
\begin{figure}[H]
		\centering
		\begin{subfigure}[t]{0.45\linewidth}
			\centering
			\begin{tikzpicture}
				\def\rad{1}
				\draw[blue,dashed] (0,0) circle(\rad);
				\draw[red,thick,->] (30:\rad) arc(30:90:\rad);\draw[red,thick] (90:\rad) arc(90:150:\rad)--(0:\rad)--(210:\rad);\draw[red,thick,->] (210:\rad) arc(210:270:\rad);\draw[red,thick] (270:\rad)  arc(270:330:\rad)--(180:\rad)--(30:\rad); 
			\end{tikzpicture}
			\caption{}
			\label{COE_CUE_no_rep_transposition}
		\end{subfigure}
		\hfill
		\begin{subfigure}[t]{0.45\linewidth}
			\centering
			\begin{tikzpicture}
				\def\rad{1}
				\draw[blue,dashed] (0,0) circle(\rad);
				\draw[red,thick,->] (30:\rad) arc(30:90:\rad);\draw[red,thick] (90:\rad) arc(90:150:\rad)--(0:\rad)--(210:\rad);\draw[red,thick,->] (210:\rad) arc(210:270:\rad);\draw[red,thick] (270:\rad)  arc(270:330:\rad)--(180:\rad)--(30:\rad); 
				\draw[fill=green] (180:\rad) circle(1.5pt);
				\draw[fill=green] (0:\rad) circle(1.5pt);
			\end{tikzpicture}
			\caption{}
			\label{COE_CUE_rep_transposition}
		\end{subfigure}
		\caption{\justifying\small Reduced diagrams of (a) and (b) are identical. The green circles in diagram (b) indicate that the states at those time steps are identical.}
		\label{red_diag_thm3_example}
	\end{figure}

	Following the rules in Sec.~\ref{COE_Rules}-\ref{CSE_rules}, the leading-order SFF determining the linear ramp can be easily derived for the $\mathcal{T}^2=1$ and without $\mathcal{T}$-symmetry cases by studying the $t$ cyclic and $t$ anticyclic variants of identity permutation, $I$,
	\begin{align}
		K_1^{(1)}(t)&=\sum_{l=0}^{t-1}\left(\mathcal{X}_{\mathcal{C}^l I}+\mathcal{X}_{\mathcal{R}\mathcal{C}^l I}\right),\\
		K_0^{(1)}(t)&=\sum_{l=0}^{t-1}\left(\mathcal{Y}_{\mathcal{C}^l I}+\mathcal{Y}_{\mathcal{R}\mathcal{C}^l I}\right),
	\end{align}
	where the superscript $(1)$ represents that the above expressions only contain the leading-order SFF, and $\mathcal{C}^l$ denotes cyclic permutation, whose action is defined as $\underline{n}_{\mathcal{C}^l(\tau)}=\underline{n}_{\tau+l}$ for $\tau=1,...,t$. We explicitly evaluate $K_1^{(1)}(t)$ and $K_0^{(1)}(t)$ in Sec.~\ref{leading_order_SFF}. However, for the $\mathcal{T}^2=-1$ case, each permutation $\pi$ has further $2^t$ variants due to configurations $\vec{\sigma}$ in Eq.~(\ref{SFF_RPA_Z_pi}). Therefore, a brute-force calculation based on the rules in Sec.~\ref{CSE_rules} is practically impossible. Instead, a new approach is required. Our new approach starts with a general analysis of arbitrary diagrams in Sec. \ref{theorem_2}, which we formulate into the following theorem.
	\\\\
	\textbf{Theorem 2}: Contributions $\mathcal{X}_\pi$, $\mathcal{X}_\pi^{\{\underline{n},\underline{n}\}}$, $\mathcal{Y}_\pi$, $\mathcal{Y}_\pi^{\{\underline{n},\underline{n}\}}$, $\mathcal{Z}_{\pi,\vec{\sigma}}$, $\mathcal{Z}_{\pi,\vec{\sigma}}^{\{\underline{n},\underline{n}\}}$, and $\mathcal{Z}_{\pi,\vec{\sigma}}^{\{\underline{n},\mathcal{T}\underline{n}\}}$ can be expressed as a sum of three kinds of terms, namely, Type \Rom{1}, Type \Rom{2}, and Type \Rom{3}. Type \Rom{1} term is $1/\mathcal{N}^\mu$, where $\mu$ is a non-negative integer. This term is universal since it is independent of Hamiltonian parameters. Type \Rom{2} and Type \Rom{3} terms are nonuniversal because they depend on Hamiltonian parameters. Furthermore, Type \Rom{3} terms decay exponentially with time, while Type \Rom{2} terms remain finite at long times.
	\\\\
	\textbf{Theorem 2} implies that only Type \Rom{1} and Type \Rom{2} terms are significant in determining SFF at long time. More specifically, when the contribution of all the permutations is added, Type \Rom{1} terms must add up to give a universal RMT form of SFF, whereas Type \Rom{2} terms must cancel each other out. Type \Rom{3} terms only contribute to the nonuniversal part of SFF and vanish beyond $t^*$. \textbf{Theorem 2} also suggests that Type \Rom{3} terms are insignificant when the goal is to understand the emergence of RMT behavior. Thus, eliminating some Type \Rom{3} terms, we define reduced diagrams (see, for example, Fig. \ref{red_diag_example}) in Sec. \ref{reduced_diagram}.
	The reduced diagrams provide further analytical control. Performing a general analysis of reduced diagrams, we discovered a meaningful pattern, which we formulated into the following theorems. For systems with $\mathcal{T}^2=1$ and in the absence of $\mathcal{T}$-symmetry, we have the following theorem:
	\\\\
	\textbf{Theorem 3}: For each reduced diagram, there exists a diagram with minimal repetition of states that has an identical reduced diagram.
    \\\\
	\textbf{Theorem 3} implies that for a given permutation $\pi$, there exists a permutation $\pi'$ such that Type \Rom{1} and Type \Rom{2} contribution of $\mathcal{X}_{\pi}$ and $\mathcal{X}_{\pi'}^{\{\underline{n},\underline{n}\}}$ (or $\mathcal{Y}_\pi$ and $\mathcal{Y}_{\pi'}^{\{\underline{n},\underline{n}\}}$) are identical as in Fig.~\ref{red_diag_thm3_example}.
	\noindent For $\mathcal{T}$-invariant systems with $\mathcal{T}^2=-1$, we have the following theorem:
    \\\\
	\textbf{Theorem 4}: For each reduced diagram, there exists either a diagram with minimal repetition of states or a diagram with two states at different time steps related by time reversal. These diagrams also have an identical reduced diagram.
	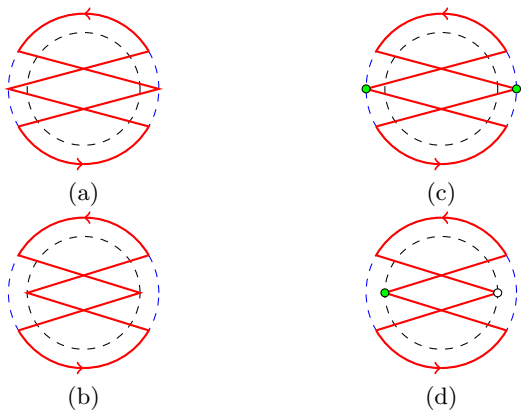
\begin{figure}[H]
		\centering
		\begin{subfigure}[t]{0.45\linewidth}
			\centering
			\begin{tikzpicture}
				\def\radib{1}
				\def\radis{0.75}
				\draw[blue,dashed] (0,0) circle(\radib);
				\draw[black,dashed] (0,0) circle(\radis);
				\draw[red,thick,->] (30:\radib) arc(30:90:\radib);\draw[red,thick] (90:\radib) arc(90:150:\radib)--(0:\radib)--(210:\radib);\draw[red,thick,->] (210:\radib) arc(210:270:\radib);\draw[red,thick] (270:\radib)  arc(270:330:\radib)--(180:\radib)--(30:\radib); 
			\end{tikzpicture}
			\caption{}
			\begin{tikzpicture}
				\def\radib{1}
				\def\radis{0.75}
				\draw[blue,dashed] (0,0) circle(\radib);
				\draw[black,dashed] (0,0) circle(\radis);
				\draw[red,thick,->] (30:\radib) arc(30:90:\radib);\draw[red,thick] (90:\radib) arc(90:150:\radib)--(0:\radis)--(210:\radib);\draw[red,thick,->] (210:\radib) arc(210:270:\radib);\draw[red,thick] (270:\radib)  arc(270:330:\radib)--(180:\radis)--(30:\radib); 
			\end{tikzpicture}
			\caption{}
		\end{subfigure}
		\hfill
		\begin{subfigure}[t]{0.45\linewidth}
			\centering
			\begin{tikzpicture}
				\def\radib{1}
				\def\radis{0.75}
				\draw[blue,dashed] (0,0) circle(\radib);
				\draw[black,dashed] (0,0) circle(\radis);
				\draw[red,thick,->] (30:\radib) arc(30:90:\radib);\draw[red,thick] (90:\radib) arc(90:150:\radib)--(0:\radib)--(210:\radib);\draw[red,thick,->] (210:\radib) arc(210:270:\radib);\draw[red,thick] (270:\radib)  arc(270:330:\radib)--(180:\radib)--(30:\radib); 
				\draw[fill=green] (180:\radib) circle(1.5pt);
				\draw[fill=green] (0:\radib) circle(1.5pt);
			\end{tikzpicture}
			\caption{}\label{CSE_rep_diag_transposition}
			\begin{tikzpicture}
				\def\radib{1}
				\def\radis{0.75}
				\draw[blue,dashed] (0,0) circle(\radib);
				\draw[black,dashed] (0,0) circle(\radis);
				\draw[red,thick,->] (30:\radib) arc(30:90:\radib);\draw[red,thick] (90:\radib) arc(90:150:\radib)--(0:\radis)--(210:\radib);\draw[red,thick,->] (210:\radib) arc(210:270:\radib);\draw[red,thick] (270:\radib)  arc(270:330:\radib)--(180:\radis)--(30:\radib); 
				\draw[fill=green] (180:\radis) circle(1.5pt);
				\draw[fill=white] (0:\radis) circle(1.5pt);
			\end{tikzpicture}
			\caption{}
			\label{CSE_Trep_diag_transposition}
		\end{subfigure}
		\caption{\justifying\small Reduced diagrams of (a) and (c) are identical. Reduced diagrams of (b) and (d) are identical. The green circles in (c) indicate that the states at those time steps are identical. The green and white circle in (d) indicate that the states at those time steps are related by time reversal.}
		\label{red_diag_thm4_example}
	\end{figure}    
	\textbf{Theorem 4} implies that for a given permutation $\pi$ and configuration $\vec{\sigma}$, there exists another permutation $\pi'$ such that Type \Rom{1} and Type \Rom{2} contribution of either $\mathcal{Z}_{\pi,\vec{\sigma}}$ and $\mathcal{Z}_{\pi',\vec{\sigma}}^{\{\underline{n},\underline{n}\}}$ or $\mathcal{Z}_{\pi,\vec{\sigma}}$ and $\mathcal{Z}_{\pi',\vec{\sigma}}^{\{\underline{n},\mathcal{T}\underline{n}\}}$ are identical as in Fig. \ref{red_diag_thm4_example}.
	Following \textbf{Theorem 3} and \textbf{Theorem 4}, we identify diagrams whose reduced diagrams are not cancelled. Such diagrams give universal RMT form of SFF. We identify such diagrams determining SFF up to second order in $t$ as shown in Sec.~\ref{leading_order_SFF}-\ref{Second_order_correction}.
	
	The SFF of physical chaotic systems takes the RMT SFF form only beyond $t^*$. To find the system-size scaling of $t^*$, we study strongly interacting fermionic chains with or without $\mathcal{T}$-symmetry in Sec. \ref{leading_order_SFF}. To study systems with $\mathcal{T}^2=1$ and in the absence of $\mathcal{T}$-symmetry, we consider a chain of spinless fermions described by
	\begin{align}
		\hat{H}_0^{0}&=\sum_{x=1}^L\epsilon_x \hat{n}_x+\sum_{x<y}\frac{U_0}{|x-y|^\alpha}\hat{n}_x\hat{n}_{y}, \label{H0}\\
		\hat{H}_1^{0}&=\sum_{x<y}-J_{xy}\hat{c}_x^\dagger \hat{c}_{y}+\Delta_{xy}c_x^\dagger c_{y}^\dagger+h.c.,
		\label{Hs0}
	\end{align}
	where the superscript $0$ represents spinless fermions, $\hat{n}_x\equiv \hat{c}^\dagger_x\hat{c}_x, \hat{c}_x$ and $\hat{c}^\dagger_x$ are, respectively, fermion occupation number, annihilation, and creation operators at site $x$. We take onsite energies $\epsilon_1,...,\epsilon_L$ as Gaussian random numbers with zero mean and finite standard deviation of $\Delta\epsilon$. The hopping amplitude, denoted by $J_{xy}$, takes real or imaginary values according to the presence or absence of $\mathcal{T}$-symmetry. The pairing amplitude $\Delta_{xy}$ takes only real values.
	
	For $\mathcal{T}$-invariant systems with $\mathcal{T}^2=-1$, we take a chain of interacting spin-1/2 fermions given as 
	\begin{align}
		\hat{H}_0^{1/2}&=\sum_{x=1}^L\epsilon_x\left(\hat{n}_{x\uparrow}+\hat{n}_{x\downarrow}\right)+\sum_{x<y}\sum_{\sigma\sigma'} \frac{U^{\sigma\sigma'}_{xy}}{|x-y|^\alpha}\hat{n}_{x\sigma}\hat{n}_{y\sigma'}\notag\\
		&+\sum_{x=1}^LU_x \hat{n}_{x\uparrow}\hat{n}_{x\downarrow}, \label{H01/2}\\
		\hat{H}_1^{1/2}&=\sum_{x=1}^L\sum_{\sigma\sigma'}J_{\sigma\sigma'}\hat{c}_{x\sigma}^\dagger \hat{c}_{x+1\sigma'}+\Delta_{\sigma\sigma'}\hat{c}_{x\sigma}^\dagger \hat{c}_{x+1\sigma'}^\dagger+h.c.,
		\label{Hs1/2}
	\end{align}
	where the superscript $1/2$ stands for spin-1/2 fermions, $\sigma=\uparrow,\downarrow$ represent the spin state along the $z$-axis, $\hat{n}_{x\uparrow} \equiv \hat{c}_{x\uparrow}^{\dagger}\hat{c}_{x\uparrow}$ and $\hat{n}_{x\downarrow} \equiv \hat{c}_{x\downarrow}^{\dagger}\hat{c}_{x\downarrow}$ are fermion occupation number operators at site $x$ for spin states $|\uparrow\rangle$ and $|\downarrow\rangle$, respectively. The operators $\hat{c}_{x\sigma}$ and $\hat{c}_{x\sigma}^\dagger$ are annihilation and creation operators of fermions at site $x$ with spin $\sigma$. The long-range interaction parameters $U_{ij}^{\sigma\sigma'}$ are all chosen as Gaussian random numbers with mean $U_0$ and standard deviation $\Delta U_0$ to ensure phases are only doubly degenerate due to the Kramers' degeneracy. The particle-particle repulsion $U_x$ at a site $x$ is also chosen as Gaussian random numbers with mean $\bar{U}$ and standard deviation $\Delta \bar{U}$. Additionally, the annihilation operators of spin-1/2 fermions transform under $\mathcal{T}$ as
	\begin{align}
		\mathcal{T}\hat{c}_{x\uparrow}\mathcal{T}^{-1}=-\hat{c}_{x\downarrow},\quad \mathcal{T}\hat{c}_{x\downarrow}\mathcal{T}^{-1}=\hat{c}_{x\uparrow}.
	\end{align}
	Thus, $\hat{H}^{1/2}_0$ and $\hat{H}^{1/2}_1$ are invariant under $\mathcal{T}$ when $U_{ij}^{\uparrow\uparrow}=U_{ij}^{\downarrow\downarrow}$, $U_{ij}^{\uparrow\downarrow}=U_{ij}^{\downarrow\uparrow}$, $J_{\downarrow\downarrow}=J_{\uparrow\uparrow}^*$, $J_{\uparrow\downarrow}=-J_{\downarrow\uparrow}^*$, $\Delta_{\downarrow\downarrow}=\Delta_{\uparrow\uparrow}^*$, and  $\Delta_{\uparrow\downarrow}=-\Delta_{\downarrow\uparrow}^*$. We also take PBC  on real space for all the models by imposing $\hat{c}_{x+L}\equiv\hat{c}_x$, $\hat{c}_{x+L\sigma}\equiv\hat{c}_{x\sigma}$, and replacing $|x-y|$ by $min(|x-y|,L-|x-y|)$ in long-range interactions.

	\section{Outline of the Results}
	\label{Outline_of_the_Results}
	We have identified the diagrams determining the universal RMT form of SFF up to two leading orders in time for all three Dyson's circular ensembles \cite{Dyson1962, Dyson_1970}. Different system-size scaling of $t^*$ have been found in recent years for various systems of COE and CUE classes with or without a $U(1)$ symmetry \cite{KosPRX2018,ChanPRL2018,Gharibyan2018, FriedmanPRL2019,RoyPRE2020,RoyPRE2022,Kumar2024}. There is hardly any result on interacting many-body quantum chaos of CSE class. We here provide a unified description of system-size scaling of $t^*$ for all three Dyson's statistics in the presence or absence of a $U(1)$ symmetry, which we summarize in Tab.~\ref{Thouless_time_table}.
	\subsection{Leading-order SFF: Linear Ramp}\label{Outline1st}
	The leading-order SFF is determined by the identity permutation and its variants. For $\mathcal{T}$-invariant systems with $\mathcal{T}^2=1$, there are $t$ cyclic and $t$ anticyclic permutations as variants of identity permutation. All these variants have identical contributions due to the symmetric nature of the matrix $V$ as summarized in Tab.~\ref{T_const}, and PBC in time. Therefore, we find the leading-order SFF for the COE class by applying the rules in Sec.~\ref{COE_Rules} as
	\begin{align}
		K_1^{(1)}(t)&=2t\;\text{tr}\mathcal{M}^t\notag\\
		&=2t\left(1+\lambda_1^t+...+\lambda_{\mathcal{N}-1}^t\right).
	\end{align}
	Here, $\mathcal{M}$ is a doubly stochastic (Markov) matrix given by $\mathcal{M}=V\bullet V^*$, where \enquote{$\bullet$} represents the Hadamard product defined as $\left(A\bullet B\right)_{i,j}=A_{i,j}B_{i,j}$. The largest eigenvalue $(\lambda_0)$ of $\mathcal{M}$ is one whereas other eigenvalues $\lambda_1,...,\lambda_{\mathcal{N}-1}$ have magnitude less than one. The leading order in $t$ contributions to the SFF can be physically interpreted using the Markov matrix as a return probability $P_t(\underline{n})=\langle \underline{n}|\mathcal{M}^t|\underline{n}\rangle$ to an initial state $|\underline{n}\rangle$ after $t$ time steps. Thus, we can write $K_1^{(1)}(t)=2t \sum_{\underline{n}}P_t(\underline{n})$.
	
	For systems in the absence of $\mathcal{T}$-symmetry, the contribution of cyclic variants differs from the anticyclic variants due to the non-symmetric nature of the matrix $V$. We thus get the leading-order SFF for the CUE class by applying the rules in Sec.~\ref{CUE_rules} as
	\begin{align}
		K_0^{(1)}(t)&=t\left(\text{tr}\mathcal{M}^t+\text{tr}\tilde{\mathcal{M}}^t\right)\notag\\
		&=t\left(1+\lambda_1^t+...+\lambda_{\mathcal{N}-1}^t+\chi_0^t+...+\chi_{\mathcal{N}-1}^t\right),
	\end{align}
	where $\tilde{\mathcal{M}}$ is a complex Hermitian matrix defined by $\tilde{\mathcal{M}}=V\bullet V^\dagger$. The eigenvalues of $\tilde{\mathcal{M}}$, $\chi_0,...,\chi_{\mathcal{N}-1}$, have magnitude less than one.
	
	\begin{table*}[t!]
		\centering
		\begin{tabular}{!{\vrule width 0.7pt}c!{\vrule width 0.7pt}c!{\vrule width 0.7pt}c!{\vrule width 0.7pt}c!{\vrule width 0.7pt}c!{\vrule width 0.7pt}c!{\vrule width 0.7pt}}
			\noalign{\hrule height 0.7pt}
			&\textbf{Parameters ($\beta$)}& \textbf{Symmetry}&\textbf{$\lambda_1$ in thermodynamic limit}& \textbf{Degeneracy, $d_1$} & \textbf{Thouless time, $t^*$}\\
			\noalign{\hrule height 0.7pt}
			\multirow{4}{*}{$\boldsymbol{\mathcal{T}^2=1}$} &$J\neq 0, \Delta =0$&$[\hat{H}(t),\hat{N}]=0$&$1-c_{\beta}/L^2$&$\mathcal{O}(L^0)$&$\mathcal{O}(L^2)$\\ \cline{2-6}
			&$J\neq \Delta\neq 0$&$[\hat{H}(t),\hat{N}] \ne 0$&$\mathcal{O}(L^0)$&$\mathcal{O}(L^0)$& $\mathcal{O}(L^0)$\\ \cline{2-6}
			&$|J|=|\Delta|\neq 0$&$[\hat{H}(t),\hat{N}] \ne 0$&$\mathcal{O}(L^0)$&$\mathcal{O}(L^2)$&$\mathcal{O}(\ln L)$\\ \cline{2-6}
			&$J=0,\Delta\neq 0$&$[\hat{H}(t),\hat{N}_s]=0$&$1-c_{\beta}/L^2$&$\mathcal{O}(L^0)$&$\mathcal{O}(L^2)$\\
			\noalign{\hrule height 0.7pt}
			\multirow{3}{*}{\makecell{Absence of\\ $\boldsymbol{\mathcal{T}}$-symmetry}} &$J,g\neq 0; \Delta,\Delta'=0$&$[\hat{H}(t),\hat{N}]=0$&$1-c_{\beta}/L^2$&$\mathcal{O}(L^0)$&$\mathcal{O}(L^2)$\\ \cline{2-6}
			&$J,g\neq 0;\Delta,\Delta'\neq 0$&$[\hat{H}(t),\hat{N}]\ne 0$&$\mathcal{O}(L^0)$&$\mathcal{O}(L^0)$&$\mathcal{O}(L^0)$\\ \cline{2-6}
			&$|J|=|\Delta|,|g|=|\Delta'|$&$[\hat{H}(t),\hat{N}] \ne 0$&$\mathcal{O}(L^0)$&$\mathcal{O}(L)$&$\mathcal{O}(\ln L)$\\
			\noalign{\hrule height 0.7pt}
			\multirow{3}{*}{$\boldsymbol{\mathcal{T}^2=-1}$} &$J_{\sigma\sigma'}\neq 0; \Delta_{\sigma\sigma'}=0$&$[\hat{H}(t),\hat{N}_t]=0$&$1-c_{\beta}/L^2$&$\mathcal{O}(L^0)$&$\mathcal{O}(L^2)$\\ \cline{2-6}
			&$J_{\sigma\sigma'}\neq 0,\Delta_{\sigma\sigma'}\neq 0$&$[\hat{H}(t),\hat{N}_t] \ne 0$&$\mathcal{O}(L^0)$&$\mathcal{O}(L^0)$&$\mathcal{O}(L^0)$\\ \cline{2-6}
			&$|J_{\sigma\sigma'}|=|\Delta_{\sigma\sigma'}|$&$[\hat{H}(t),\hat{N}_t]\ne 0$&$\mathcal{O}(L^0)$&$\mathcal{O}(L)$&$\mathcal{O}(\ln L)$\\
			\noalign{\hrule height 0.7pt}
		\end{tabular}
		\caption{\justifying\small System-size scaling of Thouless time $(t^*)$ for periodically kicked strongly interacting fermionic chains of length $L$ with or without time reversal $(\mathcal{T})$ symmetry. The operators $\hat{N}=\sum_{x=1}^L \hat{n}_x$ and $\hat{N}_t=\sum_{x=1}^L\sum_{\sigma} \hat{n}_{x\sigma}$ denote the total fermion number for spinless and spinful model, respectively, and $\hat{N}_s=\sum_{x=1}^L (-1)^x\hat{n}_x$ is the staggered fermion number for even $L$. The presence or absence of a $U(1)$ symmetry can be found by their commutation with the chain's Hamiltonian $\hat{H}(t)$.}
		\label{Thouless_time_table}
	\end{table*}
	
	For $\mathcal{T}$-invariant systems with $\mathcal{T}^2=-1$, each of the cyclic and anticyclic variant have further $2^t$ variants corresponding to different configurations $\vec{\sigma}=(\sigma_1,...,\sigma_t)$. Analyzing their reduced diagrams, we find that only diagrams with $\vec{\sigma}=(0,...,0)$ or $\vec{\sigma}=(1,...,1)$ have non-vanishing contribution at long times leading to the SFF for the CSE class at leading order in $t$ as
	\begin{align}
		K_{-1}^{(1)}(t)&=2t\left(\text{tr}\mathcal{M}^t+\text{tr}\tilde{\mathcal{M}}^t\right)\notag\\
		&=2t\left(1+\lambda_1^t+...+\lambda_{\mathcal{N}-1}^t+\chi_0^t+...+\chi_{\mathcal{N}-1}^t\right).
	\end{align}
	
	The Thouless time $t^*$ is determined by the second-largest eigenvalue $\lambda_1$ of a doubly stochastic matrix $\mathcal{M}$ for all three classes. Beyond $t^*$, we have $K_1^{(1)}\simeq 2t$, $K_0^{(1)}\simeq t$, and $K_{-1}^{(1)}(t)\simeq 2t$, which are identical to the RMT predictions for these classes up to first order in $t$, Eqs.~(\ref{COE_RMT_SFF}-\ref{CSE_RMT_SFF}). We find that $t^*$ is related to eigenvalue $\lambda_1$ and its degeneracy $d_1$ through
	\begin{align}
		t^*\simeq\frac{\ln(d_1)+1}{|\ln\lambda_1|}.
		\label{Thouless_time_result}
	\end{align}
	We determine the system-size of scaling of $t^*$ in the presence or absence of a $U(1)$ symmetry by studying the Hamiltonians Eqs.~(\ref{H0},\ref{Hs0}) for systems with $\mathcal{T}^2=1$ and in the absence of $\mathcal{T}$-symmetry. To study the former case, we choose the hopping and pairing parameters as
	\begin{align}
		J_{xy}&=\begin{cases}\label{nn_hopping}
			\text{$J$, if $|x-y|=1$},\\
			\text{0, otherwise}
		\end{cases}\\
		\label{nn_pairing}
		\Delta_{xy}&=\begin{cases}
			\text{$\Delta$, if $|x-y|=1$},\\
			\text{0, otherwise}
		\end{cases}
	\end{align}
	where $J$ and $\Delta$ are real valued parameters. We break $\mathcal{T}$-symmetry by adding an imaginary next-nearest-neighbor hopping. So we consider the following parameters to investigate $t^*$ for systems in the absence of $\mathcal{T}$-symmetry.
	\begin{align}
		J_{xy}&=\begin{cases}\label{CUE_hopping}
			\text{$J$, if $|x-y|=1$},\\
			\text{$ig$, if $|x-y|=2$},\\
			\text{0, otherwise}
		\end{cases}\\
		\label{CUE_pairing}
		\Delta_{xy}&=\begin{cases}
			\text{$\Delta$, if $|x-y|=1$},\\
			\text{$\Delta'$, if $|x-y|=2$},\\
			\text{0, otherwise}
		\end{cases}
	\end{align}
	where $J,g,\Delta$ and $\Delta'$ are real valued parameters. We explore a chain of strongly interacting spin-1/2 fermions in Eqs.~(\ref{H01/2},\ref{Hs1/2}) to determine $t^*$ for $\mathcal{T}$-invariant systems with $\mathcal{T}^2=-1$. We obtain different system-size scaling of $t^*$ for various choices of parameters in all three classes as summarized in Tab.~\ref{Thouless_time_table}. While the second-largest eigenvalue $\lambda_1$ of $\mathcal{M}$ determines when the universal RMT SFF form would emerge for all three classes, the other eigenvalues of $\mathcal{M}$ and $\tilde{\mathcal{M}}$ fix the nonuniversal behavior of the SFF at short timescales for any physical systems. We have shown a nice comparison between the SFF obtained within the RPA and that from direct numerics in Fig.~\ref{COE_SFF_RPA_1_2} both at short and long times for a $\mathcal{T}$-invariant model with $\mathcal{T}^2=1$. %{\color{red} role of eigenvalues of $\tilde{\mathcal{M}}$?}
	
	\subsection{Second-order correction}\label{Outline2nd}
	The second-order term in the RMT SFF is inversely proportional to the Hilbert space dimension $\mathcal{N}$ for $\mathcal{T}^2=\pm1$ and such term is absent in the absence of $\mathcal{T}$-symmetry. From the Type \Rom{1} term for different permutations, we find that the following permutations and their variants contribute to the second-order correction in the SFF. These are (1) single transpositions ($T$) defined as the interchange of any two states in $\{\underline{n}_1,...,\underline{n}_t\}$, (2) sub-sequence reversal ($S$) defined as reversal of the order of states between two-time steps, (3) identity permutation with a state repeated twice ($R$) and (4) sub-sequence reversal with a state repeated twice ($SR$). The last permutation reverses the order of states between the two appearances of the repeated state.	
	
	Analyzing reduced diagrams for these permutations, we find that the reduced diagrams of $T$ and $R$ are exactly identical for all three Dyson classes. This implies that the Type \Rom{1} and Type \Rom{2} contributions of these permutations exactly cancel each other. Similarly, the reduced diagrams of $S$ are canceled by the reduced diagrams of $SR$. However, there are two extra $SR$ diagrams with long red arcs. Computing all the variants of these diagrams, we obtain  the second-order RMT correction to the SFF for all three Dyson classes:
	\begin{align}
		K_1^{(2)}(t)&= -\frac{2t^2}{\mathcal{N}}+\mathcal{O}(\lambda_1^t),\\
		K_0^{(2)}(t)&= \mathcal{O}(\lambda_1^t,\chi_0^t),\\
		K_{-1}^{(2)}(t)&= \frac{2t^2}{\mathcal{N}}+\mathcal{O}(\lambda_1^t,\chi_0^t).
	\end{align}
	In an alternative diagrammatic representation, the extra $SR$ diagrams resemble the Sieber-Richter pairs of near-miss and self-crossing orbits \cite{Sieber2001,Sieber2002}. Nevertheless, there are specific differences in these diagrams as we are studying quantum systems here on a discrete Hilbert space compared to semiclassical periodic orbit theory on continuous phase space. We further notice that the nonuniversal terms at second order in $t$ of our calculated SFF do not contribute significantly to the short-time nonuniversal behavior of SFF when we compare the SFF computed within the RPA and that from direct numerics, Fig. \ref{COE_SFF_RPA_1_2}.   %{\color{red} Check}
	
	\section{Rules to calculate $\mathcal{X}_\pi$ and $\mathcal{X}_{\pi}^{\{\underline{n},\underline{n}\}}$}
	\label{COE_Rules}
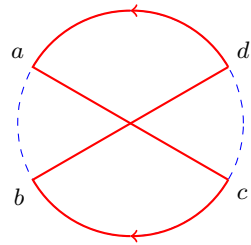
\begin{figure}[h]
		\centering
		\begin{tikzpicture}
			\def\rad{1.5}
			\draw[blue,dashed] (0,0) circle(\rad);
			\draw[red,thick,->] (30:\rad) arc(30:90:\rad);\draw[red,thick] (90:\rad) arc(90:150:\rad)--(-30:\rad);\draw[red,thick,->] (-30:\rad) arc(-30:-90:\rad);\draw[red,thick] (-90:\rad) arc(-90:-150:\rad)--(30:\rad);
			\draw (150:\rad) node[above left]{$a$};
			\draw (-150:\rad) node[below left]{$b$};
			\draw (-30:\rad) node[below right]{$c$};
			\draw (30:\rad) node[above right]{$d$};
		\end{tikzpicture}
		\caption{\justifying\small A diagram representing a sub-sequence reversal which reverses the order of states from $\tau_1$ to $\tau_2$. States at time steps $\tau_1$ and $\tau_2$ are denoted by the labels $b$ and $c$, $b\equiv\underline{n}_{\tau_1},c\equiv\underline{n}_{\tau_2}$. The other labels $a$ and $d$ denote states at the corresponding time steps, $a\equiv\underline{n}_{\tau_1-1},d\equiv\underline{n}_{\tau_2+1}$.}
		\label{Example_diag_COE}
	\end{figure}

	We now enumerate some rules, which can be applied to find $\mathcal{X}_\pi$ and $\mathcal{X}_{\pi}^{\{\underline{n},\underline{n}\}}$ for $\mathcal{T}$-invariant systems with $\mathcal{T}^2=1$. The rules are: \\
	\noindent (\romannumeral 1) First draw a diagram representing a permutation $\pi$ (e.g., a sub-sequence reversal shown in Fig.~ \ref{Example_diag_COE}).
	\\
	\noindent (\romannumeral 2) Denote a state repeated twice by green circles, \enquote{\raisebox{-0.5\height}{\tikz{\draw[fill=green] (0,0) circle(1.5pt);}}}, at the corresponding time steps (e.g., Fig.~\ref{COE_CUE_rep_transposition}).\\
	\noindent (\romannumeral 3) For a red arc of length $n$ time steps, insert a factor as \\
	\noindent \raisebox{-0.5\height}{\tikz{\draw[red,thick,->] (30:1) arc(30:90:1);\draw[red,thick] (90:1) arc(90:150:1);
			\draw (30:1) node[right]{$a$};
			\draw (150:1) node[left]{$b$};
			\draw (90:1) node[above]{\scriptsize$n$ time steps};}} $\equiv$  \raisebox{-0.5\height}{\tikz{\draw[red,thick] (30:1) arc(30:90:1);\draw[red,thick,<-] (90:1) arc(90:150:1);
			\draw (30:1) node[right]{$a$};
			\draw (150:1) node[left]{$b$};
			\draw (90:1) node[above]{\scriptsize$n$ time steps};}} 
	$\equiv \left(\mathcal{M}^{n}\right)_{a,b}$. The matrix $\mathcal{M}$ is symmetric due to the symmetric unitary property of the matrix $V$. Thus, performing eigendecomposition of $\mathcal{M}$, we obtain
	\begin{align}
		\left(\mathcal{M}^{n}\right)_{a,b}=\sum_{i=0}^{\mathcal{N}-1}\lambda_i^n\mathcal{M}^{(i)}_{a,b},
	\end{align}
	where $\lambda_i$ are eigenvalues of $\mathcal{M}$, such that $\lambda_0=1\geq |\lambda_1|\geq ...\geq |\lambda_{\mathcal{N}-1}|$, \cite{RoyPRE2020}, and $\mathcal{M}^{(i)}=|\lambda_i\rangle\langle\lambda_i|$. The state $|\lambda_i\rangle$ is an eigenvector of $\mathcal{M}$ corresponding to eigenvalue $\lambda_i$.\\
	\noindent (\romannumeral 4) \raisebox{-0.5\height}{\tikz{\draw[blue,dashed] (0:1)--(0,0);\draw[blue,dashed,<-] (0,0)--(180:1);
			\draw (0:1) node[right]{$b$};
			\draw (180:1) node[left]{$a$};}} $\equiv$ \raisebox{-0.5\height}{\tikz{\draw[blue,dashed] (-1,0)--(0,0);\draw[blue,dashed,<-] (0,0)--(1,0);
			\draw (0:1) node[right]{$b$};
			\draw (180:1) node[left]{$a$};}}
	$\equiv V_{a,b}$.\\
	\noindent (\romannumeral 5) \raisebox{-0.5\height}{\tikz{\draw[red,thick] (0:1)--(0,0);\draw[red,thick,<-] (0,0)--(180:1);
			\draw (0:1) node[right]{$b$};
			\draw (180:1) node[left]{$a$};}} $\equiv$ \raisebox{-0.5\height}{\tikz{\draw[red,thick] (-1,0)--(0,0);\draw[red,thick,<-] (0,0)--(1,0);
			\draw (0:1) node[right]{$b$};
			\draw (180:1) node[left]{$a$};}}
	$\equiv V^*_{a,b}$.\\
	\noindent (\romannumeral 6) Take a sum over all the matrix indices repeated twice or more.\\
	We derive the above rules in App.~\ref{COE_rules_der}. Applying the above rules, we evaluate the contribution $\mathcal{X}_{S}$ of a sub-sequence reversal described in Fig.~\ref{Example_diag_COE} as
	\begin{align}
		\mathcal{X}_{S}&=\left(\sum_{i=1}^{\mathcal{N}-1}\lambda_i^{\nu_1}\mathcal{M}^{(i)}_{d,a}\right)\left(\sum_{j=0}^{\mathcal{N}-1}\lambda_j^{\nu_2}\mathcal{M}^{(j)}_{b,c}\right)\notag\\
		&\times V_{a,b}V_{c,d}V^*_{a,c}V^*_{b,d},\notag\\
		&=\sum_{i,j=0}^{\mathcal{N}-1}\lambda_i^{\nu_1}\lambda_j^{\nu_2}Q^{ij}_{S},
		\label{X_pi_COE_example}
	\end{align}
	where $Q^{ij}_{S}=\mathcal{M}^{(i)}_{d,a}\mathcal{M}^{(j)}_{b,c}V_{a,b}V_{c,d}V^*_{a,c}V^*_{b,d}$, and $\nu_1=t-\tau_2+\tau_1-2$ and $\nu_2=\tau_2-\tau_1$ are the lengths of upper and lower red arcs of Fig.~\ref{Example_diag_COE} measured in units of the kicking period $\tau_p$.
	\section{Rules to calculate $\mathcal{Y}_\pi$ and $\mathcal{Y}_{\pi}^{\{\underline{n},\underline{n}\}}$}
	\label{CUE_rules}
	We here list the rules for finding $\mathcal{Y}_\pi$ and $\mathcal{Y}_{\pi}^{\{\underline{n},\underline{n}\}}$ for systems in the absence of $\mathcal{T}$-symmetry.\\
	
	\noindent (\romannumeral 1) First draw a diagram representing a permutation $\pi$ (e.g., a sub-sequence reversal shown in Fig.~\ref{Example_diag_COE}).\\
	\noindent (\romannumeral 2) Represent a state repeated twice by green circles, \enquote{\raisebox{-0.5\height}{\tikz{\draw[fill=green] (0,0) circle(1.5pt);}}}, at the corresponding time steps (e.g., Fig.~\ref{COE_CUE_rep_transposition}).\\
	\noindent (\romannumeral 3) For a red arc of length $n$ time steps with a counterclockwise arrow, insert a factor as\\
	\noindent \raisebox{-0.5\height}{\tikz{\draw[red,thick,->] (30:1) arc(30:90:1);\draw[red,thick] (90:1) arc(90:150:1);
			\draw (30:1) node[right]{$a$};
			\draw (150:1) node[left]{$b$};
			\draw (90:1) node[above]{\scriptsize$n$ time steps};}} 
	$\equiv \left(\mathcal{M}^{n}\right)_{a,b}=\sum_{i=0}^{\mathcal{N}-1}\lambda_i^n\mathcal{M}^{(i)}_{a,b}$. The doubly stochastic matrix $\mathcal{M}$ is non-symmetric due to the non-symmetric nature of the matrix $V$, thus, $\mathcal{M}^{(i)}=\prescript{}{R}{|\lambda_i\rangle}\langle\lambda_i|_L$, where $\prescript{}{R}{|\lambda_i\rangle}$ and $\langle\lambda_i|_L$ are the right and left eigenvectors of $\mathcal{M}$ with an eigenvalue $\lambda_i$.\\
	\noindent (\romannumeral 4) For a red arc of length $n$ time steps with a clockwise arrow insert\\
	\noindent \raisebox{-0.5\height}{\tikz{\draw[red,thick] (30:1) arc(30:90:1);\draw[red,thick,<-] (90:1) arc(90:150:1);
			\draw (30:1) node[right]{$a$};
			\draw (150:1) node[left]{$b$};
			\draw (90:1) node[above]{\scriptsize$n$ time steps};}} 
	$\equiv \left(\tilde{\mathcal{M}}^{n}\right)_{a,b}=\sum_{i=0}^{\mathcal{N}-1}\chi_i^n\tilde{\mathcal{M}}^{(i)}_{a,b}$. The matrix $\tilde{\mathcal{M}}$ is complex Hermitian with eigenvalues, $\chi_0,\chi_1,...,\chi_{\mathcal{N}-1}$,  satisfying $1> |\chi_0|\geq |\chi_1|\geq ...\geq |\chi_{\mathcal{N}-1}|$ (following the Ger$\check{\rm s}$gorin circle theorem as we explain in App.~\ref{Gersgorin}), and $\tilde{\mathcal{M}}^{(i)}=|\chi_i\rangle\langle\chi_i|$, where $|\chi_i\rangle$ is an eigenvector of $\tilde{\mathcal{M}}$ corresponding to eigenvalue $\chi_i$.\\
	\noindent (\romannumeral 5) \raisebox{-0.5\height}{\tikz{\draw[blue,dashed] (0:1)--(0,0);\draw[blue,dashed,<-] (0,0)--(180:1);
			\draw (0:1) node[right]{$b$};
			\draw (180:1) node[left]{$a$};}} $\equiv V_{a,b}$, \raisebox{-0.5\height}{\tikz{\draw[blue,dashed] (-1,0)--(0,0);\draw[blue,dashed,<-] (0,0)--(1,0);
			\draw (0:1) node[right]{$b$};
			\draw (180:1) node[left]{$a$};}}
	$\equiv V_{b,a}$.\\
	\noindent (\romannumeral 6) \raisebox{-0.5\height}{\tikz{\draw[red,thick] (0:1)--(0,0);\draw[red,thick,<-] (0,0)--(180:1);
			\draw (0:1) node[right]{$b$};
			\draw (180:1) node[left]{$a$};}} $\equiv V^*_{a,b}$, \raisebox{-0.5\height}{\tikz{\draw[red,thick] (-1,0)--(0,0);\draw[red,thick,<-] (0,0)--(1,0);
			\draw (0:1) node[right]{$b$};
			\draw (180:1) node[left]{$a$};}}
	$\equiv V^*_{b,a}$.\\
	\noindent (\romannumeral 7) Take sum over all the matrix indices, which are repeated twice or more.\\
	We give a derivation of the above rules in App.~\ref{T_absent_rul_der}. We can apply the above rules to evaluate the contribution $\mathcal{Y}_{S}$ of a sub-sequence reversal described in Fig.~\ref{Example_diag_COE} as
	\begin{align}
		\mathcal{Y}_{S}&=\sum_{i,j=0}^{\mathcal{N}-1}\lambda_i^{\nu_1}\chi_{j}^{\nu_2}Q^{i;j}_{S},
		\label{Y_pi_CUE_example}
	\end{align}
	where $Q^{i;j}_{S}=\mathcal{M}^{(i)}_{d,a}\tilde{\mathcal{M}}^{(j)}_{b,c}V_{a,b}V_{c,d}V^*_{a,c}V^*_{b,d}$, and $\nu_1, \nu_2$ are the same as in Sec.~\ref{COE_Rules}. As mentioned in the rules above, the contribution of red arcs with counterclockwise and clockwise arrows is different when $\mathcal{T}$-symmetry is absent, therefore, we separate their indices by a semicolon, as in $Q^{i;j}_{S}$.
	\section{Rules to calculate $\mathcal{Z}_{\pi,\vec{\sigma}},\mathcal{Z}_{\pi,\vec{\sigma}}^{\{\underline{n},\underline{n}\}}$, and $\mathcal{Z}_{\pi,\vec{\sigma}}^{\{\underline{n},\mathcal{T}\underline{n}\}}$}
	\label{CSE_rules}
\begin{figure}[h]
		\centering
		\begin{tikzpicture}
			\def\radib{1.5}
			\def\radis{1}
			\draw[blue,dashed] (0,0) circle(\radib);
			\draw[black,dashed] (0,0) circle(\radis);
			\draw[red,thick,->] (30:\radib) arc(30:90:\radib);\draw[red,thick] (90:\radib) arc(90:150:\radib)--(0:\radib)--(210:\radis);\draw[red,thick,->] (210:\radis) arc(210:270:\radis);\draw[red,thick] (270:\radis) arc(270:330:\radis)--(180:\radib)--(30:\radib);
			\draw (150:\radib) node[above left]{$a$};
			\draw (180:\radib) node[left]{$b$};
			\draw (210:\radib) node[below left]{$c$};
			\draw (-30:\radib) node[below right]{$d$};
			\draw (0:\radib) node[right]{$e$};
			\draw (30:\radib) node[above right]{$f$};
		\end{tikzpicture}
		\caption{\justifying\small A diagram representing a transposition of states at time steps $\tau_1$ and $\tau_2$ with $\vec{\sigma}=(\sigma_1,...,\sigma_t)$ such that $\sigma_{\tau}=1$ for $\tau\in[\tau_1+1,\tau_2-1]$ and $\sigma_\tau=0$ for $\tau\notin [\tau_1+1,\tau_2-1]$. The labels at the vertices denote states on the blue circle at those time steps, $a\equiv\underline{n}_{\tau_1-1},b\equiv\underline{n}_{\tau_1},c\equiv\underline{n}_{\tau_1+1},d\equiv\underline{n}_{\tau_2-1},e\equiv\underline{n}_{\tau_2},f\equiv\underline{n}_{\tau_2+1}$. The labels $b$ and $e$ represent the transposed states.}
		\label{Example_diag_CSE}
	\end{figure}
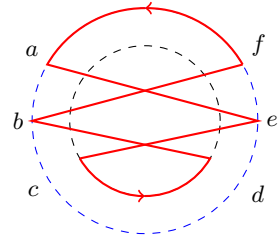

	We give the rules for evaluating $\mathcal{Z}_{\pi,\vec{\sigma}},\mathcal{Z}_{\pi,\vec{\sigma}}^{\{\underline{n},\underline{n}\}}$, and $\mathcal{Z}_{\pi,\vec{\sigma}}^{\{\underline{n},\mathcal{T}\underline{n}\}}$ for $\mathcal{T}$-invariant systems with $\mathcal{T}^2=-1$. The rules are:    
    
	\noindent (\romannumeral 1) Draw a dashed blue circle representing initial states $\{\underline{n}_1,...,\underline{n}_t\}$ and a concentric dashed black circle of smaller radius denoting the time-reversed states $\{\mathcal{T}\underline{n}_1,...,\mathcal{T}\underline{n}_t\}$. Draw next the red curve representing a permutation $\pi$ and a configuration $\vec{\sigma}$ (e.g., Fig.~\ref{Example_diag_CSE}).\\
	\noindent (\romannumeral 2) Draw green circles, \enquote{\raisebox{-0.5\height}{\tikz{\draw[fill=green] (0,0) circle(1.5pt);}}}, for a state repeated twice at the corresponding time steps (e.g., Fig.~\ref{CSE_rep_diag_transposition}). Draw green and white circles, \enquote{\raisebox{-0.5\height}{\tikz{\draw[fill=white] (0,0) circle(1.5pt);}}},  for two states related by $\mathcal{T}$ at the corresponding time steps (e.g., Fig.~\ref{CSE_Trep_diag_transposition}).\\
	\noindent (\romannumeral 3) For a red arc of length $n$ time steps with a counterclockwise arrow on the outer circle, insert a factor as \\
	\noindent \raisebox{-0.5\height}{\tikz{\draw[red,thick,->] (30:1) arc(30:90:1);\draw[red,thick] (90:1) arc(90:150:1);\draw[black,dashed] (30:0.75) arc(30:150:0.75);
			\draw (30:1) node[right]{$a$};
			\draw (150:1) node[left]{$b$};
			\draw (90:1) node[above]{\scriptsize$n$ time steps};}} 
	$\equiv \left(\mathcal{M}^{n}\right)_{a,b}=\sum_{i=0}^{\mathcal{N}-1}\lambda_i^n\mathcal{M}^{(i)}_{a,b}$. The doubly stochastic matrix $\mathcal{M}$ is non-symmetric due to the non-symmetric nature of the matrix $V$, thus, $\mathcal{M}^{(i)}=\prescript{}{R}{|\lambda_i\rangle}\langle\lambda_i|_L$, where $\prescript{}{R}{|\lambda_i\rangle}$ and $\langle\lambda_i|_L$ are respectively the right and left eigenvectors of $\mathcal{M}$ corresponding to eigenvalue $\lambda_i$.\\
	\noindent (\romannumeral 4) For a red arc of length $n$ time steps with a clockwise arrow on the outer circle, insert a factor as\\
	\noindent \raisebox{-0.5\height}{\tikz{\draw[red,thick] (30:1) arc(30:90:1);\draw[red,thick,<-] (90:1) arc(90:150:1);\draw[black,dashed] (30:0.75) arc(30:150:0.75);
			\draw (30:1) node[right]{$a$};
			\draw (150:1) node[left]{$b$};
			\draw (90:1) node[above]{\scriptsize$n$ time steps};}} 
	$\equiv \left(\tilde{\mathcal{M}}^{n}\right)_{a,b}=\sum_{i=0}^{\mathcal{N}-1}\chi_i^n\tilde{\mathcal{M}}^{(i)}_{a,b}$. The matrix $\tilde{\mathcal{M}}$ is complex Hermitian with eigenvalues $\chi_0,...,\chi_{\mathcal{N}-1}$ satisfying $1> |\chi_0|\geq |\chi_1|\geq ...\geq |\chi_{\mathcal{N}-1}|$ (again following the Ger$\check{\rm s}$gorin circle theorem as discussed in App.~\ref{Gersgorin}), and $\tilde{\mathcal{M}}^{(i)}=|\chi_i\rangle\langle\chi_i|$, where $|\chi_i\rangle$ is an eigenvector of $\tilde{\mathcal{M}}$ corresponding to eigenvalue $\chi_i$.\\
	\noindent (\romannumeral 5) For a red arc with a clockwise arrow on the inner circle, insert a factor as\\
	\noindent \raisebox{-0.5\height}{\tikz{\draw[red,thick] (30:0.75) arc(30:90:0.75);\draw[red,thick,<-] (90:0.75) arc(90:150:0.75);\draw[blue,dashed] (30:1) arc(30:150:1);
			\draw (30:1) node[right]{$a$};
			\draw (150:1) node[left]{$b$};
			\draw (90:1) node[above]{\scriptsize$n$ time steps};}} 
	$\equiv \left(\mathcal{M}^{n}\right)_{a,b}=\sum_{i=0}^{\mathcal{N}-1}\lambda_i^n\mathcal{M}^{(i)}_{a,b}$.\\
	\noindent (\romannumeral 6) For a red arc with a counterclockwise arrow on the inner circle, insert a factor as\\
	\noindent \raisebox{-0.5\height}{\tikz{\draw[red,thick,->] (30:0.75) arc(30:90:0.75);\draw[red,thick] (90:0.75) arc(90:150:0.75);\draw[blue,dashed] (30:1) arc(30:150:1);
			\draw (30:1) node[right]{$a$};
			\draw (150:1) node[left]{$b$};
			\draw (90:1) node[above]{\scriptsize$n$ time steps};}} 
	$\equiv \left(\tilde{\mathcal{M}}^{n}\right)_{a,b}=\sum_{i=0}^{\mathcal{N}-1}\chi_i^n\tilde{\mathcal{M}}^{(i)}_{a,b}$.\\
	\noindent (\romannumeral 7) \raisebox{-0.5\height}{\tikz{\draw[blue,dashed,->] (-0.75,0)--(0,0);\draw[blue,dashed] (0,0)--(0.75,0);
			\draw (-0.75,0) node[left]{$a$};
			\draw (0.75,0) node[right]{$b$};}} $\equiv V_{a,b}$, \raisebox{-0.5\height}{\tikz{\draw[blue,dashed,->] (0.75,0)--(0,0);\draw[blue,dashed] (0,0)--(-0.75,0);
			\draw (-0.75,0) node[left]{$a$};
			\draw (0.75,0) node[right]{$b$};}}
	$\equiv V_{b,a}$.\\
	\noindent (\romannumeral 8) \raisebox{-0.5\height}{\tikz{\draw[red,thick,->] (-0.75,0)--(0,0);\draw[red,thick] (0,0)--(0.75,0);
			\draw (-0.75,0) node[left]{$a$};
			\draw (0.75,0) node[right]{$b$};}} $\equiv V^*_{a,b}$, \raisebox{-0.5\height}{\tikz{\draw[red,thick,->] (0.75,0)--(0,0);\draw[red,thick] (0,0)--(-0.75,0);
			\draw (-0.75,0) node[left]{$a$};
			\draw (0.75,0) node[right]{$b$};}}
	$\equiv V^*_{b,a}$.\\
	\noindent (\romannumeral 9) Take sum over all the matrix indices, which are repeated twice or more.\\
	Since the matrix $V$ is non-symmetric in the computational basis, the rules here are identical to those in Sec.~\ref{CUE_rules} for systems in the absence of $\mathcal{T}$-symmetry except the rules (\romannumeral 5) and (\romannumeral 6), which follow naturally from the \textbf{Theorem 1}. Following the above rules, the contribution  $\mathcal{Z}_{T,\vec{\sigma}}$ of a transposition with configuration $\vec{\sigma}$ as shown in Fig.~\ref{Example_diag_CSE} can be expressed as
	\begin{align}
		\mathcal{Z}_{T,\vec{\sigma}}&=\sum_{i,j=0}^{\mathcal{N}-1}\lambda_i^{\nu_1}\chi_{j}^{\nu_2}Q^{i;;j;}_{T},
		\label{Z_pi_CSE_example}
	\end{align}
	where
	\begin{align}
		Q^{i;;j;}_{T}=\mathcal{M}^{(i)}_{f,a}\tilde{\mathcal{M}}^{(j)}_{c,d}V_{a,b}V_{b,c}V_{d,e}V_{e,f}V^*_{ae}V^*_{e,\mathcal{T}c}V^*_{\mathcal{T}d,b}V^*_{b,f},
	\end{align}
	and $\nu_1,\nu_2$ are the same as in Sec.~\ref{COE_Rules}. In general, $i_1,i_2,i_3$ and $i_4$ in the superscript of $Q_\pi^{i_1;i_2;i_3;i_4}$ represent, respectively, the indices coming from red arcs with counterclockwise arrow on the outer circle, the indices coming from red arcs with clockwise arrow on the outer circle, the indices coming from red arcs with counterclockwise arrow on the inner circle, and the indices coming from red arcs with clockwise arrow on the inner circle. Since, Fig. \ref{Z_pi_CSE_example} does not contain arcs with clockwise arrow on the inner or outer circle, $Q^{i;;j;}_T$ contains empty slots, separated by semicolon, for such arcs.
	
	\section{Proof of Theorem 2}
	\label{theorem_2}
	We now prove the \textbf{Theorem 2} in Sec.~\ref{Models_and_Observables} for all three cases of $\mathcal{T}$-symmetry.
	\subsection{$\mathcal{T}^2=1$}
	We first consider a permutation $\pi$, whose diagram contains number $r$ of red arcs of lengths $\nu_1,...,\nu_r$ with no repetition of states. Following the rules in Sec.~\ref{COE_Rules}, the contribution to SFF can be written similar to Eq.~(\ref{X_pi_COE_example}) as
	\begin{align}
		\mathcal{X}_\pi&=\sum_{\{i_q\}}\left(\prod_{q=1}^r\lambda_{i_q}^{\nu_q}\right)Q^{i_1...i_r}_\pi,
		\label{X_pi_COE_r_arcs}
	\end{align} 
	where $\sum_{\{i_q\}}=\sum_{i_1,...,i_r}$ with $i_q=0,...,\mathcal{N}-1$, $\forall q$. For a nontrivial permutation, such as the one shown in Fig.~\ref{Example_diag_COE}, some of the red line segments do not overlap with the blue circle. Let $\delta$ be the total number of such red line segments. Then, we have
	\begin{align}
		\sum_q\nu_q=t-\delta.
	\end{align}
	In Eq.~(\ref{X_pi_COE_r_arcs}), the factor $\left(\prod_{q=1}^r\lambda_{i_q}^{\nu_q}\right)$ determines the behavior of different terms at long time as $|\lambda_{i_q}|<1$ when $i_q\neq 0$ and $\lambda_{i_q}=1$ when $i_q=0$. Thus, all the terms in Eq.~(\ref{X_pi_COE_r_arcs}) can be divided into three different categories as following.\\
	\textbf{Case 1}: $i_q=0,\forall q$. Since $\lambda_{0}=1$, the right side of Eq.~(\ref{X_pi_COE_r_arcs}) gives the term $Q^{0...0}_\pi$, which is a function of $\mathcal{M}^{(0)}$, $V$ and $V^*$. Since $\mathcal{M}$ is a doubly stochastic matrix, we have $\langle\lambda_0|\equiv(1/\sqrt{\mathcal{N}})(1\; .\; .\; .\; 1)$. Therefore, all the matrix elements of $\mathcal{M}^{(0)}$ are $1/\mathcal{N}$. A study of different permutations reveals that $Q^{0...0}_\pi$ can typically be simplified to the form $1/\mathcal{N}^\mu$ (check App.~\ref{COE_2nd_full_der}), where $\mu$ is a non-negative integer. The absence of Hamiltonian parameters implies that this is a universal term. This is the Type \Rom{1} term in the \textbf{Theorem 2}. If $Q^{0...0}_\pi$ can not be simplified to a form independent of Hamiltonian parameters, the term is nonuniversal and considered as a Type \Rom{2} term because it remains finite at long times.\\
	\textbf{Case 2}: $i_q\neq 0,\forall q$. In this case, $|\lambda_{i_q}|<1,\forall q$, so $\prod_{q=1}^{r}\lambda_{i_q}^{\nu_q}$ can be interpreted as a product of $(t-\delta)$ real numbers of magnitude less than one. Thus, it decays exponentially with $t$. Therefore, such terms are Type \Rom{3}.\\
	\textbf{Case 3}: All the remaining terms involve at least  one $i_{q'}=0$ and at least one $i_{q''}\neq 0$. Thus, we can divide the set $\{1,...,r\}$ into two subsets $Q_1=\{q'_1,...,q'_{r_1}\}$ and $Q_2=\{q''_1,...,q''_{r_2}\}$ with $r_1,r_2\geq 1$ and $r_1+r_2=r$, such that, $i_{q}=0,\forall q\in Q_1$ and $i_{q}\neq 0,\forall q\in Q_2$. Therefore,
	\begin{align}
		\prod_{q=1}^{r}\lambda_{i_q}^{\nu_q}&=\left(\prod_{q\in Q_1}\lambda_{0}^{\nu_q}\right)\left(\prod_{q\in Q_2}\lambda_{i_{q}}^{\nu_q}\right)=\prod_{q\in Q_2}\lambda_{i_{q}}^{\nu_q}.
	\end{align}
	Since $\sum_{q=1}^r\nu_q=t-\delta$, $\sum_{q=1}^r\nu_q$ diverges with increasing time. We can further write $\sum_{q=1}^r\nu_q=\sum_{q\in Q_1}\nu_q+\sum_{q\in Q_2}\nu_q$. Therefore, $\sum_{q\in Q_2}\nu_q$ exhibits two possible behaviors in the long time limit: (a) it diverges, or (b) it remains finite. Consequently, $\prod_{q=1}^{r}\lambda_{i_q}^{\nu_q}$ falls exponentially at long times for case (a), and it remains finite in case (b). We substitute the long-time form of $\prod_{q=1}^{r}\lambda_{i_q}^{\nu_q}$ in Eq.~(\ref{X_pi_COE_r_arcs}) to find that the terms with case (a) behavior fall exponentially with time, and the terms with case (b) behavior diverge. Therefore, the case (a) terms are Type \Rom{3} and the case (b) terms, as they depend on the Hamiltonian parameters, are Type \Rom{2}.
	
	Therefore, $\mathcal{X}_\pi$ can be expressed as a sum of Type \Rom{1}, Type \Rom{2}, and Type \Rom{3} terms in accordance with \textbf{Theorem 2}. A similar analysis can be performed to show that \textbf{Theorem 2} is also valid for $\mathcal{X}_\pi^{\{\underline{n},\underline{n}\}}$.
	
	\subsection{Absence of $\mathcal{T}$-symmetry}
	We now consider a permutation $\pi$, whose diagram contains $r$ number of red arcs with counterclockwise arrows and lengths $\nu_1,...,\nu_r$, and $\tilde{r}$ number of red arcs with  clockwise arrows and lengths $\tilde{\nu}_1,...,\tilde{\nu}_{\tilde{r}}$. Following the rules in Sec.~\ref{CUE_rules}, the contribution to SFF can be written similar to Eq.~(\ref{Y_pi_CUE_example}):
	\begin{align}
		\mathcal{Y}_\pi&=\sum_{\{i_q\}}\sum_{\{\tilde{i}_q\}}\left(\prod_{q=1}^r\lambda_{i_q}^{\nu_q}\right)\left(\prod_{q=1}^{\tilde{r}}\chi_{\tilde{i}_q}^{\tilde{\nu}_q}\right)\notag\\
		&\qquad\times Q^{i_1...i_r;\tilde{i}_1...\tilde{i}_{\tilde{r}}}_\pi,
		\label{Y_pi_CUE_general}
	\end{align}
	where
	\begin{align}
		\sum_{q=1}^r\nu_q+\sum_{q=1}^{\tilde{r}}\tilde{\nu}_q=t-\delta.
	\end{align}
	Since $|\chi_i|<1$ for $i=0,...,\mathcal{N}-1$, $|\lambda_i|<1$ for $i=1,...,\mathcal{N}-1$, and $\lambda_0=1$, all the terms in Eq.~(\ref{Y_pi_CUE_general}) can be categorized as following.\\
	\textbf{Case 1}: $\tilde{r}=0$ and $i_q=0,\forall q$. The right-hand side of Eq.~(\ref{Y_pi_CUE_general}) gives the term $Q^{0...0;}_\pi$. Like the $\mathcal{T}^2=1$ case, typically, $Q^{0...0;}_\pi=1/\mathcal{N}^\mu$, where $\mu$ is a non-negative integer and thus it leads to a Type \Rom{1} term. If $Q^{0...0;}_\pi$ can not be simplified to a form independent of the Hamiltonian parameters, it is a Type \Rom{2} term.\\
	\textbf{Case 2}: $i_q\neq 0,\forall q$ or $r=0$. In this case, the factor $\left(\prod_{q=1}^r\lambda_{i_q}^{\nu_q}\right)\left(\prod_{q=1}^{\tilde{r}}\chi_{\tilde{i}_q}^{\tilde{\nu}_q}\right)$ in Eq.~(\ref{Y_pi_CUE_general}) is a product of $(t-\delta)$ real or complex numbers of magnitude less than one. Therefore, the corresponding terms fall exponentially with time. Such terms are Type \Rom{3}.\\
	\textbf{Case 3}: All the remaining terms have at least one $i_q=0$, and if $\tilde{r}=0$, then at least one $i_q\neq 0$. Therefore, the set $\{1,...,r\}$ can be divided into two subsets $Q_1=\{q|i_q=0\}$ and $Q_2=\{q|i_q\neq 0\}$. We have 
	\begin{align}
		\sum_{q=1}^r\nu_q+\sum_{q=1}^{\tilde{r}}\tilde{\nu}_q=\sum_{q\in Q_1}\nu_q+\sum_{q\in Q_2}\nu_q+\sum_{q=1}^{\tilde{r}}\tilde{\nu}_q=t-\delta,
	\end{align}
	and
	\begin{align}
		\left(\prod_{q=1}^r\lambda_{i_q}^{\nu_q}\right)\left(\prod_{q=1}^{\tilde{r}}\chi_{\tilde{i}_q}^{\tilde{\nu}_q}\right)&=\Big(\prod_{q\in Q_1}\lambda_{0}^{\nu_q}\prod_{q\in Q_2}\lambda_{i_q}^{\nu_q}\Big)\Big(\prod_{q=1}^{\tilde{r}}\chi_{\tilde{i}_q}^{\tilde{\nu}_q}\Big)\notag\\
		&=\left(\prod_{q\in Q_2}\lambda_{i_q}^{\nu_q}\right)\left(\prod_{q=1}^{\tilde{r}}\chi_{\tilde{i}_q}^{\tilde{\nu}_q}\right).
	\end{align}
	Thus, $\left(\prod_{q=1}^r\lambda_{i_q}^{\nu_q}\right)\left(\prod_{q=1}^{\tilde{r}}\chi_{\tilde{i}_q}^{\tilde{\nu}_q}\right)$ can be interpreted as a product of $(\sum_{q\in Q_2}\nu_q+\sum_{q=1}^{\tilde{r}}\tilde{\nu}_q)$ real or complex numbers of magnitude less than one. Therefore, when $\sum_{q\in Q_2}\nu_q+\sum_{q=1}^{\tilde{r}}\tilde{\nu}_q$ diverges with time, the corresponding terms in Eq.~(\ref{Y_pi_CUE_general}) decay exponentially with time. When $\sum_{q\in Q_2}\nu_q+\sum_{q=1}^{\tilde{r}}\tilde{\nu}_q$ remains finite at long times, the corresponding terms in Eq.~(\ref{Y_pi_CUE_general}) also remain finite. Thus, this case includes both Type \Rom{2} and Type \Rom{3} terms. The above analysis can be extended to demonstrate that \textbf{Theorem 2} is also valid for $\mathcal{Y}_\pi^{\{\underline{n},\underline{n}\}}$.
	
	\subsection{$\mathcal{T}^2=-1$}
	We here take a permutation $\pi$ and a configuration $\vec{\sigma}$, whose diagram contains $r_o$ number of red arcs on the outer circle with counterclockwise arrows and lengths $\nu_1,...,\nu_{r_o}$, $\tilde{r}_o$ number of red arcs on the outer circle with clockwise arrows and lengths $\tilde{\nu}_1,...,\tilde{\nu}_{\tilde{r}_o}$, $r_i$ number of red arcs on the inner circle with counterclockwise arrows and lengths $\xi_1,...,\xi_{r_i}$, and $\tilde{r}_i$ number of red arcs on the inner circle with clockwise arrows and lengths $\tilde{\xi}_1,...,\tilde{\xi}_{\tilde{r}_i}$. Following the rules in Sec.~\ref{CSE_rules}, the contribution to SFF can be written  following Eq.~(\ref{Z_pi_CSE_example}) as
	\begin{align}
		\mathcal{Z}_{\pi,\vec{\sigma}}&=\sum_{\{i_q\}}\sum_{\{\tilde{i}_q\}}\sum_{\{j_q\}}\sum_{\{\tilde{j}_q\}}\notag\\
		&\times\left(\prod_{q=1}^{r_o}\lambda_{i_q}^{\nu_q}\right)\left(\prod_{q=1}^{\tilde{r}_o}\chi_{\tilde{i}_q}^{\tilde{\nu}_q}\right)\left(\prod_{q=1}^{r_i}\chi_{j_q}^{\xi_q}\right)\left(\prod_{q=1}^{\tilde{r}_i}\lambda_{\tilde{j}_q}^{\tilde{\xi}_q}\right)\notag\\
		&\times Q^{i_1...i_{r_o};\tilde{i}_1...\tilde{i}_{\tilde{r}_o};j_1...j_{r_i};\tilde{j}_1...\tilde{j}_{\tilde{r}_i}}_\pi,
		\label{Z_pi_CSE_general}
	\end{align}
	where the sum of lengths of all the red arcs satisfy
	\begin{align}
		\sum_{q=1}^{r_o}\nu_q+\sum_{q=1}^{\tilde{r}_o}\tilde{\nu}_q+\sum_{q=1}^{r_i}\xi_q+\sum_{q=1}^{\tilde{r}_i}\tilde{\xi}_i=t-\delta.
		\label{CSE_arc_len_sum}
	\end{align}
	Similar to the case without $\mathcal{T}$-symmetry in Eq.~(\ref{Y_pi_CUE_general}), the different terms on the right-hand side of Eq.~(\ref{Z_pi_CSE_general}) can be categorized as follows:\\
	\textbf{Case 1}: $\tilde{r}_o=0$, $r_i=0$, and $i_q=0$ and $j_q=0$, $\forall q$. The right-hand side of Eq.~(\ref{Z_pi_CSE_general}) gives the term $Q^{0...0;;;0...0}_\pi$. Similar to $\mathcal{T}^2=1$ and without $\mathcal{T}$-symmetry cases, typically $Q^{0...0;;;0...0}_\pi=1/\mathcal{N}^\mu$, where $\mu$ is a non-negative integer. Therefore, this is a Type \Rom{1} term. If $Q^{0...0;;;0...0}_\pi$ can not be simplified to a form independent of Hamiltonian parameters, it is a Type \Rom{2} term.\\
	\textbf{Case 2}: $i_q\neq 0$ and $j_q\neq 0$, $\forall q$ or $r_o=\tilde{r}_i=0$. The factor $\left(\prod_{q=1}^{r_o}\lambda_{i_q}^{\nu_q}\right)\left(\prod_{q=1}^{\tilde{r}_o}\chi_{\tilde{i}_q}^{\tilde{\nu}_q}\right)\left(\prod_{q=1}^{r_i}\chi_{j_q}^{\xi_q}\right)\left(\prod_{q=1}^{\tilde{r}_i}\lambda_{\tilde{j}_q}^{\tilde{\xi}_q}\right)$ can be interpreted as a product of $(t-\delta)$ real or complex numbers of magnitude less than one. Therefore, the corresponding terms fall exponentially with time. Consequently, such terms are Type \Rom{3}.\\ 
	\textbf{Case 3}: All the remaining terms have at least one $i_q=0$ or at least one $\tilde{j}_q=0$, and if $\tilde{r}_o=r_i=0$ then at least one $i_q\neq 0$ or at least one $\tilde{j}_q\neq 0$. Therefore, the set $\{1,...,r_o\}$ can be divided into two subsets $Q_{o,1}=\{q|i_q=0\}$ and $Q_{o,2}=\{q|i_q\neq 0\}$. Similarly, the set $\{1,...,\tilde{r}_i\}$ can be divided into two subsets $Q_{i,1}=\{q|\tilde{j}_q=0\}$ and $Q_{i,2}=\{q|\tilde{j}_q\neq 0\}$. Thus, we have
	\begin{align}
		&\left(\prod_{q=1}^{r_o}\lambda_{i_q}^{\nu_q}\right)\left(\prod_{q=1}^{\tilde{r}_o}\chi_{\tilde{i}_q}^{\tilde{\nu}_q}\right)\left(\prod_{q=1}^{r_i}\chi_{j_q}^{\xi_q}\right)\left(\prod_{q=1}^{\tilde{r}_i}\lambda_{\tilde{j}_q}^{\tilde{\xi}_q}\right)\notag\\
		&=\left(\prod_{q\in Q_{o,1}}\lambda_{0}^{\nu_q}\prod_{q\in Q_{o,2}}\lambda_{i_q}^{\nu_q}\right)\left(\prod_{q=1}^{\tilde{r}_o}\chi_{\tilde{i}_q}^{\tilde{\nu}_q}\right)\left(\prod_{q=1}^{r_i}\chi_{j_q}^{\xi_q}\right)\notag\\
		&\quad\times\left(\prod_{q\in Q_{i,1}}\lambda_{0}^{\tilde{\xi}_q}\prod_{q\in Q_{i,2}}\lambda_{\tilde{j}_q}^{\tilde{\xi}_q}\right)\notag\\
		&=\left(\prod_{q\in Q_{o,2}}\lambda_{i_q}^{\nu_q}\right)\left(\prod_{q=1}^{\tilde{r}_o}\chi_{\tilde{i}_q}^{\tilde{\nu}_q}\right)\left(\prod_{q=1}^{r_i}\chi_{j_q}^{\xi_q}\right)\left(\prod_{q\in Q_{i,2}}\lambda_{\tilde{j}_q}^{\tilde{\xi}_q}\right).
		\label{Z_pi_CSE_prod_general}
	\end{align}
	The right-hand side in Eq.~(\ref{Z_pi_CSE_prod_general}) can be interpreted as a product of $t'$ real or complex numbers of magnitude less than one, where
	\begin{align}
		t'=\sum_{q\in Q_{o,2}}\nu_q+\sum_{q=1}^{\tilde{r}_o}\tilde{\nu}_q+\sum_{q=1}^{r_i}\xi_q+\sum_{q\in Q_{i,2}}\tilde{\xi}_q.
	\end{align}
	According to Eq.~(\ref{CSE_arc_len_sum}), $t'$ can either diverge or remain finite at long times, resulting in Type \Rom{3} and Type \Rom{2} terms, respectively. Thus, $\mathcal{Z}_{\pi,\vec{\sigma}}$ can be expressed as sum of Type \Rom{1}, Type \Rom{2}, and Type \Rom{3} terms in accordance with \textbf{Theorem 2}. A similar analysis indicates that \textbf{Theorem 2} is also valid for $\mathcal{Z}_{\pi,\vec{\sigma}}^{\{\underline{n},\underline{n}\}}$ and $\mathcal{Z}_{\pi,\vec{\sigma}}^{\{\underline{n},\mathcal{T}\underline{n}\}}$.
	
	The above analysis for all three cases of $\mathcal{T}$-symmetry provides further insights into the contribution of different diagrams. More specifically, red arcs whose length diverges with increasing time lead to Type \Rom{3} terms, whereas red arcs whose length remains finite with increasing time lead to Type \Rom{2} terms.
	
	\section{Reduced Diagrams}
	\label{reduced_diagram}
	Computing $\mathcal{X}_\pi$, $\mathcal{X}_\pi^{\{\underline{n},\underline{n}\}}$, $\mathcal{Y}_\pi$, $\mathcal{Y}_\pi^{\{\underline{n},\underline{n}\}}$, $\mathcal{Z}_{\pi,\vec{\sigma}}$, $\mathcal{Z}_{\pi,\vec{\sigma}}^{\{\underline{n},\underline{n}\}}$, and $\mathcal{Z}_{\pi,\vec{\sigma}}^{\{\underline{n},\mathcal{T}\underline{n}\}}$ for all permutations using the rules in Secs.~\ref{COE_Rules}, \ref{CUE_rules}, and \ref{CSE_rules} is practically impossible. However, \textbf{Theorem 2} has shown a general pattern in each of them. More specifically, each of them can be expressed as a sum of a universal Type \Rom{1} term, and nonuniversal Type \Rom{2} and Type \Rom{3} terms. Since Type \Rom{3} terms decay exponentially with time, they don't determine the SFF beyond $t^*$. Therefore, the Type \Rom{1} and Type \Rom{2} terms are only significant for studying emergence of RMT behavior beyond $t^*$. With this motivation, we now define a reduced diagram. In the proof of \textbf{Theorem 2} in Sec.~\ref{theorem_2}, \textbf{Case 2} terms are all Type \Rom{3} in each case of $\mathcal{T}$-symmetry. A reduced diagram eliminates \textbf{Case 2} terms. A precise definition of a reduced diagram is different for $\mathcal{T}^2=1$, in the absence of $\mathcal{T}$-symmetry, and $\mathcal{T}^2=-1$ cases. Thus, we present them explicitly for each case.\\
	
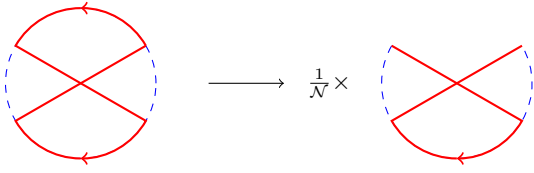
\begin{figure}[h]
		\centering
		\begin{tikzpicture}
			\def\rad{1}
			\draw[blue,dashed] (0,0) circle(\rad);
			\draw[red,thick,->] (30:\rad) arc(30:90:\rad);\draw[red,thick,->] (90:\rad) arc(90:150:\rad)--(-30:\rad) arc(-30:-90:\rad);\draw[red,thick] (-90:\rad) arc(-90:-150:\rad)--(30:\rad);
			\draw[->] (1.7,0)--(2.7,0);
			\begin{scope}[shift={(5,0)}]
				\draw[blue,dashed] (150:\rad) arc(150:390:\rad);
				\draw[red,thick,->] (150:\rad)--(-30:\rad) arc(-30:-90:\rad);\draw[red,thick] (-90:\rad) arc(-90:-150:\rad)--(30:\rad);
				\draw (180:{1.7*\rad}) node[]{$\frac{1}{\mathcal{N}}\times$};
			\end{scope}
		\end{tikzpicture}
		\caption{\justifying\small A diagram and the corresponding reduced diagram obtained by inserting a factor of $1/\mathcal{N}$ for the upper arc with a counterclockwise arrow. The lower arc has a clockwise arrow and can not be removed to get a reduced diagram.}
		\label{Example_red_diag_CUE}
	\end{figure}

	\noindent\boldsymbol{$\mathcal{T}^2=1$}: 
	In Eq.~(\ref{X_pi_COE_r_arcs}), \textbf{Case 2} terms can be eliminated by fixing any one of $i_{q'}=0$ while allowing $i_q\in\{0,1,...,\mathcal{N}-1\}$ when $q\neq q'$. This is diagrammatically equivalent to inserting a factor of $\lambda_0^{\nu_{q'}}\mathcal{M}^{(0)}_{a,b} (=1/\mathcal{N})$ for a red arc of length $\nu_{q'}$ as in Fig.~\ref{Example_red_diag_COE}, instead of $\sum_{i_{q'}}\lambda_{i_{q'}}^{\nu_{q'}}\mathcal{M}^{(i_{q'})}_{a,b}$ as mentioned in Sec.~\ref{COE_Rules}.
	\begin{figure}[h]
		\centering
		\begin{tikzpicture}
			\def\rad{1}
			\draw[blue,dashed] (0,0) circle(\rad);
			\draw[red,thick,->] (30:\rad) arc(30:90:\rad);\draw[red,thick,->] (90:\rad) arc(90:150:\rad)--(0:\rad)--(210:\rad) arc(210:270:\rad);\draw[red,thick] (270:\rad) arc(270:330:\rad)--(180:\rad)--(30:\rad);
			\draw[->] (1.7,0)--(2.7,0);
			\begin{scope}[shift={(5,0)}]
				\draw[blue,dashed] (150:\rad) arc(150:390:\rad);
				\draw[red,thick,->] (150:\rad)--(0:\rad)--(210:\rad) arc(210:270:\rad);\draw[red,thick] (270:\rad) arc(270:330:\rad)--(180:\rad)--(30:\rad);
				\draw (180:{1.7*\rad}) node[]{$\frac{1}{\mathcal{N}}\times$};
			\end{scope}
		\end{tikzpicture}
		\caption{\justifying\small A diagram and the corresponding reduced diagram obtained by inserting a factor of $1/\mathcal{N}$ for the upper arc. The reduced diagram eliminates some of the Type \Rom{3} terms, which do not contribute to the SFF beyond $t^*$.}
		\label{Example_red_diag_COE}
	\end{figure}
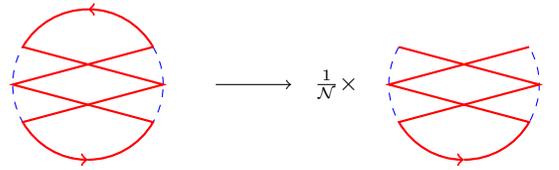\\
	\noindent\textbf{Absence of} $\boldsymbol{\mathcal{T}}$\textbf{-symmetry}: In Eq.~(\ref{Y_pi_CUE_general}), \textbf{Case 2} terms can be eliminated by fixing any one of $i_{q'}=0$ while allowing $i_q\in\{0,1,...,\mathcal{N}-1\}$ when $q\neq q'$. This is diagrammatically equivalent to inserting a factor of $\lambda_0^{\nu_{q'}}\mathcal{M}^{(0)}_{a,b} (=1/\mathcal{N})$ for a red arc with a counterclockwise arrow and length $\nu_{q'}$ as in Fig.~\ref{Example_red_diag_CUE}, instead of $\sum_{i_{q'}}\lambda_{i_{q'}}^{\nu_{q'}}\mathcal{M}^{(i_{q'})}_{a,b}$ as given in Sec.~\ref{CUE_rules}.
	\\
	\noindent\boldsymbol{$\mathcal{T}^2=-1$}: In Eq.~(\ref{Z_pi_CSE_general}), \textbf{Case 2} terms can be eliminated by fixing any one of $i_{q'}=0$ while allowing $i_q\in\{0,1,...,\mathcal{N}-1\}$ when $q\neq q'$ or fixing any one of $\tilde{j}_{q'}=0$ while allowing $\tilde{j}_q\in\{0,1,...,\mathcal{N}-1\}$ when $q\neq q'$. This is diagrammatically equivalent to inserting a factor of $\lambda_0^{\nu_{q'}}\mathcal{M}^{(0)}_{a,b}(=1/\mathcal{N})$ for a red arc with a counterclockwise arrow and length $\nu_{q'}$ on the outer circle or $\lambda_0^{\tilde{\xi}_{q'}}\mathcal{M}^{(0)}_{a,b}(=1/\mathcal{N})$ for a red arc with a clockwise arrow and length $\tilde{\xi}_{q'}$ on the inner circle as shown in Fig.~\ref{Example_red_diag_CSE}, instead of $\sum_{i_{q'}}\lambda_{i_{q'}}^{\nu_{q'}}\mathcal{M}^{(i_{q'})}_{a,b}$ and $\sum_{\tilde{j}_{q'}}\lambda_{\tilde{j}_{q'}}^{\tilde{\xi}_{q'}}\mathcal{M}^{(\tilde{j}_{q'})}_{a,b}$, respectively, as mentioned in Sec.~\ref{CSE_rules}.
	\begin{figure}[h]
		\centering
		\begin{tikzpicture}
			\def\radib{1}
			\def\radis{0.75}
			\draw[blue,dashed] (0,0) circle(\radib);
			\draw[black,dashed] (0,0) circle(\radis);
			\draw[red,thick,->] (30:\radib) arc(30:90:\radib);\draw[red,thick,->] (90:\radib) arc(90:150:\radib)--(-30:\radis) arc(-30:-90:\radis);\draw[red,thick] (-90:\radis) arc(-90:-150:\radis)--(30:\radib);
			\draw[->] (1.7,0)--(2.7,0);
			\begin{scope}[shift={(5,0)}]
				\draw[blue,dashed] (150:\radib) arc(150:390:\radib);
				\draw[black,dashed] (150:\radis) arc(150:390:\radis);
				\draw[red,thick,->] (150:\radib)--(-30:\radis) arc(-30:-90:\radis);\draw[red,thick] (-90:\radis) arc(-90:-150:\radis)--(30:\radib);
				\draw (180:{1.7*\radib}) node[]{$\frac{1}{\mathcal{N}}\times$};
			\end{scope}
		\end{tikzpicture}
		\caption{\justifying \small A diagram and a corresponding reduced diagram obtained by inserting a factor of $1/\mathcal{N}$ for the upper arc with counterclockwise arrow on the outer circle. The lower arc is present on the inner circle and has a clockwise arrow, therefore, unlike the $\mathcal{T}$ absent case, Fig.~\ref{Example_red_diag_CUE}, a reduced diagram can be obtained by removing this arc as well.}
		\label{Example_red_diag_CSE}
	\end{figure}
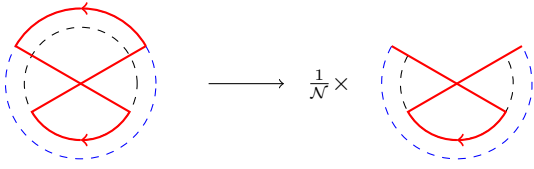\\
	
	The reduced diagrams have some natural properties as follows:\\
	\textbf{Property 1}: When $\mathcal{T}^2=1$, a diagram containing $r$ number of red arcs has $r$ different reduced diagrams. In the absence of $\mathcal{T}$-symmetry, a diagram containing $r$ number of red arcs with counterclockwise arrows has $r$ different reduced diagrams. When $\mathcal{T}^2=-1$, a diagram containing $r_o$ number of red arcs with counterclockwise arrows on the outer circle and $\tilde{r}_i$ number of red arcs with clockwise arrows on the inner circle has $r_o+\tilde{r}_i$ different reduced diagrams.\\
	\textbf{Property 2}: If Type \Rom{1} term exists then each reduced diagram gives the same Type \Rom{1} term when evaluated using the rules in Secs.~\ref{COE_Rules}-\ref{CSE_rules}.\\
	\textbf{Property 3}: A reduced diagram does not give Type \Rom{2} terms if the size of each arc increases with time.
	
\begin{figure}[H]
		\centering
		\begin{subfigure}[t]{0.4\linewidth}
			\centering
			\begin{tikzpicture}
				\def\radis{0.75}
				\def\radib{1}
				\draw[blue,dashed] (60:\radib) arc(60:180:\radib);
				\draw[black,dashed] (60:\radis) arc(60:180:\radis);
				\draw[red,thick,->] (60:\radib) arc(60:120:\radib);\draw[red,thick] (120:\radib) arc(120:150:\radib)--(180:\radis);
				\draw (150:\radib) node[above left]{$a$};
				\draw (180:\radib) node[left]{$b$};\draw (180:\radis) node[right]{$\mathcal{T} b$};
			\end{tikzpicture}
			\caption{}
			\label{CSE_portion_of_a_diag}
		\end{subfigure}
		\hfill
		\begin{subfigure}[t]{0.5\linewidth}
			\centering
			\begin{tikzpicture}
				\def\radis{0.75}
				\def\radib{1}
				\draw[blue,dashed] (150:\radib) arc(150:180:\radib);
				\draw[black,dashed] (60:\radis) arc(60:180:\radis);
				\draw[red,thick] (150:\radib)--(180:\radis);
				\draw (150:\radib) node[above left]{$a$};
				\draw (180:\radib) node[left]{$b$};\draw (180:\radis) node[right]{$\mathcal{T} b$};
				\draw (180:{1.5*\radib}) node[left]{$\frac{1}{\mathcal{N}}\times$};
				\draw (0:{1.2*\radib}) node[]{$\propto \delta_{b,\mathcal{T}b}=0$};
			\end{tikzpicture}
			\caption{}
			\label{CSE_red_of_portion}
		\end{subfigure}
		\caption{\justifying\small (a) Portion of an allowed diagram containing a red arc with a counterclockwise arrow on the outer circle and a jump to the inner circle at one end of the arc. (b) The reduced diagram with respect to this arc has a vanishing contribution since $b$ and $\mathcal{T}b$ are orthogonal.}
		\label{CSE_red_diag_vanish_thm}
	\end{figure}
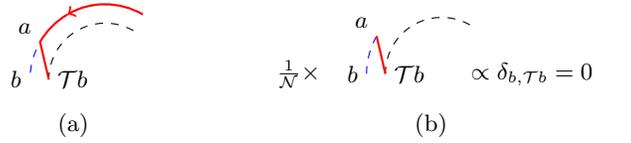

	When $\mathcal{T}^2=-1$, the basis states related by time reversal are orthogonal to each other as described in Tab.~\ref{T_const}. This property results in the following theorem for the corresponding reduced diagrams.\\\\	
	\noindent\textbf{Theorem 5}: For a diagram, containing a red arc with a counterclockwise arrow on the outer circle or a clockwise arrow on the inner circle such that the arc makes a jump to the other circle at one end, the reduced diagram with respect to the same arc has a vanishing contribution.\\\\
	\noindent\textbf{Proof}: Fig.~\ref{CSE_portion_of_a_diag} shows a portion of an arbitrary diagram containing a red arc with a counterclockwise arrow on the outer circle. The arc follows the outer circle until state $a$, then instead of connecting the next state $b$ on the same circle, it connects to the time reversed state $\mathcal{T}b$ on the inner circle. A reduced diagram can be obtained by removing this arc and inserting a factor of $1/\mathcal{N}$ as in Fig.~\ref{CSE_red_of_portion}. Applying the rules in Sec.~\ref{CSE_rules} to the blue and red line segments connected to the state $a$ and summing over $a$ implies that the reduced diagram has a vanishing contribution.
	
	In the other case, when arc is on the inner circle, it can be pushed to the outer circle by applying \textbf{Theorem 1} as in Fig.~\ref{CSE_corollary}. Following the arguments presented above, the reduced diagram vanishes once again.
	\begin{figure}[H]
		\centering
		\begin{tikzpicture}
			\def\radis{0.75}
			\def\radib{1}
			\draw[blue,dashed] (60:\radib) arc(60:180:\radib);
			\draw[black,dashed] (60:\radis) arc(60:180:\radis);
			\draw[red,thick] (60:\radis) arc(60:120:\radis);\draw[red,thick,<-] (120:\radis) arc(120:150:\radis)--(180:\radib);
			\draw (150:\radib) node[above left]{$a$};
			\draw (180:\radib) node[below left]{$b$};\draw (180:\radis) node[below right]{$\mathcal{T} b$};
			\begin{scope}[shift={(1.5cm,0)}]
				\draw (0,0.25cm) node[]{Theorem 1};
				\draw[->] (-1,0)--(1,0);
			\end{scope}
			\begin{scope}[shift={(4cm,0)}]
				\draw[blue,dashed] (60:\radib) arc(60:180:\radib);
				\draw[black,dashed] (60:\radis) arc(60:180:\radis);
				\draw[red,thick,->] (60:\radib) arc(60:120:\radib);\draw[red,thick] (120:\radib) arc(120:150:\radib)--(180:\radis);
				\draw (150:\radib) node[above left]{$a$};
				\draw (180:\radib) node[left]{$b$};\draw (180:\radis) node[right]{$\mathcal{T} b$};
			\end{scope}
		\end{tikzpicture}
		\caption{}
		\label{CSE_corollary}
	\end{figure}
	\section{Exact cancellation of reduced diagrams}
	\label{Exact_cancellation_of_reduced_diagrams}
	In Eq.~(\ref{SFF_RPA_X_pi}), $\mathcal{X}_\pi$ denotes the contribution of a diagram, where the states at different time steps are considered different even though there is an implicit repetition of states. However, $\mathcal{X}_\pi^{\{\underline{n},\underline{n}\}}$ represents the contribution of a diagram, where we explicitly consider a state appearing at two different time steps. We find that the two kinds of diagrams have identical reduced diagrams. The same is true for the other two cases of $\mathcal{T}$-symmetry. Since $K_1(t)$ in Eq.~(\ref{SFF_RPA_X_pi}) is a difference between the contribution of these two kinds of diagrams $(\mathcal{X}_\pi,\mathcal{X}_\pi^{\{\underline{n},\underline{n}\}})$, the identical reduced diagrams cancel each other. Consequently, their Type \Rom{1} and Type \Rom{2} contributions also cancel. This is a crucial detail for finding the diagrams responsible for the universal RMT behavior. We already presented the precise statement for cancellation of reduced diagrams as \textbf{Theorem 3} and \textbf{Theorem 4} in Sec.~\ref{Models_and_Observables}. In the following subsections, we present their proofs.
	\subsection{$\mathcal{T}^2=1$ and absence of $\mathcal{T}$-symmetry}
	\noindent\textbf{Proof of Theorem 3}: In an arbitrary diagram, e.g., in Fig.~\ref{Example_gen_diag_COE}, a red arc starts from a state $d$ following the blue circle up to another state $a$, and then connects to a state $c$, other than the immediate next state $b$ on the blue circle. The rest of the red curve is irrelevant for the proof. A reduced diagram can be obtained by removing this arc and inserting a factor of $1/\mathcal{N}$ as given in Fig.~\ref{Example_gen_diag_COE}. We apply the rules in Sec.~\ref{COE_Rules} or Sec.~\ref{CUE_rules} to the blue and red line segments, respectively,  between the states $a$ and $b$ and the states $a$ and $c$, and sum over $a$ to find
	\begin{align}
		\mathcal{X}_{\pi}(\text{reduced})&\propto V_{a,b}V^*_{a,c}=\delta_{b,c}.
	\end{align}
	Thus, the states $b$ and $c$ are identical in the reduced diagram in Fig.~\ref{Example_gen_diag_COE}. It suggests that an original diagram in Fig.~\ref{Example_rep_gen_diag_COE} with identical states $b$ and $c$, whose contribution is denoted by $\mathcal{X}_\pi^{\{\underline{n},\underline{n}\}}$, also has a similar reduced diagram with respect to the same arc. This diagram in the left of Fig.~\ref{Example_rep_gen_diag_COE} represents a permutation of $t$ states, two of which are identical.
	\begin{figure}[H]
		\centering
		\begin{tikzpicture}
			\def\rad{1}
			\draw[blue,dashed] (0,0) circle(\rad);
			\draw[red,thick,->] (30:\rad) arc(30:90:\rad);\draw[red,thick] (90:\rad) arc(90:150:\rad)--(-30:\rad);
			\draw (150:\rad) node[above left]{$a$};
			\draw (180:\rad) node[left]{$b$};
			%\draw (210:\rad) node[below left]{$c$};
			\draw (-30:\rad) node[below right]{$c$};
			%\draw (0:\rad) node[right]{$e$};
			\draw (30:\rad) node[above right]{$d$};
                        \draw (270:{\rad+0.2}) node[below]{(a)};
			\draw[->] (1.7,0)--(2.7,0);
			\begin{scope}[shift={(5,0)}]
				\draw[blue,dashed] (150:\rad) arc(150:390:\rad);
				\draw[red,thick] (150:\rad)--(-30:\rad);
				\draw (150:\rad) node[above left]{$a$};
				\draw (180:\rad) node[left]{$b$};
				\draw (-30:\rad) node[below right]{$c$};
				\draw (30:\rad) node[above right]{$d$};
				\draw (180:{1.7*\rad}) node[]{$\frac{1}{\mathcal{N}}\times$};
                                \draw (270:{\rad+0.2}) node[below]{(b)};
			\end{scope}
		\end{tikzpicture}
		\caption{\justifying\small A diagram with only a portion of the red path emphasizing the general structure at one end of a red arc along with a reduced diagram. A sum over the state $a$ implies that the states $b$ and $c$ are identical in the reduced diagram.}
		\label{Example_gen_diag_COE}
		\begin{tikzpicture}
			\def\rad{1}
			\draw[blue,dashed] (0,0) circle(\rad);
			\draw[red,thick,->] (30:\rad) arc(30:90:\rad);\draw[red,thick] (90:\rad) arc(90:150:\rad)--(-30:\rad);
			\draw (150:\rad) node[above left]{$a$};
			\draw[fill=green] (180:\rad) circle(1.5pt) node[left]{$b$};
			%\draw (210:\rad) node[below left]{$c$};
			\draw[fill=green] (-30:\rad) circle(1.5pt) node[below right]{$b$};
			%\draw (0:\rad) node[right]{$e$};
			\draw (30:\rad) node[above right]{$d$};
			\draw[->] (1.7,0)--(2.7,0);
			\begin{scope}[shift={(5,0)}]
				\draw[blue,dashed] (150:\rad) arc(150:390:\rad);
				\draw[red,thick] (150:\rad)--(-30:\rad);
				\draw (150:\rad) node[above left]{$a$};
				\draw[fill=green] (180:\rad) circle(1.5pt) node[left]{$b$};
				\draw[fill=green] (-30:\rad) circle(1.5pt) node[below right]{$b$};
				\draw (30:\rad) node[above right]{$d$};
				\draw (180:{1.7*\rad}) node[]{$\frac{1}{\mathcal{N}}\times$};
			\end{scope}
		\end{tikzpicture}
		\caption{\justifying \small A diagram with the states $b$ and $c$ of the diagram in Fig.~\ref{Example_gen_diag_COE}a being identical from the beginning. This diagram also has a reduced diagram identical to that in Fig.~\ref{Example_gen_diag_COE}b.}
		\label{Example_rep_gen_diag_COE}
	\end{figure}
	\subsection{$\mathcal{T}^2=-1$}
	\noindent\textbf{Proof of Theorem 4}: In an arbitrary diagram, a red arc with a counterclockwise arrow on the outer circle starts from a state $d$ following the blue path up to another state $a$, and then connects to a state $c$ on the outer circle as in Fig.~\ref{Example_gen_diag_no_rep_CSE} or a state $\mathcal{T}c$ on the inner circle as in Fig.~\ref{CSE_c_inner}, other than the immediate next state $b$ on the outer circle. When state $c$ is connected as in Fig.~\ref{Example_gen_diag_no_rep_CSE}, an analysis similar to the proof of \textbf{Theorem 3} implies that there exists an identical diagram with the state $c$ same as the state $b$ as shown in Fig.~\ref{Example_gen_diag_rep_CSE}. These two diagrams in Figs.~\ref{Example_gen_diag_no_rep_CSE},\ref{Example_gen_diag_rep_CSE} have identical reduced diagrams. When the state $\mathcal{T}c$ is connected in the diagram in Fig.~\ref{CSE_c_inner}, another diagram with an identical reduced diagram has a state and its time reversed state at two different time steps on the outer circle as in Fig.~\ref{CSE_b_Tb}.
	\begin{figure}[H]
		\centering
		\begin{subfigure}[t]{0.45\linewidth}
			\centering
			\begin{tikzpicture}
				\def\rad{1}
				\def\radis{0.75}
				\draw[blue,dashed] (0,0) circle(\rad);
				\draw[black,dashed] (0,0) circle(\radis);
				\draw[red,thick,->] (30:\rad) arc(30:90:\rad);\draw[red,thick] (90:\rad) arc(90:150:\rad)--(-30:\rad);
				\draw (150:\rad) node[above left]{$a$};
				\draw (180:\rad) node[left]{$b$};
				%\draw (210:\rad) node[below left]{$c$};
				\draw (-30:\rad) node[below right]{$c$};
				%\draw (0:\rad) node[right]{$e$};
				\draw (30:\rad) node[above right]{$d$};
			\end{tikzpicture}
			\caption{}
			\label{Example_gen_diag_no_rep_CSE}
		\end{subfigure}
		\hfill
		\begin{subfigure}[t]{0.45\linewidth}
			\centering
			\begin{tikzpicture}
				\def\rad{1}
				\def\radis{0.75}
				\draw[blue,dashed] (0,0) circle(\rad);
				\draw[black,dashed] (0,0) circle(\radis);
				\draw[red,thick,->] (30:\rad) arc(30:90:\rad);\draw[red,thick] (90:\rad) arc(90:150:\rad)--(-30:\rad);
				\draw (150:\rad) node[above left]{$a$};
				\draw[fill=green] (180:\rad) circle(1.5pt) node[left]{$b$};
				%\draw (210:\rad) node[below left]{$c$};
				\draw[fill=green] (-30:\rad) circle(1.5pt) node[below right]{$b$};
				%\draw (0:\rad) node[right]{$e$};
				\draw (30:\rad) node[above right]{$d$};
			\end{tikzpicture}
			\caption{}
			\label{Example_gen_diag_rep_CSE}
		\end{subfigure}
		\caption{}
	\end{figure}
	\begin{figure}[h]
		\centering
		\begin{subfigure}[t]{0.45\linewidth}
			\centering
			\begin{tikzpicture}
				\def\rad{1}
				\def\radis{0.75}
				\draw[blue,dashed] (0,0) circle(\rad);
				\draw[black,dashed] (0,0) circle(\radis);
				\draw[red,thick,->] (30:\rad) arc(30:90:\rad);\draw[red,thick] (90:\rad) arc(90:150:\rad)--(-30:\radis);
				\draw (150:\rad) node[above left]{$a$};
				\draw (180:\rad) node[left]{$b$};
				%\draw (210:\rad) node[below left]{$c$};
				\draw (-30:{\rad}) node[below right]{$c$};
				%\draw (0:\rad) node[right]{$e$};
				\draw (30:\rad) node[above right]{$d$};
			\end{tikzpicture}
			\caption{}
			\label{CSE_c_inner}
		\end{subfigure}
		\hfill
		\begin{subfigure}[t]{0.45\linewidth}
			\centering
			\begin{tikzpicture}
				\def\rad{1}
				\def\radis{0.75}
				\draw[blue,dashed] (0,0) circle(\rad);
				\draw[black,dashed] (0,0) circle(\radis);
				\draw[red,thick,->] (30:\rad) arc(30:90:\rad);\draw[red,thick] (90:\rad) arc(90:150:\rad)--(-30:\radis);
				\draw (150:\rad) node[above left]{$a$};
				\draw[fill=green] (180:\rad) circle(1.5pt) node[left]{$b$};
				\draw[fill=white] (-30:\rad) circle(1.5pt) node[below right]{$\mathcal{T}b$};
				\draw (30:\rad) node[above right]{$d$};
			\end{tikzpicture}
			\caption{}
			\label{CSE_b_Tb}
		\end{subfigure}
		\caption{}
		\label{Example_rep_gen_diag_CSE}
	\end{figure}
	
	\section{Leading-order SFF: Linear Ramp}
	\label{leading_order_SFF}
	In all three cases of $\mathcal{T}$-symmetry, the leading-order SFF is determined by the identity permutation and its variants. 
	\begin{figure*}[t!]
		\centering
		\includegraphics[width=\linewidth]{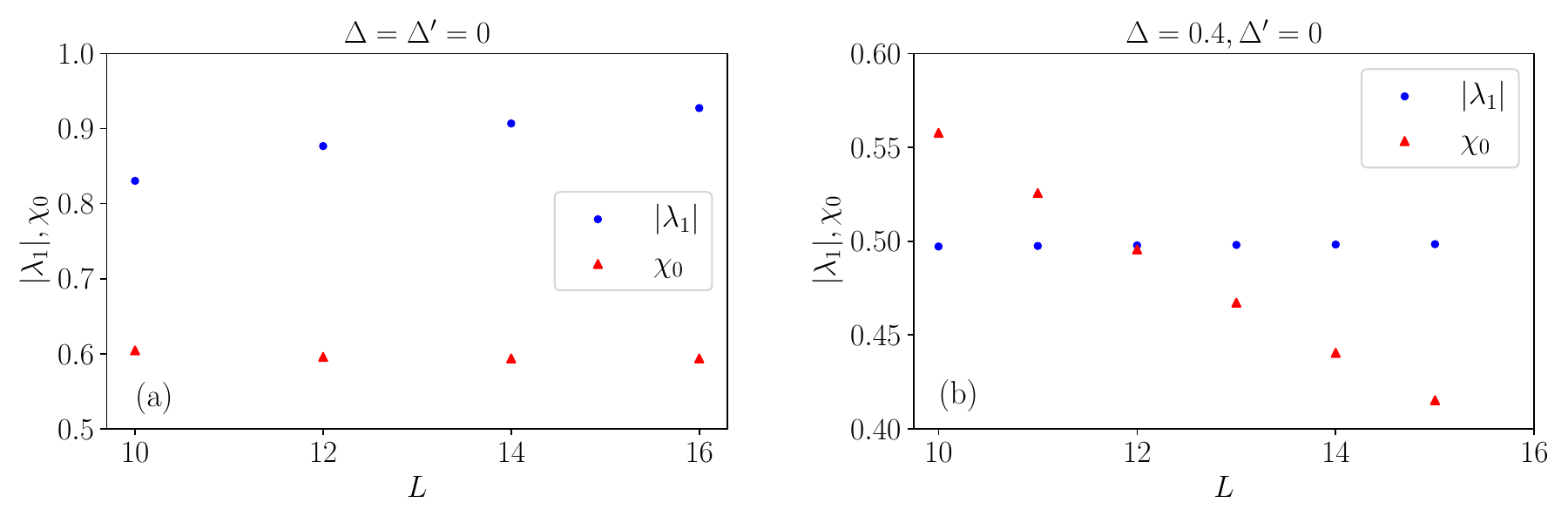}
		\caption{\justifying\small Second-largest eigenvalue $\lambda_1$ of $\mathcal{M}$ and largest eigenvalue $\chi_0$ of $\tilde{\mathcal{M}}$ for different system sizes $L$ for a kicked strongly interacting spinless fermionic chain in the absence of $\mathcal{T}$-symmetry, and (a) with $(\Delta=\Delta'=0)$ and (b) without $(\Delta=0.4,\Delta'=0)$ particle-number conservation. We use the Hamiltonian in Eq.~(\ref{Hs0}) with parameters in Eqs.~(\ref{CUE_hopping},\ref{CUE_pairing}). We apply PBC in real space, and take half filling for $\Delta=\Delta' = 0$. Here, $J=0.5$ and $g=0.25$.}
		\label{M_Second_vs_Mp_largest_eval}
	\end{figure*}
	\subsection{$\mathcal{T}^2=1$}
	For $\mathcal{T}$-invariant systems with $\mathcal{T}^2=1$, the identity permutation has $t$ cyclic and $t$ anticyclic variants, all of which have the same contribution due to PBC in time and the symmetric nature of the matrix $V$. Therefore,
	\begin{align}
		K_1^{(1)}(t)&=\sum_{l=0}^{t-1}\mathcal{X}_{\mathcal{C}^lI}+\mathcal{X}_{\mathcal{R}\mathcal{C}^lI}=2t\;\mathcal{X}_{I},
	\end{align}
	where the subscript $I$ denotes the identity permutation, which is described diagrammatically by overlapping blue and red circles. Therefore, we find by applying the rules in Sec.~\ref{COE_Rules}:
	\begin{align}
		\mathcal{X}_{I}&=\text{tr}\mathcal{M}^t,
	\end{align}
	which gives the leading-order SFF:
	\begin{align}
		K_1^{(1)}(t)&=2t\;\text{tr}\mathcal{M}^t=2t(1+\lambda_1^t+...+\lambda_{\mathcal{N}-1}^t).
		\label{SFF_COE_leading_order}
	\end{align}
	Since $|\lambda_i|<1$ for $i=1,...,\mathcal{N}-1$, we find at long times $K_1^{(1)}(t)\simeq 2t$, which is the COE result up to the first order in time. We notice that all the Type \Rom{3} terms are exponentially decaying in accordance with \textbf{Theorem 2}. The last Type \Rom{3} term to vanish is $\lambda_1^t$. This term determines $t^*$. We define $t^*$ as the solution of the equation:
	\begin{align}
		d_1\lambda_1^{t^*}\simeq\frac{1}{e},
		\label{Thouless_time_equation}
	\end{align}
	where $d_1$ is the degeneracy of eigenvalue $\lambda_1$. We find $t^*$ by solving Eq.~(\ref{Thouless_time_equation}):
	\begin{align}
		t^*\simeq\frac{\ln(d_1)+1}{|\ln\lambda_1|}.
		\label{Thouless_time}
	\end{align}
	Following \textcite{RoyPRE2020}, we study a periodically kicked strongly interacting spinless fermion chain in Eqs.~(\ref{H0},\ref{Hs0}) with the parameters in Eqs.~(\ref{nn_hopping},\ref{nn_pairing}) to determine $t^*$ in the presence and absence of a $U(1)$ symmetry. When $\Delta=0$, the Hamiltonians in Eqs.~(\ref{H0},\ref{Hs0}) commute with $\hat{N}$ in Tab.~\ref{Thouless_time_table}. Therefore, the model has $U(1)$ symmetry for $\Delta=0$. Our numerical study reveals that the matrix $\mathcal{M}$ is $SU(2)$ invariant, which implies the existence of $SU(2)$ multiplets, with descendant states in different particle-number sectors. The eigenvalue $\lambda_1$ is independent of the total number of particles. Therefore, the corresponding eigenstate is a descendant state. We can find $\lambda_1$ numerically for very large system sizes $L$ for one-particle sector. In thermodynamic limit of $L \to \infty$, we have
	\begin{align}
		\lambda_1\simeq 1-\frac{c_{\beta}}{L^2},
		\label{COE_lambda_1_U(1)_Sym}
	\end{align}
	where $c_{\beta}$ is a constant, which depends on hopping $J$ (see Tab.~\ref{Thouless_time_table}). Substituting Eq.~(\ref{COE_lambda_1_U(1)_Sym}) and $d_1=2$ (as our numerical study reveals that $\lambda_1$ is doubly degenerate) into Eq.~(\ref{Thouless_time}), we can find $t^*$ as \cite{RoyPRE2020, RoyPRE2022},
	\begin{align}
		t^*\propto L^2.
	\end{align}
	We can find $\lambda_1$ and $d_1$ analytically by mapping the matrix $\mathcal{M}$ to a many-body spin $1/2$ Hamiltonian in the Trotter regime of small parameters $|J|,|\Delta|\ll 1/\tau_p=1$ \cite{RoyPRE2020, RoyPRE2022}.
	\begin{align}
		\mathcal{M} &\simeq (1-c_1 L)\mathds{1}_\mathcal{N}+\sum_{x=1}^L\sum_\nu c_\nu s_x^\nu s_{x+1}^\nu,
		\label{mapped_M}
	\end{align}
	where $\nu=1,2,3$, $c_1=(|J|^2+|\Delta|^2)/2$, $c_2=c_3=(|J|^2-|\Delta|^2)/2$, and $s_x^\nu$ are the Pauli matrices at site $x$. In the presence of $U(1)$ symmetry when $\Delta=0$, $\mathcal{M}$ becomes an $XXX$-Heisenberg spin $1/2$ chain, which gives $\lambda_1\simeq 1-2\pi^2 J^2/L^2$ for pseudo-momenta $k=2\pi/L, 2\pi-2\pi/L$ in the thermodynamic limit. Consequently, $t^*\propto L^2$. When $\Delta\neq 0$, $U(1)$ symmetry is absent and the Eq.~(\ref{mapped_M}) becomes an $XXZ$-Heisenberg chain, which is well known to have a gapped spectrum. Therefore, $\lambda_1 \simeq \mathcal{O}(L^0)$ and $d_1=2$ due to reflection symmetry. Thus, $t^*\propto L^0$. The above system-size scalings of $t^*$ following a similar analysis were studied in \cite{RoyPRE2020}. However, two other regimes of parameters were missed in \cite{RoyPRE2020}. When $|J|=|\Delta|$, the Eq.~(\ref{mapped_M}) becomes an Ising Hamiltonian, leading to $\lambda_1=1-4J^2$ and $d_1=L(L+1)/2$. Thus, we have $t^*\propto \ln(L)$, which matches with the system-size scalings of $t^*$ for a local kicking model studied in \textcite{KosPRX2018}. Further, for $J=0$ but $\Delta\neq 0$, the Eq.~(\ref{mapped_M}) is equivalent to an $XXZ$-Heisenberg Hamiltonian with an anisotropy parameter $-1$. For even $L$, alternating site spins of such Hamiltonian can be rotated to transform $\mathcal{M}$ back to an $XXX$-Heisenberg Hamiltonian, which implies $t^*\propto L^2$. In this case, the conserved charge is $\hat{N}_s$ in Tab.~\ref{Thouless_time_table}. Thus, the model is still $U(1)$ symmetric, and we expect $t^* \propto L^2$. The above findings suggest that the various $L$-scalings of $\lambda_1$ and its degeneracy $d_1$ lead to different $L$-scaling of $t^*$. We here provide a unified description of such scalings.
    \begin{figure*}[t!]
		\centering
		\includegraphics[width=\linewidth]{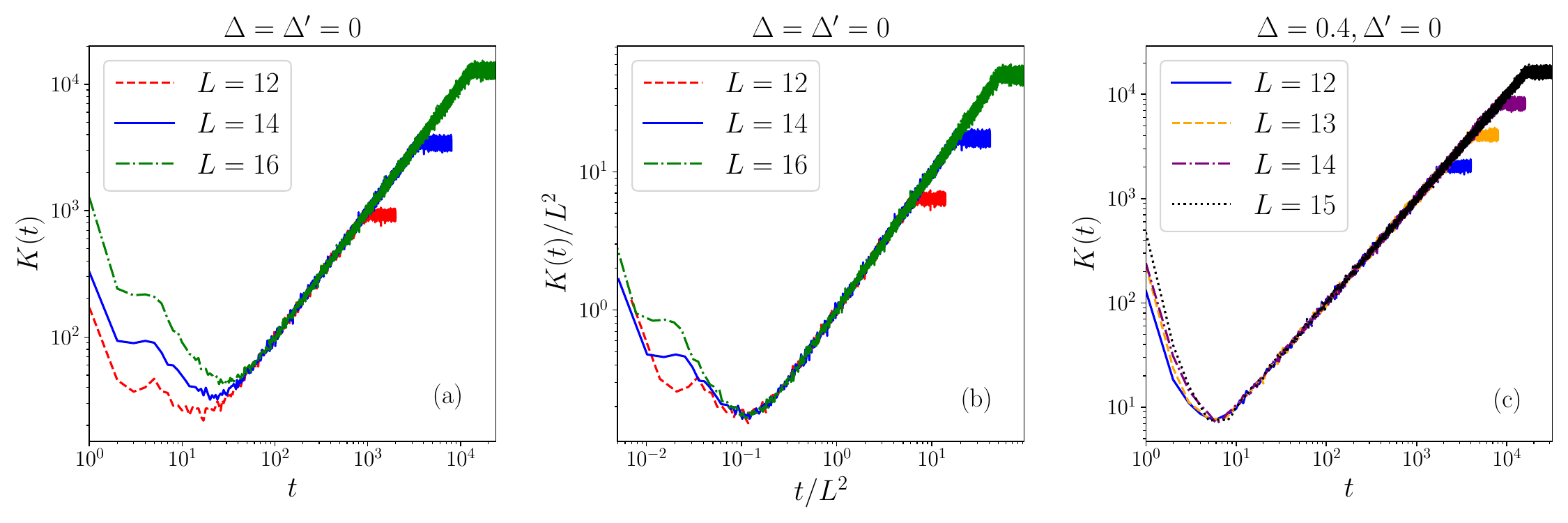}
		\caption{\justifying\small Spectral form factor $K(t)$ for different system sizes $L$ calculated numerically using Eq.~(\ref{SFF}) for a kicked strongly interacting spinless fermionic chain in the absence of $\mathcal{T}$-symmetry, and (a), (b) with $(\Delta=\Delta'=0)$ and (c) without $(\Delta=0.4,\Delta'=0)$ particle-number conservation. We use the Hamiltonians in Eqs.~(\ref{H0},\ref{Hs0}) with parameters in Eqs.~(\ref{CUE_hopping},\ref{CUE_pairing}). Here, $\Delta\epsilon=1,U_0=15,\alpha=1.5,J=0.5,g=0.25$. We apply PBC in real space, and take half filling $N/L = 1/2$ for $\Delta=\Delta' = 0$.  In (b), we show the data collapse in scaled time $t/L^2$. An averaging over 400 realizations of disorder is performed in each case.}
		\label{SFF_CUE_f_nnn_Direct}
	\end{figure*}
	\subsection{Absence of $\mathcal{T}$-symmetry}
	In this case, the contribution of $t$ cyclic variants and $t$ anticyclic variants are different due to the non-symmetric nature of the matrix $V$. Therefore, the leading-order SFF is
	\begin{align}
		K_0^{(1)}(t)&=\sum_{l=0}^{t-1}\mathcal{Y}_{\mathcal{C}^lI}+\mathcal{Y}_{\mathcal{R}\mathcal{C}^lI}=t\left[\mathcal{Y}_{I}+\mathcal{Y}_{\mathcal{R}I}\right].
		\label{CUE_leading_order_SFF}
	\end{align}
	Applying the rules in Sec.~\ref{CUE_rules} to the cyclic and anticyclic variants of the identity permutation, we obtain
	\begin{align}
		\mathcal{Y}_{I}&=\text{tr}\mathcal{M}^t,
		\label{identity_cyclic}\\
		\mathcal{Y}_{\mathcal{R}I}&=\text{tr}\tilde{\mathcal{M}}^t.
		\label{identity_anticyclic}
	\end{align}
	Substituting Eqs.~(\ref{identity_cyclic}, \ref{identity_anticyclic}) in Eq.~(\ref{CUE_leading_order_SFF}), we find
	\begin{align}
		K_0^{(1)}(t)&=t\left(\text{tr}\mathcal{M}^t+\text{tr}\tilde{\mathcal{M}}^t\right)\notag\\
		&=t\left(1+\lambda_1^t+...+\lambda_{\mathcal{N}-1}^t+\chi_0^t+...+\chi_{\mathcal{N}-1}^t\right).
	\end{align}
	Beyond $t^*$, determined by $max(\lambda_1,\chi_0)$, the SFF takes form $K_0^{(1)}(t)\simeq t$, which is the CUE prediction. We next study the Hamiltonians in Eqs.~(\ref{H0},\ref{Hs0}) with the parameters in Eqs.~(\ref{CUE_hopping},\ref{CUE_pairing}) to determine $t^*$ for a physical model in the absence of $\mathcal{T}$-symmetry. Our numerical study with different parameters shows that $\lambda_1$ of $\mathcal{M}$ is larger than the largest eigenvalue $\chi_0$ of $\tilde{\mathcal{M}}$ for longer system sizes as shown in Figs.~\ref{M_Second_vs_Mp_largest_eval}a, \ref{M_Second_vs_Mp_largest_eval}b. Thus, $\lambda_1$ determines $t^*$. When $\Delta=\Delta'=0$, the Hamiltonians in Eqs.~(\ref{H0},\ref{Hs0}) commute with $\hat{N}$ in Tab.~\ref{Thouless_time_table} indicating a $U(1)$ symmetry of the model. We also numerically find that the matrix $\mathcal{M}$ is $SU(2)$ invariant for arbitrary $J$ and $g$, which indicates the existence of descendant states. Our numerical study further reveals that $\lambda_1$ is independent of total number of particles, and the corresponding eigenstate is a descendant state. We numerically determine $\lambda_1$ for very long system sizes from one-particle sector. In thermodynamic limit of $L \to \infty$, we find
	\begin{align}
		\lambda_1\simeq 1-\frac{c_{\beta}}{L^2},
		\label{CUE_lambda_1_U(1)_Sym}
	\end{align}
	where $c_{\beta}$ is a constant and it depends on hopping $J$ and $g$. Substituting Eq.~(\ref{CUE_lambda_1_U(1)_Sym}) and $d_1=2$ (as a numerical study shows that $\lambda_1$ is doubly degenerate) in Eq.~(\ref{Thouless_time}), we obtain 
	\begin{align}
		t^*\propto L^2.
	\end{align}
	We can find an analytical expression for $\lambda_1$ by mapping the matrix $\mathcal{M}$ in the Trotter regime ($J$, $g$, $\Delta$, $\Delta'$ $\ll 1/\tau_p=1$),  to an $XXZ$-Heisenberg spin $1/2$ Hamiltonian with nearest-neighbor and next-nearest-neighbor couplings as
	\begin{align}
		\mathcal{M} &\simeq (1-(c_1+c'_1) L)\mathds{1}_\mathcal{N}+\sum_{x=1}^L\sum_\nu c_\nu s_x^\nu s_{x+1}^\nu+c'_\nu s_x^\nu s_{x+2}^\nu,
		\label{mapped_M_CUE}
	\end{align}
	where $c_{\nu}$ are the same as in the previous sub-section and $c'_1=(|g|^2+|\Delta'|^2)/2$, and $c'_2=c'_3=(|g|^2-|\Delta'|^2)/2$. For $\Delta=\Delta'=0$, Eq.~(\ref{mapped_M_CUE}) becomes a $SU(2)$ symmetric $XXX$-Heisenberg chain with nearest-neighbor and next-nearest-neighbor couplings, which leads to $\lambda_1\simeq 1-(2\pi^2J^2+8\pi^2g^2)/L^2$ in the thermodynamic limit. It gives $t^*\propto L^2$ \cite{FriedmanPRL2019}. When $|\Delta|=|J|$ and $|\Delta'|=|g|$, Eq.~(\ref{mapped_M_CUE}) becomes an Ising Hamiltonian with nearest-neighbor and next-nearest-neighbor couplings, leading to $\lambda_1=1-4|J|^2-4|g|^2$ and $d_1=L$. Thus, we have $t^*\propto \ln L$, which matches with the system size scalings of $t^*$ for a quantum circuit model in Chan \textit{et al.} \cite{ChanPRL2018}. When $|\Delta|\neq|J|$ or $|\Delta'|\neq|g|$, our numerical study in Fig. \ref{M_Second_vs_Mp_largest_eval}b shows that $\lambda_1$ is a constant as $L$ increases, and $d_1=1$, which gives $t^*\propto L^0$. Therefore, the system size scalings of $t^*$ do not change for systems without $\mathcal{T}$-symmetry from those $\mathcal{T}$-invariant systems with $\mathcal{T}^2=1$. We confirm our analytical predictions of $t^* \propto L^2$ in Figs.~\ref{SFF_CUE_f_nnn_Direct}a,\ref{SFF_CUE_f_nnn_Direct}b and $t^* \propto L^0$ in Fig.~\ref{SFF_CUE_f_nnn_Direct}c by numerically computing the SFF directly using Eq.~(\ref{SFF}) without the RPA.
\subsection{$\mathcal{T}^2=-1$}
\begin{figure*}[t!]
		\centering
		\includegraphics[width=\linewidth]{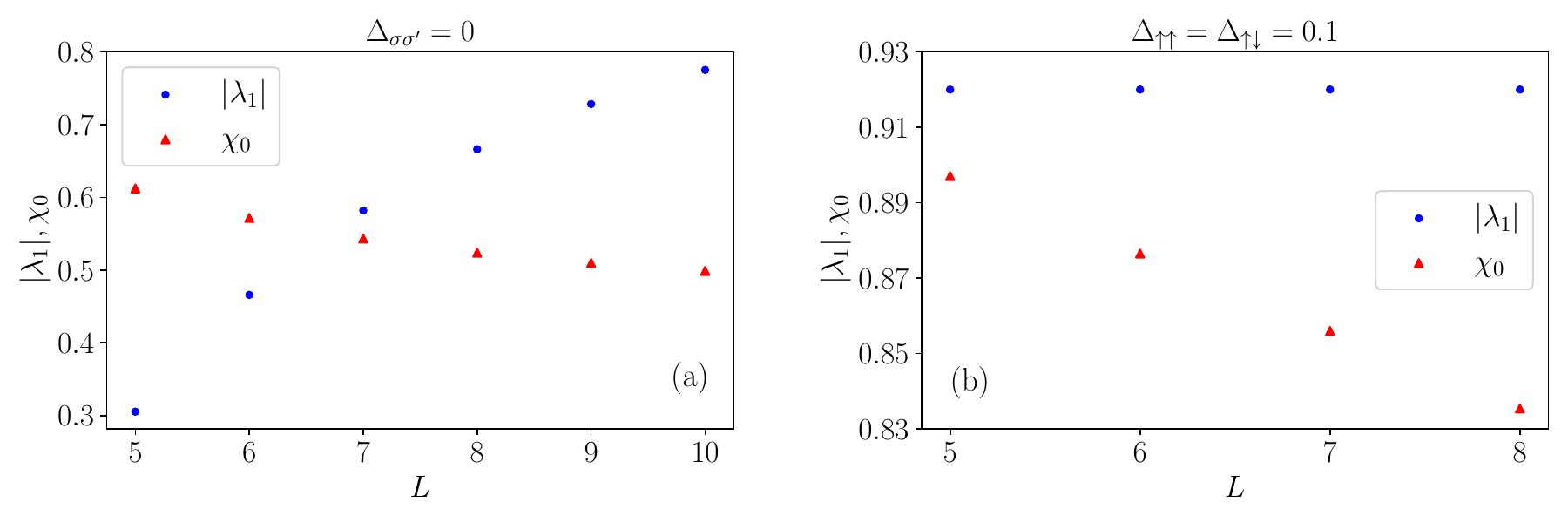}
		\caption{\justifying\small Second-largest eigenvalue $|\lambda_1|$ of $\mathcal{M}$ and largest eigenvalue $\chi_0$ of $\tilde{\mathcal{M}}$ for different system sizes $L$ for a kicked spinful $\mathcal{T}$-invariant fermionic chain with $\mathcal{T}^2=-1$, and (a) with ($\Delta_{\sigma\sigma'}=0$) and (b) without ($\Delta_{\uparrow\uparrow}=\Delta_{\uparrow\downarrow}=0.1$) particle-number conservation. We use the Hamiltonian in Eq.~(\ref{Hs1/2}), and apply PBC in real space. In (a), we take $J_{\uparrow\uparrow}=0.5+0.25i,J_{\uparrow\downarrow}=0.5+0.25i$ and a fixed total number of particles ($N_t=5$), and in (b), we take $J_{\uparrow\uparrow}=0.1+0.1i,J_{\uparrow\downarrow}=0.1+0.05i$.}
		\label{M_vs_Mp_CSE}
    \end{figure*}

Following Eq.~(\ref{SFF_RPA_Z_pi}), the leading-order SFF can be expressed as a sum of contributions of the variants of identity permutation as follows:
\begin{align}
    K_{-1}^{(1)}&=\sum_{\vec{\sigma}}\sum_{l=0}^{t-1}\left(\mathcal{Z}_{\mathcal{C}^lI,\vec{\sigma}}+\mathcal{Z}_{\mathcal{R}\mathcal{C}^lI,\vec{\sigma}}\right).
\end{align}
Similar to the systems without $\mathcal{T}$-symmetry, the matrix $V$ is also non-symmetric for $\mathcal{T}$-invariant systems with $\mathcal{T}^2=-1$. Further, due to PBC in time, the $t$ cyclic variants for each $\vec{\sigma}$ give identical contributions. Similarly, the $t$ anticyclic variants for each $\vec{\sigma}$ also give identical contributions but differ from the cyclic variants due to the non-symmetric nature of the matrix $V$. Therefore,
\begin{align}
    K_{-1}^{(1)}&=t\sum_{\vec{\sigma}}\left(\mathcal{Z}_{I,\vec{\sigma}}+\mathcal{Z}_{\mathcal{R}I,\vec{\sigma}}\right).
\end{align}
To find which diagrams determine the leading-order SFF, we consider the diagram of an arbitrary variant as shown in Fig.~\ref{CSE_identity_variant}. We notice that the diagram contains red arcs on the inner or outer circle connected via jumps between the two circles. \textbf{Theorem 5} implies that all the reduced diagrams of such variants have vanishing contributions. Therefore, such diagrams do not contribute to the SFF in the ergodic phase. \textbf{Theorem 5} also implies that Type \Rom{1} contribution can only come from diagrams with no jump to the inner or outer circle as shown in Figs.~\ref{CSE_identity_a} and \ref{CSE_identity_b}, which correspond to $\vec{\sigma}=(0,...,0)$ and $\vec{\sigma}=(1,...,1)$, respectively. Thus, the leading-order SFF is
	\begin{figure}[H]
		\centering
		\begin{tikzpicture}
			\def\radis{0.75}
			\def\radib{1}
			\draw[blue,dashed] (0,0) circle(\radib);
			\draw[black,dashed] (0,0) circle(\radis);
			\draw[red,thick,->] (30:\radib) arc(30:90:\radib);\draw[red,thick] (90:\radib) arc(90:120:\radib)--(125:\radis)--(130:\radib)--(135:\radis)--(140:\radib) arc(140:180:\radib)--(185:\radis) arc(185:270:\radis)--(275:\radib) arc(275:390:\radib);
		\end{tikzpicture}
		\caption{}
		\label{CSE_identity_variant}
	\end{figure}
	\begin{figure}[H]
		\centering
		\begin{subfigure}[t]{0.45\linewidth}
			\centering
			\begin{tikzpicture}
				\def\radis{0.75}
				\def\radib{1}
				\draw[blue,dashed] (0,0) circle(\radib);
				\draw[black,dashed] (0,0) circle(\radis);
				\draw[red,thick,->] (0:\radib) arc(0:90:\radib);\draw[red,thick] (90:\radib) arc(90:360:\radib);
			\end{tikzpicture}
			\caption{}
			\label{CSE_identity_a}
		\end{subfigure}
		\hfill
		\begin{subfigure}[t]{0.45\linewidth}
			\centering
			\begin{tikzpicture}
				\def\radis{0.75}
				\def\radib{1}
				\draw[blue,dashed] (0,0) circle(\radib);
				\draw[black,dashed] (0,0) circle(\radis);
				\draw[red,thick,->] (0:\radis) arc(0:90:\radis);\draw[red,thick] (90:\radis) arc(90:360:\radis);
			\end{tikzpicture}
			\caption{}
			\label{CSE_identity_b}
		\end{subfigure}
		\caption{}
		\label{CSE_identity_circles}
	\end{figure}
	\begin{align}
		K_{-1}^{(1)}(t)&=t\Big[\mathcal{Z}_{I, (0,...,0)}+\mathcal{Z}_{\mathcal{R}I,(0,...,0)}+\mathcal{Z}_{I, (1,...,1)}\notag\\
		&+\mathcal{Z}_{\mathcal{R}I,(1,...,1)}\Big].
		\label{CSE_leading_order_SFF}
	\end{align}
    \begin{figure*}[t!]
	\centering
	\includegraphics[width=\linewidth]{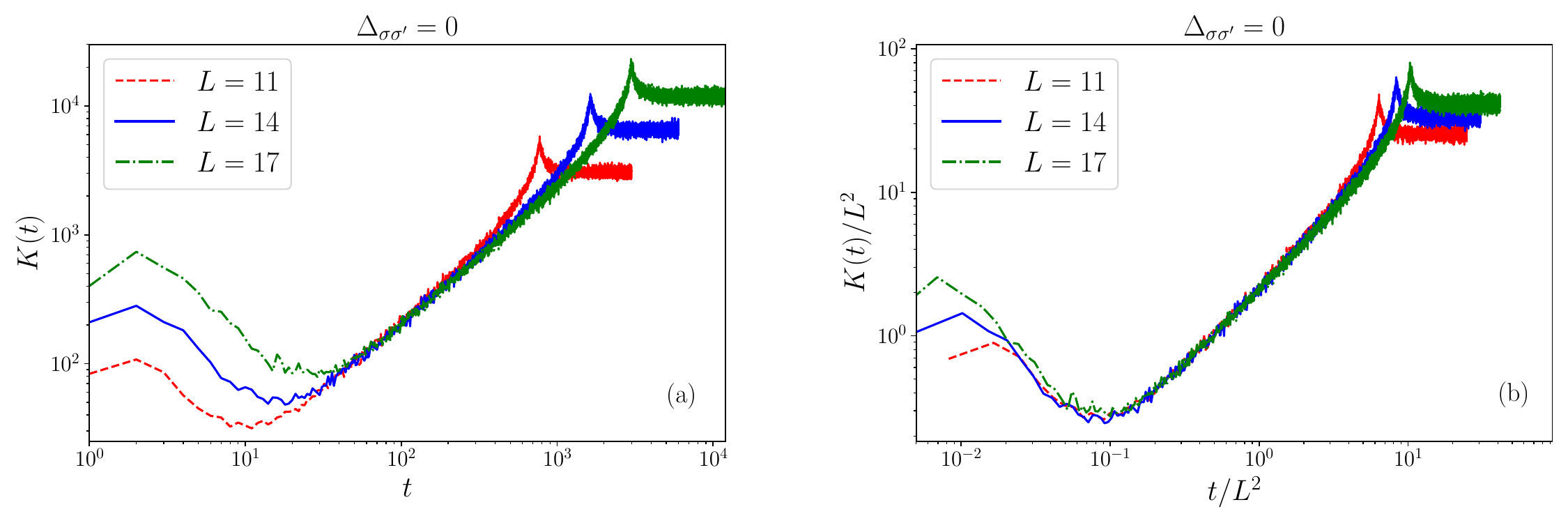}
	\caption{\justifying\small Spectral form factor $K(t)$ for different system sizes $L$ calculated numerically using Eq.~(\ref{SFF}) for a kicked strongly interacting spinful fermionic chain in the presence of $\mathcal{T}$-symmetry ($\mathcal{T}^2=-1$), and particle-number conservation. We use the Hamiltonians in Eqs.~(\ref{H01/2},\ref{Hs1/2}). Here, $\Delta\epsilon=1,U_0=25, \Delta U_0=3,\alpha=1.3, \bar{U}=20,\Delta\bar{U}=0.1,J_{\uparrow\uparrow}=0.5+0.25i,J_{\uparrow\downarrow}=0.5+0.25i$. We apply PBC in real space, and take a fixed total number of particles $(N_t = 3)$. In (b), we show the data collapse in scaled time $t/L^2$. An averaging over 320 realizations of disorder is performed in each case.}\label{directSFF_CSE}
    \end{figure*}
	Applying the rules in Sec. \ref{CSE_rules}, we obtain
	\begin{align}
		\mathcal{Z}_{I, (0,...,0)}&=\mathcal{Z}_{\mathcal{R}I, (1,...,1)}=\text{tr}\mathcal{M}^t,
		\label{CSE_identity_cyclic}\\
		\mathcal{Z}_{\mathcal{R}I,(0,...,0)}&=\mathcal{Z}_{I,(1,...,1)}=\text{tr}\tilde{\mathcal{M}}^t.
		\label{CSE_identity_anticyclic}
	\end{align}
	Substituting Eqs.~(\ref{CSE_identity_cyclic},\ref{CSE_identity_anticyclic}) in Eq.~(\ref{CSE_leading_order_SFF}), we find
	\begin{align}
		K_{-1}^{(1)}(t)&=2t\left(\text{tr}\mathcal{M}^t+\text{tr}\tilde{\mathcal{M}}^t\right)\notag\\
		&=2t\left(1+\lambda_1^t+...+\lambda_{\mathcal{N}-1}^t+\chi_0^t+...+\chi_{\mathcal{N}-1}^t\right).
	\end{align}
        The SFF becomes $K_{-1}^{(1)}\simeq 2t$ beyond $t^*$, which is determined by $max(\lambda_1,\chi_0)$. This is the leading-order CSE result in Eq. (\ref{CSE_RMT_SFF}). We study the Hamiltonians in Eqs.~(\ref{H01/2},\ref{Hs1/2}) to determine $t^*$ in a physical system. We find numerically that $\lambda_1>\chi_0$ for longer $L$ and different parameters as shown in Fig.~\ref{M_vs_Mp_CSE}. So $\lambda_1$ determines $t^*$ in this case also. When $\Delta_{\sigma\sigma'}=0$, the Hamiltonians in Eqs.~(\ref{H01/2},\ref{Hs1/2}) commute with $\hat{N}_t$ in Tab. \ref{Thouless_time_table}. Therefore, the model has $U(1)$ symmetry. Once again, we find that the matrix $\mathcal{M}$ is $SU(2)$ invariant for arbitrary values of $J_{\sigma\sigma'}$, which indicates the existence of descendant states. Thus, $\lambda_1$ is independent of total number of particles. We obtain numerically $\lambda_1$ for large $L$'s in the one-particle sector. We find that in thermodynamic limit of $L \to  \infty$:
	\begin{align}
		\lambda_1\simeq 1-\frac{c_\beta}{L^2},
		\label{CSE_lambda_1_U(1)_Sym}
	\end{align}
	where $c_\beta$ is a constant, which depends on hopping $J_{\sigma\sigma'}$. We substitute Eq.~(\ref{CSE_lambda_1_U(1)_Sym}) and $d_1=2$ (as our numerical study reveals that $\lambda_1$ is doubly degenerate) in Eq.~(\ref{Thouless_time}) to find 
	\begin{align}
		t^*\propto L^2.
	\end{align}
We can analytically determine the eigenvalues of $\mathcal{M}$ by mapping it to a spin $1/2$ Hamiltonian in the Trotter regime ( $|J_{\sigma\sigma'}|,|\Delta_{\sigma\sigma'}|\ll 1/\tau_p=1$):
	\begin{align}
		\mathcal{M}\simeq (1-L\sum_{\sigma\sigma'}c_1^{\sigma\sigma'})\mathds{1}_\mathcal{N}+\sum_{x=1}^L\sum_{\sigma\sigma'}\sum_\nu c_\nu^{\sigma\sigma'}s_{x,\sigma}^\nu s_{x+1,\sigma'}^\nu,
	\end{align}
where $c_{1}^{\sigma\sigma'}=(|J_{\sigma\sigma'}|^2+|\Delta_{\sigma\sigma'}|^2)/2$, $c_2^{\sigma\sigma'}=c_3^{\sigma\sigma'}=(|J_{\sigma\sigma'}|^2-|\Delta_{\sigma\sigma'}|^2)/2$, $\nu=1,2,3$, and $s^\nu_{x,\sigma}$ are the Pauli matrices at site $x$ for each $\sigma$. In the presence of $U(1)$ symmetry ($\Delta_{\sigma\sigma'}=0$), $\mathcal{M}$ is a $SU(2)$ invariant spin $1/2$ ladder Hamiltonian. The doubly-degenerate second-largest eigenvalue calculated from the one-excitation sector reads as $\lambda_1\simeq1-4\pi^2(|J_{\uparrow\uparrow}|^2+|J_{\uparrow\downarrow}|^2)/L^2$ in the limit of $L \to \infty$. Thus, $t^*\propto L^2$. We confirm our analytical predictions of $t^* \propto L^2$ in Fig.~\ref{directSFF_CSE} for the systems of CSE class by numerically computing the SFF directly using Eq.~(\ref{SFF}) without the RPA. When $|J_{\sigma\sigma'}|=|\Delta_{\sigma\sigma'}|$, $\mathcal{M}$ becomes a spin ladder with Ising-type interaction, which gives $\lambda_1=1-4|J_{\uparrow\uparrow}|^2-4|J_{\uparrow\downarrow}|^2$ with a degeneracy $d_1\simeq L$. Therefore, $t^*\propto \ln(L)$. When $|J_{\sigma\sigma'}|\neq|\Delta_{\sigma\sigma'}|\neq 0$, a numerical study of the spectra of the matrix $\mathcal{M}$ reveals that $\lambda_1$ remains constant as $L$ increases, as shown in Fig.~(\ref{M_vs_Mp_CSE})b, and $d_1=1$. Therefore, $t^*\propto L^0$. The above $L$-scalings of $t^*$ for strongly interacting systems of the CSE class have not been explored earlier.

	\section{Second-order correction}
	\label{Second_order_correction}
	As discussed in Sec.~\ref{Outline2nd}, the second-order correction to the SFF is determined by the permutations, whose contributions contain Type \Rom{1} term of form $1/\mathcal{N}$. In Sec.~\ref{Outline2nd}, we introduce these permutations contributing to the second-order term of SFF. In the following subsections, we compute their contribution for each case of $\mathcal{T}$-symmetry.
	
	%       is inversely proportional to the Hilbert space dimension $\mathcal{N}$ for $\mathcal{T}^2=\pm1$ and such term is absent in the absence of $\mathcal{T}$-symmetry. From the Type \Rom{1} term for different permutations, we find that the following permutations and their variants contribute to the second-order term in SFF. These are (1) single transpositions ($T$) defined as the interchange of any two states in $\{\underline{n}_1,...,\underline{n}_t\}$, (2) sub-sequence reversal ($S$) defined as reversal of the order of states between two-time steps, (3) identity permutation with a state repeated twice ($R$) and (4) sub-sequence reversal with a state repeated twice ($SR$). The last permutation reverses the order of states between the two appearances of the repeated state.

	%       Second order correction to SFF is determined by permutations whose contributions contain Type \Rom{1} term of form $1/\mathcal{N}$. Studying different permutations reveal that the following permutations and their variants contribute at second order.\\ 	($\pi_1$) Single transpositions ($T$) defined as interchange of any two states in $\{\underline{n}_1,...,\underline{n}_t\}$.\\ 	($\pi_2$) Sub-sequence reversal ($S$) defined as reversal of the order of states between two time steps.\\ 	($\pi_3$) Identity permutation with a state repeated twice ($R$).\\ 	($\pi_4$) Sub-sequence reversal with a state repeated twice ($SR$). This permutation reverses the order of states between the two appearances of the repeated state.
	
	\subsection{$\mathcal{T}^2=1$}
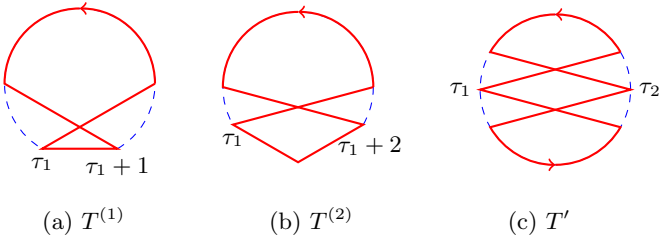
\begin{figure}[h]
		\centering
		\begin{subfigure}[b]{0.3\linewidth}
			\centering
			\begin{tikzpicture}
				\def\rad{1}
				\draw[blue,dashed] (300:\rad) arc(300:600:\rad);
				\draw[red,thick,->] (0:\rad) arc(0:90:\rad);\draw[red,thick] (90:\rad) arc(90:180:\rad)--(300:\rad)--(240:\rad)--(0:\rad);
				\draw (240:\rad) node[below]{$\tau_1$};\draw (300:\rad) node[below]{$\tau_1+1$};
			\end{tikzpicture}
			\vspace{0.1cm}
			\caption{$T^{(1)}$}
			\label{COE_T1_diagram}
		\end{subfigure}
		\hfill
		\begin{subfigure}[b]{0.3\linewidth}
			\centering
			\begin{tikzpicture}
				\def\rad{1}
				\draw[blue,dashed] (-30:\rad) arc(-30:210:\rad);
				\draw[red,thick,->] (0:\rad) arc(0:90:\rad);\draw[red,thick] (90:\rad) arc(90:180:\rad)--(330:\rad)--(270:\rad)--(210:\rad)--(0:\rad);
				\draw (210:\rad) node[below]{$\tau_1$};\draw (330:{1.1*\rad}) node[below]{$\tau_1+2$};
			\end{tikzpicture}
			\vspace{0cm}
			\caption{$T^{(2)}$}
			\label{COE_T2_diagram}
		\end{subfigure}
		\hfill
		\begin{subfigure}[b]{0.3\linewidth}
			\centering
			\begin{tikzpicture}
				\def\rad{1}
				\draw[blue,dashed] (0,0) circle(\rad);
				\draw[red,thick,->] (30:\rad) arc(30:90:\rad);\draw[red,thick,->] (90:\rad) arc(90:150:\rad)--(0:\rad)--(-150:\rad) arc(-150:-90:\rad); \draw[red,thick] (-90:\rad) arc(-90:-30:\rad)--(180:\rad)--(30:\rad);
				\draw (180:\rad) node[left]{$\tau_1$};\draw (0:\rad) node[right]{$\tau_2$};
			\end{tikzpicture}
			\vspace{0cm}
			\caption{$T'$}
			\label{COE_T'_diagram}
		\end{subfigure}
		\caption{\justifying\small Transposition $(T)$ diagrams of : (a) two nearest-neighbor states at time steps $\tau_1$ and $\tau_1+1$, (b) two next-nearest-neighbor states at time steps $\tau_1$ and $\tau_1+2$, and (c) two states at time steps $\tau_1$ and $\tau_2$, where $\tau_2-\tau_1>2$.}
		\label{COE_T_diagram}
	\end{figure}

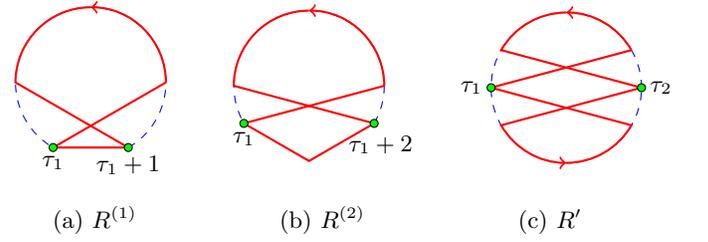
\begin{figure}[h]
		\centering
		\begin{subfigure}[b]{0.3\linewidth}
			\centering
			\begin{tikzpicture}
				\def\rad{1}
				\draw[blue,dashed] (300:\rad) arc(300:600:\rad);
				\draw[red,thick,->] (0:\rad) arc(0:90:\rad);\draw[red,thick](90:\rad) arc(90:180:\rad)--(300:\rad)--(240:\rad)--(0:\rad);
				\draw[fill=green] (240:\rad) circle(1.5pt) node[below]{$\tau_1$};\draw[fill=green] (300:\rad) circle(1.5pt) node[below]{$\tau_1+1$};
			\end{tikzpicture}
			\vspace{0.1cm}
			\caption{$R^{(1)}$}
			\label{COE_R1_diagram}
		\end{subfigure}
		\hfill
		\begin{subfigure}[b]{0.3\linewidth}
			\centering
			\begin{tikzpicture}
				\def\rad{1}
				\draw[blue,dashed] (-30:\rad) arc(-30:210:\rad);
				\draw[red,thick,->] (0:\rad) arc(0:90:\rad);\draw[red,thick] (90:\rad) arc(90:180:\rad)--(330:\rad)--(270:\rad)--(210:\rad)--(0:\rad);
				\draw[fill=green] (210:\rad) circle(1.5pt) node[below]{$\tau_1$};\draw (330:{1.1*\rad}) node[below]{$\tau_1+2$};
				\draw[fill=green] (330:\rad) circle(1.5pt);
			\end{tikzpicture}
			\vspace{0.001cm}
			\caption{$R^{(2)}$}
			\label{COE_R2_diagram}
		\end{subfigure}
		\hfill
		\begin{subfigure}[b]{0.3\linewidth}
			\centering
			\begin{tikzpicture}
				\def\rad{1}
				\draw[blue,dashed] (0,0) circle(\rad);
				\draw[red,thick,->] (30:\rad) arc(30:90:\rad);\draw[red,thick,->] (90:\rad) arc(90:150:\rad)--(0:\rad)--(-150:\rad) arc(-150:-90:\rad);\draw[red,thick] (-90:\rad) arc(-90:-30:\rad)--(180:\rad)--(30:\rad);
				\draw[fill=green] (180:\rad) circle(1.5pt) node[left]{$\tau_1$};\draw[fill=green] (0:\rad) circle(1.5pt) node[right]{$\tau_2$};
			\end{tikzpicture}
			\vspace{0.01cm}
			\caption{$R'$}
			\label{COE_R'_diagram}
		\end{subfigure}
		\caption{\justifying\small Diagrams whose reduced diagrams are identical to the reduced diagrams of transpositions in Fig. \ref{COE_T_diagram}: (a) the reduced diagram of $R^{(1)}$ obtained by removing the upper arc cancels the reduced diagram of $T^{(1)}$ obtained by removing the same arc, (b) the reduced diagram of $R^{(2)}$ obtained by removing the upper arc cancels the reduced diagram of $T^{(2)}$ obtained by removing the same arc, and (c) both the reduced diagrams of $R'$ obtained by removing either the upper or the lower arc cancel the corresponding reduced diagrams of $T'$.}
		\label{COE_R_diagram}
	\end{figure}

\begin{figure}[h]
		\centering
		\begin{subfigure}[b]{0.3\linewidth}
			\centering
			\begin{tikzpicture}
				\def\rad{1}
				\draw[blue,dashed] (300:\rad) arc(300:600:\rad);
				\draw[red,thick,->] (-60:\rad) arc(-60:90:\rad);\draw[red,thick] (90:\rad) arc(90:240:\rad)--(300:\rad);
				\draw[fill=green] (240:\rad) circle(1.5pt) node[below]{$\tau_1$};\draw[fill=green] (300:\rad) circle(1.5pt) node[below]{$\tau_1+1$};
			\end{tikzpicture}
			\vspace{0.1cm}
			\caption{$R^{(1)}$}
			\label{COE_R1_R_diagram}
		\end{subfigure}
		\hfill
		\begin{subfigure}[b]{0.3\linewidth}
			\centering
			\begin{tikzpicture}
				\def\rad{1}
				\draw[blue,dashed] (-30:\rad) arc(-30:210:\rad);
				\draw[red,thick,->] (-30:\rad) arc(-30:90:\rad);\draw[red,thick] (90:\rad) arc(90:210:\rad)--(270:\rad)--(330:\rad);
				\draw[fill=green] (210:\rad) circle(1.5pt) node[below]{$\tau_1$};\draw (330:{1.1*\rad}) node[below]{$\tau_1+2$};
				\draw[fill=green] (330:\rad) circle(1.5pt);
			\end{tikzpicture}
			\vspace{0.1cm}
			\caption{$R^{(2)}$}
			\label{COE_R2_R_diagram}
		\end{subfigure}
		\hfill
		\begin{subfigure}[b]{0.3\linewidth}
			\centering
			\begin{tikzpicture}
				\def\rad{1}
				\draw[blue,dashed] (0,0) circle(\rad);
				\draw[red,thick,->] (-90:\rad) arc(-90:90:\rad);\draw[red,thick,->] (90:\rad) arc(90:270:\rad);
				\draw[fill=green] (180:\rad) circle(1.5pt) node[left]{$\tau_1$};\draw[fill=green] (0:\rad) circle(1.5pt) node[right]{$\tau_2$};
			\end{tikzpicture}
			\vspace{0.1cm}
			\caption{$R'$}
			\label{COE_R'_R_diagram}
		\end{subfigure}
		\caption{\justifying\small Diagrams representing an identity permutation with (a) a repeated state at time steps $\tau_1$ and $\tau_1+1$, (b) a repeated state at time steps $\tau_1$ and $\tau_1+2$, and (c) a repeated state at time steps $\tau_1$ and $\tau_2$.}
		\label{COE_R_R_diagram}
	\end{figure}
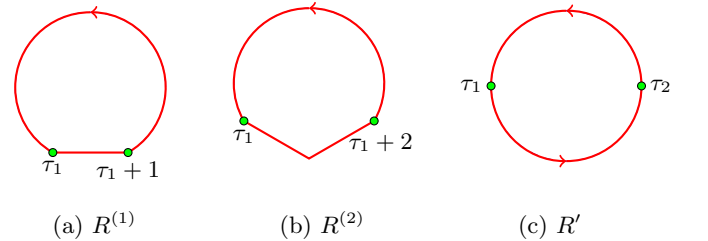

	Following Eq.~(\ref{SFF_RPA_X_pi}), the second-order correction can be expressed in terms of the contributions of these permutations, along with their cyclic and anticyclic variants as follows:
	\begin{align}
		K_1^{(2)}&=\sum_{l=0}^{t-1}\Big[\sum_{T}\left(\mathcal{X}_{\mathcal{C}^lT}+\mathcal{X}_{\mathcal{R}\mathcal{C}^lT}\right)+\sum_{S}\left(\mathcal{X}_{\mathcal{C}^lS}+\mathcal{X}_{\mathcal{R}\mathcal{C}^lS}\right)\notag\\
		&-\sum_R (\mathcal{X}^{\{\underline{n},\underline{n}\}}_{\mathcal{C}^lI}+\mathcal{X}^{\{\underline{n},\underline{n}\}}_{\mathcal{R}\mathcal{C}^lI})-\sum_{SR}(\mathcal{X}^{\{\underline{n},\underline{n}\}}_{\mathcal{C}^lS}+\mathcal{X}^{\{\underline{n},\underline{n}\}}_{\mathcal{R}\mathcal{C}^lS})\Big].
	\end{align}
	Since, the cyclic and anticyclic variants give identical contribution due to PBC in time and the symmetric nature of the matrix $V$ for systems with $\mathcal{T}^2=1$, we obtain
	\begin{align}
		K_1^{(2)}&=2t\Big[\sum_{T}\mathcal{X}_{T}+\sum_{S}\mathcal{X}_{S}-\sum_R\mathcal{X}^{\{\underline{n},\underline{n}\}}_{I}-\sum_{SR}\mathcal{X}^{\{\underline{n},\underline{n}\}}_{S}\Big].
		\label{COE_second_order}
	\end{align}
	The right-hand side terms in Eq.~(\ref{COE_second_order}) can be explicitly calculated using the rules in Sec.~\ref{COE_Rules} (check App.~\ref{COE_2nd_full_der}). However, we proceed with a much insightful approach based on the reduced diagrams, since, only Type \Rom{1} and Type \Rom{2} terms can contribute to the SFF at long times. We begin by analyzing transpositions. We observe that single transpositions can be further classified into three distinct categories based on the different shapes of the diagrams representing them: (a) nearest-neighbor transposition ($T^{(1)}$), where the interchanged states are adjacent on the blue circle, (b) next-nearest-neighbor transposition ($T^{(2)}$), where the interchanged states are separated by two time steps on the blue circle, and (c) all other transpositions ($T'$). All these transpositions are shown diagrammatically in Fig.~\ref{COE_T_diagram}.
	\noindent From \textbf{Theorem 3}, we find that the reduced diagrams of the diagrams with a state repeated twice in Figs.~\ref{COE_R1_diagram}, \ref{COE_R2_diagram}, and \ref{COE_R'_diagram} are idenitical to the reduced diagrams of the diagrams in Figs.~\ref{COE_T1_diagram}, \ref{COE_T2_diagram}, and \ref{COE_T'_diagram}, respectively.
	\noindent Since the diagrams in Figs.~\ref{COE_R1_diagram}, \ref{COE_R2_diagram}, and \ref{COE_R'_diagram} represent the transposition of a repeated state, they are identical to the identity permutation with a state repeated twice. Therefore, these are the $R$ diagrams in Fig.~\ref{COE_R_R_diagram}. Since there are equal numbers of $T$ and $R$ diagrams, as we need to choose two states from $t$ states for both operations, and since the reduced diagrams of $T$ and $R$ diagrams exactly cancel, we obtain
	\begin{align}
		\sum_{T}\mathcal{X}_{T}-\sum_R\mathcal{X}^{\{\underline{n},\underline{n}\}}_{I}=\text{Type \Rom{3} terms}.
	\end{align}
	Because Type \Rom{3} terms decay exponentially in time, we conclude that the $T$ and $R$ diagrams together do not contribute to the SFF in the universal regime.
	Now, we consider an $S$ diagram with the order of states reversed from $\tau_1$ to $\tau_2$ in Fig.~\ref{COE_S_diagram}. When a reduced diagram is obtained by removing the upper arc, \textbf{Theorem 3} implies that an $SR$ diagram with a repeated state at time steps $\tau_1$ and $\tau_2$ (see Fig.~\ref{COE_SR_diagram}) also has an identical reduced diagram after the upper arc is removed. However, when a reduced diagram is obtained by removing the lower arc of the $S$ diagram in Fig.~\ref{COE_S_diagram}, a different $SR$ diagram results in an identical reduced diagram. This $SR$ diagram, shown in Fig.~\ref{COE_S_SR_lower_reduced_diagram}, has identical states at time steps $\tau_1-1$ and $\tau_2+1$.
	\begin{figure}[h]
		\centering
		\begin{subfigure}[t]{0.45\linewidth}
			\centering
			\begin{tikzpicture}
				\def\rad{1}
				\draw[dashed,blue] (0,0) circle(\rad);
				\draw[red,thick,->] (30:\rad) arc(30:90:\rad);\draw[red,thick,->] (90:\rad) arc(90:150:\rad)--(-30:\rad) arc(-30:-90:\rad);\draw[red,thick] (-90:\rad) arc(-90:-150:\rad)--(30:\rad);
				\draw (210:\rad) node[below left]{$\tau_1$};
				\draw (330:\rad) node[below right]{$\tau_2$};
			\end{tikzpicture}
			\caption{$S$}
			\label{COE_S_diagram}
		\end{subfigure}
		\hfill
		\begin{subfigure}[t]{0.45\linewidth}
			\centering
			\begin{tikzpicture}
				\def\rad{1}
				\draw[dashed,blue] (0,0) circle(\rad);
				\draw[red,thick,->] (30:\rad) arc(30:90:\rad);\draw[red,thick,->] (90:\rad) arc(90:150:\rad)--(-30:\rad) arc(-30:-90:\rad);\draw[red,thick] (-90:\rad) arc(-90:-150:\rad)--(30:\rad);
				\draw[fill=green] (210:\rad) circle(1.5pt) node[below left]{$\tau_1$};
				\draw[fill=green] (330:\rad) circle(1.5pt) node[below right]{$\tau_2$};
			\end{tikzpicture}
			\caption{$SR$}
			\label{COE_SR_diagram}
		\end{subfigure}
		\caption{\justifying\small Diagrams representing (a) a sub-sequence reversal $(S)$, and (b) a sub-sequence reversal with a state repeated twice $(SR)$.}
		\label{COE_S_SR_diagram}
	\end{figure}
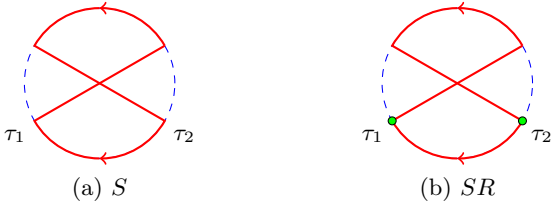
	\begin{figure}[h]
		\centering
		\begin{tikzpicture}
			\def\rad{1}
			\draw[dashed,blue] (0,0) circle(\rad);
			\draw[red,thick,->] (30:\rad) arc(30:90:\rad);\draw[red,thick,->] (90:\rad) arc(90:150:\rad)--(-30:\rad) arc(-30:-90:\rad);\draw[red,thick] (-90:\rad) arc(-90:-150:\rad)--(30:\rad);
			\draw[fill=green] (150:\rad) circle(1.5pt) node[below left]{$\tau_1-1$};
			\draw[fill=green] (30:\rad) circle(1.5pt) node[below right]{$\tau_2+1$};
			\draw (0:{2.25*\rad}) node[]{$\equiv$};
			\begin{scope}[shift={(4,0)}]
				\draw[dashed,blue] (0,0) circle(\rad);
				\draw[red,thick,->] (30:\rad) arc(30:90:\rad);\draw[red,thick,->] (90:\rad) arc(90:150:\rad)--(-30:\rad) arc(-30:-90:\rad);\draw[red,thick] (-90:\rad) arc(-90:-150:\rad)--(30:\rad);
				\draw[fill=green] (210:\rad) circle(1.5pt) node[below left]{$\tau_1-1$};
				\draw[fill=green] (-30:\rad) circle(1.5pt) node[below right]{$\tau_2+1$};
			\end{scope}
		\end{tikzpicture}
		\caption{}
		%\caption{\justifying\small Two equivalent $SR$ diagrams whose reduced diagram obtained by removing the lower arc cancel the reduced diagram obtained by removing the lower arc of the $S$ diagram in Fig.~\ref{COE_S_diagram}.}
		\label{COE_S_SR_lower_reduced_diagram}
	\end{figure}
 Unlike the $T$ and $R$ diagrams, the total number of $SR$ diagrams is two more than that of $S$ diagrams. It can be understood by noticing that the minimum value of $\tau_2-\tau_1$ for an $S$ diagram is $3$ to avoid overlap with nearest-neighbor and next-nearest-neighbor transpositions, and its maximum value is $t-5$ as the minimum length of the upper arc is also 3 for the same reason. However, for an $SR$ diagram in Fig.~\ref{COE_SR_diagram}, the minimum value of $\tau_2-\tau_1$ is 3 as a repeated state must be at least three time steps far for a sub-sequence reversal to give a different permutation ($aa\rightarrow aa, aba\rightarrow aba, abca\rightarrow acba$) and its maximum value is $t-3$ as the minimum length of the upper arc can be 1. Thus, there are two extra $SR$ diagrams compared to $S$ diagrams. These extra $SR$ diagrams must determine the second-order correction to the SFF in the universal regime. To calculate their contributions, we must know the lengths of the red arcs in these diagrams.\\
 \textbf{Statement}: The lengths of both red arcs in the extra two $SR$ diagrams diverge with increasing time.\\
 \textbf{Proof}: Consider the $SR$ diagram in Fig.~\ref{COE_SR_diagram}. The sum of the lengths of both red arcs is $(t-2)$. Therefore, the length of at least one red arc diverges with increasing time. Now, consider the case where the lower arc in Fig.~\ref{COE_SR_diagram} has fixed length, i.e., $(\tau_2-\tau_1)$ remains constant with respect to $t$. As mentioned at the end of Sec.\ref{theorem_2}, only red arcs of finite length determine Type \Rom{2} terms, while a Type \Rom{1} term is identical for each reduced diagram, according to \textbf{Property 1}. Therefore, it suffices to study a single reduced diagram of the $SR$ diagram in Fig.~\ref{COE_SR_diagram}, which contains the fixed length red arc. In Fig.~\ref{COE_SR_diagram}, such a reduced diagram can be obtained by removing the upper red arc. As explained previously, there exists an $S$ diagram, shown in Fig.~\ref{COE_S_diagram}, with a lower red arc of length $(\tau_2-\tau_1)$. The reduced diagram of this $S$ diagram, obtained by removing the upper red arc, cancels out the corresponding reduced diagram of the $SR$ diagram. Similarly, when length of the upper red arc, $(t-\tau_2+\tau_1-2)$, of an $SR$ diagram in Fig.~\ref{COE_SR_diagram} remains constant with respect to time, there exists an $S$ diagram with the length of the upper red arc $(t-\tau_2+\tau_1)$. The reduced diagrams of such $S$ and $SR$ diagrams, obtained by removing the lower arc, cancel each other out.
 
 We have shown that for each $SR$ diagram containing at least one red arc whose size remains finite with increasing time, there exists an $S$ diagram such that they cancel the Type \Rom{1} and Type \Rom{2} terms in each other's contributions in Eq.~(\ref{COE_second_order}). Therefore, the lengths of the red arcs in the extra two $SR$ diagrams must diverge with increasing time.
  
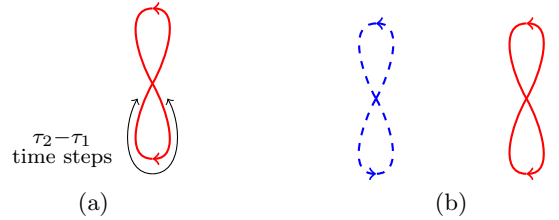
\begin{figure}[h]
		\centering
		\begin{subfigure}[t]{0.45\linewidth}
			\centering
			\begin{tikzpicture}
				\def\rad{1}
				\draw[red,thick,->] ({2.6*\rad},0) to[out=60,in=0] ({2.6*\rad},\rad);\draw[red,thick] ({2.6*\rad},\rad) to[out=180,in=120] ({2.6*\rad},0);
				\draw[red,thick,->] ({2.6*\rad},0) to[out=-60,in=0] ({2.6*\rad},-\rad);\draw[red,thick] ({2.6*\rad},-\rad) to[out=180,in=240] ({2.6*\rad},0);
				\draw[<-] ({2.6*\rad+0.2},-0.2) to[out=-60,in=0] ({2.6*\rad},{-\rad-0.2});\draw[->] ({2.6*\rad},{-\rad-0.2}) to[out=180,in=240] ({2.6*\rad-0.2},-0.2);
				\draw ({2.6*\rad-1.2},{-0.5}) node[below] {\large$\substack{\tau_2-\tau_1\\ \text{time steps}}$};
			\end{tikzpicture}
			\caption{}
			\label{COE_SR_blue_red_overlap_alter_diagram}
		\end{subfigure}
		\hfill
		\begin{subfigure}[t]{0.45\linewidth}
			\centering
			\begin{tikzpicture}
				\def\rad{1}
				\draw[blue,dashed,->,thick] ({2.6*\rad},0) to[out=60,in=0] ({2.6*\rad},\rad);\draw[blue,dashed,thick] ({2.6*\rad},\rad) to[out=180,in=120] ({2.6*\rad},0);
				\draw[blue,dashed,thick] ({2.6*\rad},0) to[out=-60,in=0] ({2.6*\rad},-\rad);\draw[blue,dashed,<-,thick] ({2.6*\rad},-\rad) to[out=180,in=240] ({2.6*\rad},0);
				\begin{scope}[shift={(2,0)}]
					\draw[red,thick,->] ({2.6*\rad},0) to[out=60,in=0] ({2.6*\rad},\rad);\draw[red,thick] ({2.6*\rad},\rad) to[out=180,in=120] ({2.6*\rad},0);
					\draw[red,thick,->] ({2.6*\rad},0) to[out=-60,in=0] ({2.6*\rad},-\rad);\draw[red,thick] ({2.6*\rad},-\rad) to[out=180,in=240] ({2.6*\rad},0);
				\end{scope}
			\end{tikzpicture}
			\caption{}
			\label{COE_SR_as_Seiber_Richter_pair}
		\end{subfigure}
		\caption{\justifying\small (a) An $SR$ diagram in an alternative diagrammatic representation, where the two points representing a repeated state are merged. (b) Its blue and red curves are separated to show resemblance with the Sieber-Richter pairs.}
		\label{COE_SR_alter_diagram}
	\end{figure}

  Additionally, instead of representing a repeated state as two separate points, we can denote it by a single point by merging the points. This leads to an alternative diagrammatic representation. In this representation, the two extra $SR$ diagrams can be drawn as in Fig.~\ref{COE_SR_blue_red_overlap_alter_diagram}. We draw the blue and red curves separately in Fig.~\ref{COE_SR_as_Seiber_Richter_pair} to find that these curves, respectively, resemble near-miss and self-crossing orbits of the Sieber-Richter pairs \cite{Sieber2001} of a semiclassical analysis. Applying the rules in Sec.~\ref{COE_Rules} to this diagram in Fig.~\ref{COE_SR_blue_red_overlap_alter_diagram}, we get
	
\begin{figure}[h]
		\centering
		\includegraphics[width=\linewidth]{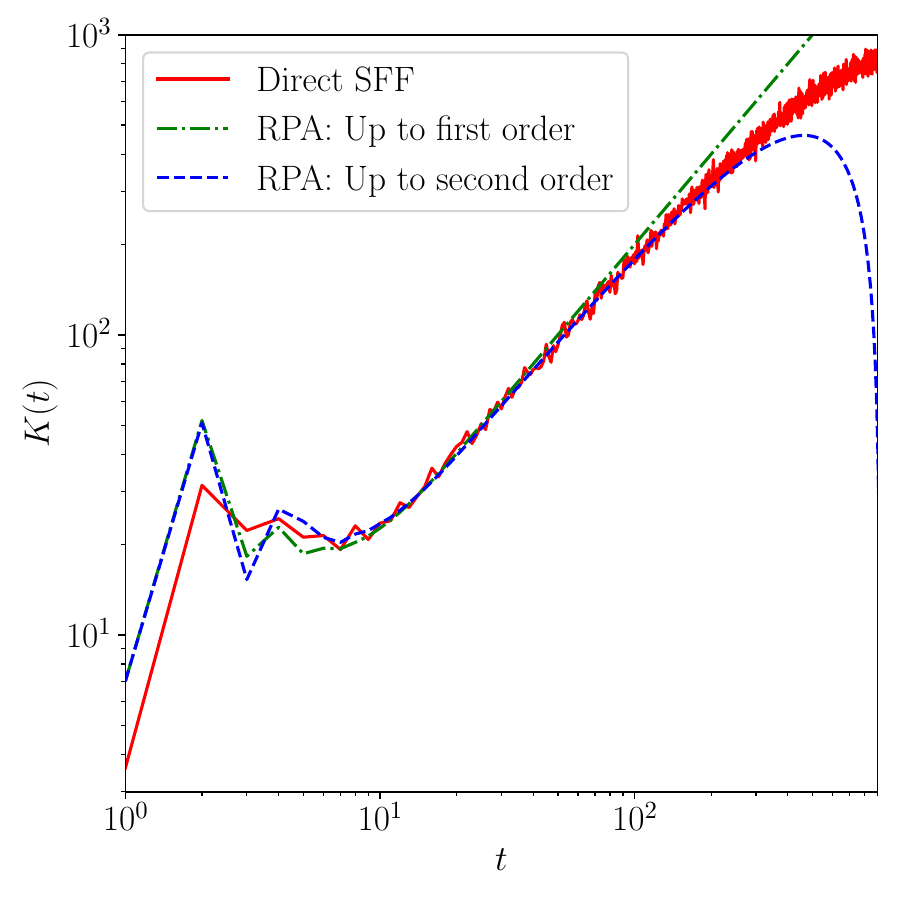}
		\caption{\justifying\small A comparison of directly calculated SFF, $K(t)$, using Eq.~(\ref{SFF}) with that obtained using the RPA up to first and second order in time including both the universal and nonuniversal terms for a kicked spinless fermionic chain with $\mathcal{T}^2=1$ and particle-number conservation. We use the Hamiltonians in Eqs.~(\ref{H0},\ref{Hs0}) with parameters in Eqs.~(\ref{nn_hopping},\ref{nn_pairing}). We take $L=12,N=6,J=1,\langle\epsilon\rangle=0,\Delta\epsilon=1,U_0=22,\alpha=1.5$. An averaging over 400 realizations of disorder is performed for the direct simulation of $K(t)$.}
		\label{COE_SFF_RPA_1_2}
	\end{figure}

	\begin{align}
		K_{1}^{(2)}(t)&=-\frac{2t}{2}\sum_{\tau_1=1}^t\sum_{\tau_2=\tau_1+\nu,\tau_1+\nu'}\sum_{a}\mathcal{M}^{\tau_2-\tau_1}_{a,a}\mathcal{M}^{t-\tau_2+\tau_1}_{a,a}\notag\\
		&=-t^2\sum_a\left(\mathcal{M}^{\nu}_{a,a}\mathcal{M}^{t-\nu}_{a,a}+\mathcal{M}^{\nu'}_{a,a}\mathcal{M}^{t-\nu'}_{a,a}\right)\notag\\
		&=-t^2\sum_a\sum_{ij}\left(\lambda_i^{\nu}\lambda_j^{t-\nu}+\lambda_i^{\nu'}\lambda_j^{t-\nu'}\right)\mathcal{M}^{(i)}_{a,a}\mathcal{M}^{(j)}_{a,a},
	\end{align}
	where $a$ represents a repeated state. The factor $2t$ in the first line accounts for the $t$ cyclic and $t$ anticyclic variants, which contribute identically. Furthermore, the extra factor $1/2$ is included to avoid double counting in summing over $\tau_1,\tau_2$ since a change in the order of $\tau_1,\tau_2$ does not lead to a new permutation when both cyclic and anticyclic variants have been considered (check App.~\ref{COE_2nd_full_der}). %Further, there are $t$ cyclic and $t$ anticyclic variants with identical contribution. We add the contribution of each variant by including a factor of $2t$.
        As the arc sizes $t-\nu$, $t-\nu'$, $\nu$, and $\nu'$ diverge with time for these extra $SR$ diagrams, we obtain at long times
	\begin{align}
		K_1^{(2)}(t)&=-t^2(1+1)\sum_a\mathcal{M}^{(0)}_{a,a}\mathcal{M}^{(0)}_{a,a}+\mathcal{O}(\lambda_1^\nu)\notag\\
		&=-\frac{2t^2}{\mathcal{N}}+\mathcal{O}(\lambda_1^\nu).
		\label{SFF_COE_2nd_order}
	\end{align}
        The first term on the right side of Eq.~(\ref{SFF_COE_2nd_order}) only survives for $t>t^*$, which is the second-order correction for the COE class. We can further interpret this contribution as $K_1^{(2)}(t)=-2t^2 \sum_{\underline{n}}P^{(2)}_t(\underline{n})$, where $P^{(2)}_t(\underline{n}) \equiv {\rm tr} \mathcal{M}^{\nu}{\rm tr} \mathcal{M}^{t-\nu}$ is the probability of returning to an initial state $|\underline{n}\rangle$ twice in time $t$ with the constraint that the interval $\nu$ between two returns diverges with $t$. Fig.~\ref{COE_SFF_RPA_1_2} shows a comparison between the SFF computed using Eq.~(\ref{SFF}) and that obtained using the RPA up to the first- and second-order universal terms along with all the Type \Rom{3} terms (check App.~\ref{COE_2nd_full_der}). It shows a good match beyond $t^*$ upto a longer time when the second-order universal term is included. The difference between the SFF computed directly and that obtained using the RPA in the nonuniversal regime of $t<t^*$ in Fig.~\ref{COE_SFF_RPA_1_2} is due to the RPA and the exclusion of all higher-order contributions beyond second order. 
%Spectral form factor $K(t)$ for different system sizes $L$ calculated numerically using Eq.~(\ref{SFF}) for a kicked strongly interacting spinless fermionic chain in the absence of $\mathcal{T}$-symmetry, and (a), (b) with $(\Delta=\Delta'=0)$ and (c) without $(\Delta=0.4,\Delta'=0)$ particle-number conservation. We use the Hamiltonians in Eqs.~(\ref{H0},\ref{Hs0}) with parameters in Eqs.~(\ref{CUE_hopping},\ref{CUE_pairing}). Here, $\Delta\epsilon=1,U_0=15,\alpha=1.5,J=0.5,g=0.25$. We apply PBC in real space, and take half filling $N/L = 1/2$ for $\Delta=\Delta' = 0$.  In (b), we show the data collapse in scaled time $t/L^2$. An averaging over 400 realizations of disorder is performed in each case.
        
\subsection{Absence of $\mathcal{T}$-symmetry}
In the absence of $\mathcal{T}$-symmetry, matrix $V$ is not symmetric. Therefore, the cyclic and anticyclic variants contribute differently. Consequently, the second-order correction can be expressed as follows:
\begin{align}
	K_0^{(2)}&=t\Big[\sum_{T}\left(\mathcal{Y}_{T}+\mathcal{Y}_{\mathcal{R}T}\right)+\sum_{S}\left(\mathcal{Y}_{S}+\mathcal{Y}_{\mathcal{R}S}\right)\notag\\
	&-\sum_R (\mathcal{Y}^{\{\underline{n},\underline{n}\}}_{I}+\mathcal{Y}^{\{\underline{n},\underline{n}\}}_{\mathcal{R}I})-\sum_{SR}(\mathcal{Y}^{\{\underline{n},\underline{n}\}}_{S}+\mathcal{Y}^{\{\underline{n},\underline{n}\}}_{\mathcal{R}S})\Big].
	\label{CUE_second_order}
\end{align}
We get Type \Rom{1} and Type \Rom{2} terms from a diagram only if it has a reduced diagram, which can be figured out from Eq.~(\ref{Y_pi_CUE_general}). Thus, the anticyclic variants of $T$ and $R$ diagrams can only contribute Type \Rom{3} terms, as they contain red arcs with clockwise arrows. Therefore,
\begin{align}
	\sum_T\mathcal{Y}_{\mathcal{R}T}&=\text{Type \Rom{3} terms},\\
	\sum_R\mathcal{Y}_{\mathcal{R}I}^{\{\underline{n},\underline{n}\}}&=\text{Type \Rom{3} terms}.
\end{align}
Furthermore, like the $\mathcal{T}^2=1$ case, the cyclic variants of $T$ and $R$ diagrams cancel out their corresponding reduced diagrams. Therefore,
\begin{align}
	\sum_T\mathcal{Y}_{T}-\sum_R\mathcal{Y}^{\{\underline{n},\underline{n}\}}_{I}&=\text{Type \Rom{3} terms}.
\end{align}
In addition, all the reduced diagrams of the $S$ diagrams are canceled by the reduced diagrams of the $SR$ diagrams. Once again, there are two extra $SR$ diagrams in which length of the red arcs diverges with increasing time. Therefore, Eq.~(\ref{CUE_second_order}) gives the following expression by ignoring the Type \Rom{3} terms:
\begin{align}
		K_0^{(2)}(t)&=-t\sum_{\substack{\text{extra $SR$}\\ \text{diagrams}}}\left(\mathcal{Y}^{\{\underline{n},\underline{n}\}}_{S}+\mathcal{Y}^{\{\underline{n},\underline{n}\}}_{\mathcal{R}S}\right).
		\label{K2_0}
\end{align}
Since one of the red arcs has a clockwise arrow and the other has a counterclockwise arrow as in Fig.~\ref{COE_SR_blue_red_overlap_alter_diagram}, we find following the rules in Sec. \ref{CUE_rules}:
	\begin{align}
		&\sum_{\substack{\text{extra $SR$}\\ \text{diagrams}}}\mathcal{Y}^{\{\underline{n},\underline{n}\}}_{S}=\frac{1}{2}\sum_{\tau_1=1}^t\sum_{\substack{\tau_2=\tau_1+\nu,\\ \tau_1+\nu'}}\sum_{a}\tilde{\mathcal{M}}^{\tau_2-\tau_1}_{a,a}\mathcal{M}^{t-\tau_2+\tau_1}_{a,a}\notag\\
		&=\frac{t}{2}\sum_{a}\left(\tilde{\mathcal{M}}^{\nu}_{a,a}\mathcal{M}^{t-\nu}_{a,a}+\tilde{\mathcal{M}}^{\nu'}_{a,a}\mathcal{M}^{t-\nu'}_{a,a}\right)\notag\\
		&=\frac{t}{2}\sum_a\sum_{ij}\left(\chi_i^{\nu}\lambda_j^{t-\nu}+\chi_i^{\nu'}\lambda_j^{t-\nu'}\right)\tilde{\mathcal{M}}^{(i)}_{a,a}\mathcal{M}^{(j)}_{a,a}.
	\end{align}
As the arc lengths $\nu$, $\nu'$, $t-\nu$, $t-\nu'$ diverge with increasing time, we find that at long times:
	\begin{align}
		\sum_{\substack{\text{extra $SR$}\\ \text{diagrams}}}\mathcal{Y}^{\{\underline{n},\underline{n}\}}_{S}&=\mathcal{O}(\lambda_1^{t-\nu},\chi_0^\nu).
		\label{Y_nn_S}
	\end{align}
	Similarly, 
	\begin{align}
		\sum_{\substack{\text{extra $SR$}\\ \text{diagrams}}}\mathcal{Y}^{\{\underline{n},\underline{n}\}}_{\mathcal{R}S}&=\frac{1}{2}\sum_{\tau_1=1}^t\sum_{\substack{\tau_2=\tau_1+\nu,\\ \tau_1+\nu'}}\sum_{a}\mathcal{M}^{\tau_2-\tau_1}_{a,a}\tilde{\mathcal{M}}^{t-\tau_2+\tau_1}_{a,a}\notag\\
		&=\mathcal{O}(\lambda_1^\nu,\chi_0^{t-\nu}).
		\label{Y_nn_RS}
	\end{align}
	Substituting Eqs.~(\ref{Y_nn_S}) and (\ref{Y_nn_RS}) in Eq.~(\ref{K2_0}), we obtain
	\begin{align}
		K_0^{(2)}(t)&=\mathcal{O}(\lambda_1^{t-\nu},\chi_0^\nu).
		\label{K20_CUE}
	\end{align}
In App. \ref{CUE_2nd_full_der}, we show an exact calculation based on the rules in Sec. \ref{CUE_rules} also leads to the same result. Therefore, unlike the $\mathcal{T}^2=1$ case, there is no universal term in second order in time, which is identical to the result of the CUE class.
	\subsection{$\mathcal{T}^2=-1$}
	In this case also, the matrix $V$ is not symmetric. Thus, the second-order correction to the SFF following Eq.~(\ref{SFF_RPA_Z_pi}) reads as below:
	\begin{align}
		K_{-1}^{(2)}&=t\sum_{\vec{\sigma}}\Big[\sum_{T}(\mathcal{Z}_{T,\vec{\sigma}}+\mathcal{Z}_{\mathcal{R}T,\vec{\sigma}})+\sum_{S}(\mathcal{Z}_{S,\vec{\sigma}}+\mathcal{Z}_{\mathcal{R}S,\vec{\sigma}})\notag\\
		&-\sum_R(\mathcal{Z}^{\{\underline{n},\underline{n}\}}_{I,\vec{\sigma}}+\mathcal{Z}^{\{\underline{n},\underline{n}\}}_{\mathcal{R}I,\vec{\sigma}})-\sum_{SR}(\mathcal{Z}^{\{\underline{n},\underline{n}\}}_{S,\vec{\sigma}}+\mathcal{Z}^{\{\underline{n},\underline{n}\}}_{\mathcal{R}S,\vec{\sigma}})\notag\\
		&-\sum_R (\mathcal{Z}^{\{\underline{n},\mathcal{T}\underline{n}\}}_{I,\vec{\sigma}}+\mathcal{Z}^{\{\underline{n},\mathcal{T}\underline{n}\}}_{\mathcal{R}I,\vec{\sigma}})\notag\\
		&-\sum_{SR}(\mathcal{Z}^{\{\underline{n},\mathcal{T}\underline{n}\}}_{S,\vec{\sigma}}+\mathcal{Z}^{\{\underline{n},\mathcal{T}\underline{n}\}}_{\mathcal{R}S,\vec{\sigma}})\Big].
		\label{CSE_second_order}
	\end{align}
	In Eq.~(\ref{CSE_second_order}), the sum over configurations $\vec{\sigma}$ leads to $2^t$ different variants for each permutation. We analyze the diagrams of different variants using \textbf{Theorem 1}, \textbf{Theorem 4} and \textbf{Theorem 5} to identify the diagrams responsible for the second-order correction. We start by analyzing the $T$ and $R$ diagrams followed by the $S$ and $SR$ diagrams.
	\subsubsection{$T$ and $R$}
	An arbitrary variant of a transposition diagram is shown in Fig.~\ref{CSE_T_variant}.
	\begin{figure}[h]
		\centering
		\begin{tikzpicture}
			\def\radis{0.75}
			\def\radib{1}
			\draw[blue,dashed] (0,0) circle(\radib);
			\draw[dashed] (0,0) circle(\radis);
			\draw[red,thick,->] (30:\radib) arc(30:90:\radib);\draw[red,thick] (90:\radib) arc(90:100:\radib)--(105:\radis)--(110:\radib) arc(110:150:\radib);
			\draw[red,thick] (210:\radib) arc(210:220:\radib)--(225:\radis) arc(225:235:\radis)--(240:\radib);\draw[red,thick,->] (240:\radib) arc(240:270:\radib);\draw[red,thick] (270:\radib) arc(270:320:\radib)--(330:\radis);
			\draw[red,thick] (150:\radib)--(0:\radis)--(210:\radib);
			\draw[red,thick] (330:\radis)--(180:\radib)--(30:\radib);
		\end{tikzpicture}
		\caption{}
		\label{CSE_T_variant}
	\end{figure}
	A necessary requirement for a variant to have Type \Rom{1} or Type \Rom{2} terms in its contribution is the existence of non-vanishing reduced diagrams. From the definition of reduced diagrams in Sec.~\ref{reduced_diagram}, a diagram must contain either a red arc with a counterclockwise arrow on the outer circle or a red arc with a clockwise arrow on the inner circle. We study the former case in Fig.~\ref{CSE_T_variant}, as the latter case is related to the former by \textbf{Theorem 1}. \textbf{Theorem 5} implies that the reduced diagrams of such variants give a non-vanishing contribution when at least one of the arcs does not have a jump to the inner circle at its ends, as shown in Fig.~\ref{CSE_T_var_12}. We obtain the reduced diagram in Fig.~\ref{CSE_T_var_red_vanishing} by removing the lower red arc from Fig.~\ref{CSE_T_var_12}. This reduced diagram has a vanishing contribution due to the time reversal of one of the transposed states. To ensure that the reduced diagram has a non-vanishing contribution, either both of the transposed states must be time-reversed, or neither should be time-reversed. This is diagrammatically equivalent to keeping both transposed states either on the outer circle or on the inner circle, as shown in Fig.~\ref{CSE_T_var_non_vanishing}.
	\begin{figure}[h]
		\centering
		\begin{subfigure}[t]{0.4\linewidth}
			\centering
			\vtop{\vspace*{\fill}\begin{tikzpicture}
					\def\radis{0.75}
					\def\radib{1}
					\draw[blue,dashed] (0,0) circle(\radib);
					\draw[dashed] (0,0) circle(\radis);
					\draw[red,thick,->] (30:\radib) arc(30:90:\radib);\draw[red,thick] (90:\radib) arc(90:100:\radib)--(105:\radis)--(110:\radib) arc(110:150:\radib);
					\draw[red,thick,->] (210:\radib) arc(210:270:\radib);\draw[red,thick] (270:\radib) arc(270:330:\radib);
					\draw[red,thick] (150:\radib)--(0:\radis)--(210:\radib);
					\draw[red,thick] (330:\radib)--(180:\radib)--(30:\radib);
			\end{tikzpicture}}
			\vspace{0.6cm}
			\caption{}
			\label{CSE_T_var_12}
		\end{subfigure}
		\hfill
		\begin{subfigure}[t]{0.5\linewidth}
			\centering
			\vtop{\vspace*{\fill}\begin{tikzpicture}
					\def\radis{0.75}
					\def\radib{1}
					\draw[blue,dashed] (-30:\radib) arc(-30:210:\radib);
					\draw[dashed] (-30:\radis) arc(-30:210:\radis);
					\draw[red,thick,->] (30:\radib) arc(30:90:\radib);\draw[red,thick] (90:\radib) arc(90:100:\radib)--(105:\radis)--(110:\radib) arc(110:150:\radib);
					\draw[red,thick] (150:\radib)--(0:\radis)--(210:\radib);
					\draw[red,thick] (330:\radib)--(180:\radib)--(30:\radib);
					\draw (180:{1.3*\radib}) node[left]{$\frac{1}{\mathcal{N}}\times$};
					\draw (210:\radib) node[below left]{$a$};
					\draw (330:\radib) node[below right]{$b$};
					\draw (180:\radib) node[left]{$c$};
					\draw (0:\radib) node[right]{$d$};
					\draw (270:\radib) node[below]{$\propto \delta_{c,\mathcal{T}d}\delta_{c,d}=0$};
			\end{tikzpicture}}
			\caption{}
			\label{CSE_T_var_red_vanishing}
		\end{subfigure}
		\caption{}
		\label{CSE_T_var_12_red}
	\end{figure}
	\begin{figure}[h]
		\centering
		\begin{subfigure}[t]{0.45\linewidth}
			\centering
			\begin{tikzpicture}
				\def\radis{0.75}
				\def\radib{1}
				\draw[blue,dashed] (0,0) circle(\radib);
				\draw[dashed] (0,0) circle(\radis);
				\draw[red,thick,->] (30:\radib) arc(30:90:\radib);\draw[red,thick] (90:\radib) arc(90:100:\radib)--(105:\radis)--(110:\radib) arc(110:150:\radib);
				\draw[red,thick,->] (210:\radib) arc(210:270:\radib);\draw[red,thick] (270:\radib) arc(270:330:\radib);
				\draw[red,thick] (150:\radib)--(0:\radib)--(210:\radib);
				\draw[red,thick] (330:\radib)--(180:\radib)--(30:\radib);
			\end{tikzpicture}
                \caption{}
		\end{subfigure}
		\hfill
		\begin{subfigure}[t]{0.45\linewidth}
			\centering
			\begin{tikzpicture}
				\def\radis{0.75}
				\def\radib{1}
				\draw[blue,dashed] (0,0) circle(\radib);
				\draw[dashed] (0,0) circle(\radis);
				\draw[red,thick,->] (30:\radib) arc(30:90:\radib);\draw[red,thick] (90:\radib) arc(90:100:\radib)--(105:\radis)--(110:\radib) arc(110:150:\radib);
				\draw[red,thick,->] (210:\radib) arc(210:270:\radib);\draw[red,thick] (270:\radib) arc(270:330:\radib);
				\draw[red,thick] (150:\radib)--(0:\radis)--(210:\radib);
				\draw[red,thick] (330:\radib)--(180:\radis)--(30:\radib);
			\end{tikzpicture}
                \caption{}
		\end{subfigure}
		\caption{}
		\label{CSE_T_var_non_vanishing}
	\end{figure}
	\noindent \textbf{Theorem 4} implies that the diagrams in Fig.~\ref{CSE_T_var_non_vanishing_red} containing a state repeated twice or a state and its time-reversed state cancel the reduced diagrams of the diagrams in Fig.~\ref{CSE_T_var_non_vanishing}. The diagrams in Fig.~\ref{CSE_T_var_non_vanishing_red} are equivalent to Fig.~\ref{CSE_R_var_non_vanishing}. We note that Fig.~\ref{CSE_R_variant_nn} is a variant of identity permutation with a state repeated twice, and Fig.~\ref{CSE_R_variant_nTn} is a variant of identity permutation with a state and its time-reversed state appearing at two different time steps. We refer to Fig.~\ref{CSE_R_variant_nn} and Fig.~\ref{CSE_R_variant_nTn} collectively as variants of the $R$ diagrams. Similarly to the $\mathcal{T}^2=1$ and absence of $\mathcal{T}$-symmetry cases, the $T$ and $R$ diagrams cancel the reduced diagrams of each other in Eq.~(\ref{SFF_RPA_Z_pi}). Therefore, the $T$ and $R$ diagrams together once again do not contribute to the SFF in the universal regime,
	\begin{align}
		\sum_{\vec{\sigma}}\Big[&\sum_{T}\left(\mathcal{Z}_{T,\vec{\sigma}}+\mathcal{Z}_{\mathcal{R}T,\vec{\sigma}}\right)-\sum_R\left(\mathcal{Z}^{\{\underline{n},\underline{n}\}}_{I,\vec{\sigma}}+\mathcal{Z}^{\{\underline{n},\underline{n}\}}_{\mathcal{R}I,\vec{\sigma}}\right)\notag\\
		&-\sum_R\left(\mathcal{Z}^{\{\underline{n},\mathcal{T}\underline{n}\}}_{I,\vec{\sigma}}+\mathcal{Z}^{\{\underline{n},\mathcal{T}\underline{n}\}}_{\mathcal{R}I,\vec{\sigma}}\right)\Big]=\text{Type \Rom{3} terms}.
	\end{align}
	\begin{figure}[h]
		\centering
		\begin{subfigure}[t]{0.45\linewidth}
			\centering
			\begin{tikzpicture}
				\def\radis{0.75}
				\def\radib{1}
				\draw[blue,dashed] (0,0) circle(\radib);
				\draw[dashed] (0,0) circle(\radis);
				\draw[red,thick,->] (30:\radib) arc(30:90:\radib);\draw[red,thick] (90:\radib) arc(90:100:\radib)--(105:\radis)--(110:\radib) arc(110:150:\radib);
				\draw[red,thick,->] (210:\radib) arc(210:270:\radib);\draw[red,thick] (270:\radib) arc(270:330:\radib);
				\draw[red,thick] (150:\radib)--(0:\radib)--(210:\radib);
				\draw[red,thick] (330:\radib)--(180:\radib)--(30:\radib);
				\draw[fill=green] (180:\radib) circle(1.5pt);
				\draw[fill=green] (0:\radib) circle(1.5pt);
			\end{tikzpicture}
			\caption{}
			\label{CSE_rep_state_diag}
		\end{subfigure}
		\hfill
		\begin{subfigure}[t]{0.45\linewidth}
			\centering
			\begin{tikzpicture}
				\def\radis{0.75}
				\def\radib{1}
				\draw[blue,dashed] (0,0) circle(\radib);
				\draw[dashed] (0,0) circle(\radis);
				\draw[red,thick,->] (30:\radib) arc(30:90:\radib);\draw[red,thick] (90:\radib) arc(90:100:\radib)--(105:\radis)--(110:\radib) arc(110:150:\radib);
				\draw[red,thick,->] (210:\radib) arc(210:270:\radib);\draw[red,thick] (270:\radib) arc(270:330:\radib);
				\draw[red,thick] (150:\radib)--(0:\radis)--(210:\radib);
				\draw[red,thick] (330:\radib)--(180:\radis)--(30:\radib);
				\draw[fill=green] (180:\radib) circle(1.5pt);
				\draw[fill=white] (0:\radib) circle(1.5pt);
			\end{tikzpicture}
			\caption{}
			\label{CSE_Kramer_pair}
		\end{subfigure}
		\caption{}
		\label{CSE_T_var_non_vanishing_red}
	\end{figure}
	
	\begin{figure}[h]
		\centering
		\begin{subfigure}[t]{0.45\linewidth}
			\begin{tikzpicture}
				\def\radis{0.75}
				\def\radib{1}
				\draw[blue,dashed] (0,0) circle(\radib);
				\draw[dashed] (0,0) circle(\radis);
				\draw[red,thick,->] (0:\radib) arc(0:90:\radib);\draw[red,thick] (90:\radib) arc(90:100:\radib)--(105:\radis)--(110:\radib) arc(110:180:\radib);
				\draw[red,thick,->] (180:\radib) arc(180:270:\radib);\draw[red,thick] (270:\radib) arc(270:360:\radib);
				\draw[fill=green] (180:\radib) circle(1.5pt) node[left]{$b$};
				\draw[fill=green] (0:\radib) circle(1.5pt) node[right]{$b$};
			\end{tikzpicture}
			\caption{}
			\label{CSE_R_variant_nn}
		\end{subfigure}
		\hfill
		\begin{subfigure}[t]{0.45\linewidth}
			\begin{tikzpicture}
				\def\radis{0.75}
				\def\radib{1}
				\draw[blue,dashed] (0,0) circle(\radib);
				\draw[dashed] (0,0) circle(\radis);
				\draw[red,thick,->] (0:\radib) arc(0:90:\radib);\draw[red,thick] (90:\radib) arc(90:100:\radib)--(105:\radis)--(110:\radib) arc(110:180:\radib);
				\draw[red,thick,->] (180:\radib) arc(180:270:\radib);\draw[red,thick] (270:\radib) arc(270:360:\radib);
				\draw[fill=green] (180:\radib) circle(1.5pt) node[left]{$b$};
				\draw[fill=white] (0:\radib) circle(1.5pt) node[right]{$\mathcal{T}b$};
			\end{tikzpicture}
			\caption{}
			\label{CSE_R_variant_nTn}
		\end{subfigure}
		\caption{}
		\label{CSE_R_var_non_vanishing}
	\end{figure}
	
	\subsubsection{S and SR}
As in the case of transposition, the variants of a $S$ diagram containing a red arc with a counterclockwise arrow on the outer circle or a red arc with a clockwise arrow on the inner circle can contribute Type  \Rom{1} and Type \Rom{2} terms. Also, both the endpoints of the reversed part should lie on the same circle as in Fig.~\ref{CSE_S_variants}. Otherwise, the reduced diagrams have vanishing contribution similar to a variant of the $T$ diagram in Fig.~\ref{CSE_T_var_12_red}.
	\begin{figure}[h]
		\centering
		\begin{subfigure}[t]{0.45\linewidth}
			\centering
			\begin{tikzpicture}
				\def\radis{0.75}
				\def\radib{1}
				\draw[blue,dashed] (0:\radib) arc(0:360:\radib);
				\draw[black,dashed] (0:\radis) arc(0:360:\radis);
				\draw[red,thick,->] (30:\radib) arc(30:90:\radib);\draw[red,thick] (90:\radib) arc(90:150:\radib);\draw[red,thick] (150:\radib)--(-30:\radib);
				\draw[red,thick,->] (-30:\radib)--(-35:\radib)--(-40:\radis)--(-45:\radib) arc(-45:-90:\radib);\draw[red,thick] (-90:\radib) arc(-90:-100:\radib)--(-105:\radis) arc(-105:-115:\radis)--(-120:\radib) arc(-120:-150:\radib)--(30:\radib);
				\draw (150:\radib) node[above left]{$a$};
				\draw (210:\radib) node[left]{$b$};
				\draw (330:\radib) node[right]{$c$};
				\draw (30:\radib) node[above right]{$d$};
			\end{tikzpicture}
			\caption{}
			\label{CSE_S_variant_out}
		\end{subfigure}
		\hfill
		\begin{subfigure}[t]{0.45\linewidth}
			\centering
			\begin{tikzpicture}
				\def\radis{0.75}
				\def\radib{1}
				\draw[blue,dashed] (0:\radib) arc(0:360:\radib);
				\draw[black,dashed] (0:\radis) arc(0:360:\radis);
				\draw[red,thick,->] (30:\radib) arc(30:90:\radib);\draw[red,thick] (90:\radib) arc(90:150:\radib);\draw[red,thick] (150:\radib)--(-30:\radis);
				\draw[red,thick,->] (-30:\radis)--(-35:\radib)--(-40:\radis)--(-45:\radib) arc(-45:-90:\radib);\draw[red,thick] (-90:\radib) arc(-90:-100:\radib)--(-105:\radis) arc(-105:-115:\radis)--(-120:\radib) arc(-120:-145:\radib)--(-150:\radis)--(30:\radib);
				\draw (150:\radib) node[above left]{$a$};
				\draw (210:\radib) node[above left]{$b$};
				\draw (330:\radib) node[above right]{$c$};
				\draw (30:\radib) node[above right]{$d$};
			\end{tikzpicture}
			\caption{}
			\label{CSE_S_variant_in}
		\end{subfigure}
		\caption{}
		\label{CSE_S_variants}
	\end{figure}
	Applying \textbf{Theorem 4}, the reduced diagram of the $SR$ diagram in Fig.~\ref{CSE_SR_1} obtained by removing the upper red arc cancels out the reduced diagram of the $S$ diagram in Fig.~\ref{CSE_S_variant_out} obtained by removing the upper red arc. Furthermore, the reduced diagram of the $SR$ diagram in Fig.~\ref{CSE_SR_2} obtained by removing the upper red arc cancels out the reduced diagram of the $S$ diagram in Fig.~\ref{CSE_S_variant_in} obtained by removing the upper red arc.
	\begin{figure}[h]
		\centering
		\begin{subfigure}[t]{0.45\linewidth}
			\centering
			\begin{tikzpicture}
				\def\radis{0.75}
				\def\radib{1}
				\draw[blue,dashed] (0:\radib) arc(0:360:\radib);
				\draw[black,dashed] (0:\radis) arc(0:360:\radis);
				\draw[red,thick] (210:\radib)--(30:\radib);\draw[red,thick,->] (30:\radib) arc(30:90:\radib);\draw[red,thick] (90:\radib) arc(90:150:\radib)--(-30:\radib);
				\draw[red,thick,->] (-30:\radib)--(-35:\radib)--(-40:\radis)--(-45:\radib) arc(-45:-90:\radib);\draw[red,thick] (-90:\radib) arc(-90:-100:\radib)--(-105:\radis) arc(-105:-115:\radis)--(-120:\radib) arc(-120:-150:\radib);
				\draw[fill=green] (210:\radib) circle(1.5pt) node[left]{$b$};
				\draw[fill=green] (-30:\radib) circle(1.5pt) node[right]{$b$};
			\end{tikzpicture}
			\caption{}
			\label{CSE_SR_1}
		\end{subfigure}
		\hfill
		\begin{subfigure}[t]{0.45\linewidth}
			\centering
			\begin{tikzpicture}
				\def\radis{0.75}
				\def\radib{1}
				\draw[blue,dashed] (0:\radib) arc(0:360:\radib);
				\draw[black,dashed] (0:\radis) arc(0:360:\radis);
				\draw[red,thick] (210:\radis)--(30:\radib);\draw[red,thick,->] (30:\radib) arc(30:90:\radib);\draw[red,thick] (90:\radib) arc(90:150:\radib)--(-30:\radis);
				\draw[red,thick,->] (-30:\radis)--(-35:\radib)--(-40:\radis)--(-45:\radib) arc(-45:-90:\radib);\draw[red,thick] (-90:\radib) arc(-90:-100:\radib)--(-105:\radis) arc(-105:-115:\radis)--(-120:\radib) arc(-120:-145:\radib)--(-150:\radis);
				\draw (210:\radib) node[left]{$b$};
				\draw (330:\radib) node[right]{$\mathcal{T}b$};
				\draw[fill=green] (210:\radis) circle(1.5pt);
				\draw[fill=white] (330:\radis) circle(1.5pt);
			\end{tikzpicture}
			\caption{}
			\label{CSE_SR_2}
		\end{subfigure}
		\caption{}
		\label{CSE_SR_variants}
	\end{figure}
        As in the cases of $\mathcal{T}^2=1$ and the absence of $\mathcal{T}$-symmetry, there are two additional $SR$ diagrams for each case in Fig.~\ref{CSE_SR_variants}. According to Eq.~(\ref{Z_pi_CSE_general}), only diagrams containing red arcs with counterclockwise arrows on the outer circle or red arcs with clockwise arrows on the inner circle can contribute in the ergodic phase. Therefore, all red arcs with clockwise arrows in Fig.~\ref{CSE_SR_variants} must be pushed to the inner circle. Further, in Fig.~\ref{CSE_SR_1}, the endpoints of the reversed parts must lie on the outer circle, whereas in Fig.~\ref{CSE_SR_2}, they must lie on the inner circle. Thus, only the diagrams in Fig.~\ref{CSE_SR_variants_erg} contribute to the SFF in the ergodic phase. Due to jumps at the ends of the lower arc in Fig.~\ref{CSE_SR_1_erg}, the reduced diagram obtained by removing the lower arc vanishes according to \textbf{Theorem 5}. According to \textbf{Property 2} in Sec.~\ref{reduced_diagram}, all reduced diagrams of a given diagram have identical Type  \Rom{1} terms. Therefore, if even one reduced diagram vanishes, the original diagram does not contribute Type \Rom{1} term. Thus, Fig.~\ref{CSE_SR_1_erg} does not contribute to the SFF in the ergodic phase. We evaluate Fig.~\ref{CSE_SR_2_erg} by applying the rules in Sec. \ref{CSE_rules}. The symbols $\mathcal{T}c$ and $c$ in Fig.~\ref{CSE_SR_2_erg} represent the states at time steps $\tau_1$ and $\tau_2$. Therefore, Eq.~\ref{CSE_second_order} gives the below expression by ignoring the Type III terms:
 %       Similar to the $\mathcal{T}^2=1$ and $\mathcal{T}$ absent case, there are two extra $SR$ diagrams of each of the case in Fig.~\ref{CSE_SR_variants}. Since, only variants with red arcs of counterclockwise arrow on the outer circle or clockwise arrow on the inner circle can have non-vanishing contribution in the ergodic phase, Eq.~(\ref{Z_pi_CSE_general}), all the red arcs with clockwise arrows in Fig.~\ref{CSE_SR_variants} must be pushed to the inner circle. In Fig.~\ref{CSE_SR_1} the end points of the reversed parts must be on the outer circle whereas in Fig.~\ref{CSE_SR_2}, they must be on the inner circle, thus, only diagrams in Fig.~\ref{CSE_SR_variants_erg} contribute to SFF in the ergodic phase. Due to jumps at the ends of lower arc in the Fig.~\ref{CSE_SR_1_erg} the reduced diagram with respect to lower arc vanish according to \textbf{Theorem 5}. According to \textbf{Property 2}, all reduced diagrams of a given diagram have identical Type \Rom{1} term. Therefore, vanishing contribution of even one reduced diagram is sufficient to conclude that an original diagram does not contribute a Type \Rom{1} term. Thus, Fig.~\ref{CSE_SR_1_erg} does not contribute to SFF in the ergodic phase. We evaluate Fig.~\ref{CSE_SR_2_erg} by applying rules in Sec. \ref{CSE_rules} and find
	\begin{figure}[h]
		\centering
		\begin{subfigure}[t]{0.45\linewidth}
			\centering
			\begin{tikzpicture}
				\def\radis{0.75}
				\def\radib{1}
				\draw[blue,dashed] (0:\radib) arc(0:360:\radib);
				\draw[black,dashed] (0:\radis) arc(0:360:\radis);
				\draw[red,thick] (210:\radib)--(30:\radib);\draw[red,thick,->] (30:\radib) arc(30:90:\radib);\draw[red,thick] (90:\radib) arc(90:150:\radib)--(-30:\radib);
				\draw[red,thick,->] (-30:\radib)--(-40:\radis)--(-45:\radis) arc(-45:-90:\radis);\draw[red,thick] (-90:\radis) arc(-90:-100:\radis)--(-105:\radis) arc(-105:-115:\radis)--(-120:\radis) arc(-120:-140:\radis)--(-150:\radib);
				\draw[fill=green] (210:\radib) circle(1.5pt) node[left]{$a$};
				\draw[fill=green] (-30:\radib) circle(1.5pt) node[right]{$a$};
			\end{tikzpicture}
			\caption{}
			\label{CSE_SR_1_erg}
		\end{subfigure}
		\hfill
		\begin{subfigure}[t]{0.45\linewidth}
			\centering
			\begin{tikzpicture}
				\def\radis{0.75}
				\def\radib{1}
				\draw[blue,dashed] (0:\radib) arc(0:360:\radib);
				\draw[black,dashed] (0:\radis) arc(0:360:\radis);
				\draw[red,thick] (210:\radis)--(30:\radib);\draw[red,thick,->] (30:\radib) arc(30:90:\radib);\draw[red,thick] (90:\radib) arc(90:150:\radib)--(-30:\radis);
				\draw[red,thick,->] (-30:\radis) arc(-30:-90:\radis);\draw[red,thick] (-90:\radis) arc(-90:-150:\radis);
				\draw (210:\radib) node[left]{$\mathcal{T}c$};
				\draw (330:\radib) node[right]{$c$};
				\draw (30:\radib) node[right]{$d$};
				\draw (150:\radib) node[left]{$e$};
				\draw[fill=green] (210:\radis) circle(1.5pt);
				\draw[fill=white] (330:\radis) circle(1.5pt);
			\end{tikzpicture}
			\caption{}
			\label{CSE_SR_2_erg}
		\end{subfigure}
		\caption{}
		\label{CSE_SR_variants_erg}
	\end{figure}
	\begin{widetext}
		\begin{align}
			K_{-1}^{(2)}(t)&=-\frac{t}{2}\sum_{\tau_1=1}^t\sum_{\substack{\tau_2=\tau_1+\nu,\\ \tau_1+\nu'}}\sum_{ij}\lambda_i^{t-\tau_2+\tau_1-2}\lambda_j^{\tau_2-\tau_1}\mathcal{M}^{(i)}_{d,e}\mathcal{M}^{(j)}_{\mathcal{T}c,c} V_{e,\mathcal{T}c}V_{c,d}V^*_{e,\mathcal{T}c}V^*_{-c,d}\notag\\
			&=-\frac{t^2}{2}\sum_{ij}\left(\lambda_i^{t-\nu-2}\lambda_j^{\nu}+\lambda_i^{t-\nu'-2}\lambda_j^{\nu'}\right)\mathcal{M}^{(i)}_{d,e}\mathcal{M}^{(j)}_{\mathcal{T}c,c} V_{e,\mathcal{T}c}V_{c,d}V^*_{e,\mathcal{T}c}V^*_{-c,d}\notag\\
			&=-\frac{t^2}{\mathcal{N}^2}V_{e,\mathcal{T}c}V_{c,d}V^*_{e,\mathcal{T}c}V^*_{-c,d}-\frac{t^2}{2\mathcal{N}}\sum_{i\neq 0}\left(\left(\lambda_j^{\nu}+\lambda_j^{\nu'}\right)\mathcal{M}^{(i)}_{\mathcal{T}c,c}+\left(\lambda_i^{t-\nu-2}+\lambda_i^{t-\nu'-2}\right)\mathcal{M}^{(i)}_{d,e}\right) V_{e,\mathcal{T}c}V_{c,d}V^*_{e,\mathcal{T}c}V^*_{-c,d}\notag\\
            &-\frac{t^2}{2}\sum_{\substack{i\neq 0,\\j\neq 0}}\left(\lambda_i^{t-\nu-2}\lambda_j^{\nu}+\lambda_i^{t-\nu'-2}\lambda_j^{\nu'}\right)\mathcal{M}^{(i)}_{d,e}\mathcal{M}^{(j)}_{\mathcal{T}c,c} V_{e,\mathcal{T}c}V_{c,d}V^*_{e,\mathcal{T}c}V^*_{-c,d}.
			\label{CSE_2nd_uni_SFF}
		\end{align}
	\end{widetext}
We find by summing over matrix indices using the unitary property of the matrix $V$:	\begin{align}
		V_{e,\mathcal{T}c}V_{c,d}V^*_{e,\mathcal{T}c}V^*_{-c,d}&=-\mathcal{N}.
		\label{CSE_V_matrices_contraction}
	\end{align}
	We substitute Eq.~(\ref{CSE_V_matrices_contraction}) in Eq.~(\ref{CSE_2nd_uni_SFF}) and use the fact that all the arc sizes $\nu$, $\nu'$, $t-\nu-2$, and $t-\nu'$ diverge with increasing time to get
	\begin{align}
		K_{-1}^{(2)}(t)&=\frac{t^2}{\mathcal{N}}+\mathcal{O}(\lambda_1^\nu).
	\end{align}
	According to \textbf{Theorem 1}, each diagram has a time-reversed counterpart that contributes identically. Thus, we must include a factor of 2, which leads to
	\begin{align}
		K_{-1}^{(2)}(t)&=\frac{2t^2}{\mathcal{N}}+\mathcal{O}(\lambda_1^\nu).
	\end{align}
	The last result is the second-order correction to the SFF for the CSE class of RMT in Eq.~(\ref{CSE_RMT_SFF}). Further, the extra SR diagrams for this case do not resemble near-miss and self-crossing orbits of the Sieber-Richter pairs of a semiclassical analysis \cite{Stefan_Heusler_2001}. Unlike $\mathcal{T}^2=1$, we could not give a simple probabilistic interpretation of the second-order correction in this, which probably indicates a fully quantum origin of this contribution.   
	
	\section{Higher Dimensions}
	\label{Higher_dimensions}
	Previous studies using RPA have explored the emergence of a universal RMT form of SFF, mainly in 1D physical models of strongly interacting fermions, bosons, qubits, and their mixtures \cite{KosPRX2018,RoyPRE2020,RoyPRE2022,Kumar2024}. Nevertheless, as we mentioned in Sec.~\ref{Intro}, it would be easier to experimentally probe our many results in higher dimensions with less control. Further, higher dimensional models are abundant in nature contrary to strictly 1D models, which are more of a simplification. An example below shows that the above formalism to derive SFF analytically works in higher dimensions. Particularly, the Eqs.~(\ref{SFF_RPA_X_pi})-(\ref{SFF_RPA_Z_pi}) are equally valid for higher dimensional many-body systems. We take a two-dimensional (2D) version of the $\mathcal{T}$-invariant system of interacting spinless fermions with $\mathcal{T}^2=1$ given in Eq.~(\ref{Hs0}):
	\begin{align}
		\hat{H}^0_0&=\sum_{\vec{x}}\epsilon_{\vec{x}}\hat{n}_{\vec{x}}+\frac{1}{2}\sum_{\vec{x}\neq \vec{y}}\frac{U_0}{|\vec{x}-\vec{y}|^\alpha}\hat{n}_{\vec{x}}\hat{n}_{\vec{y}},\\
		\hat{H}^0_1&=\sum_{\langle \vec{x},\vec{y}\rangle}-J\hat{c}^\dagger_{\vec{x}}\hat{c}_{\vec{y}}+\Delta \hat{c}^\dagger_{\vec{x}}\hat{c}^\dagger_{\vec{y}}+h.c.,
	\end{align}
	where $\vec{x}\equiv (x_1,x_2)$ is the position of a site on the 2D square lattice with $1\leq x_1\leq L_1,1\leq x_2\leq L_2$, and $\langle\vec{x},\vec{y}\rangle$ denotes nearest-neighbor sites. We use PBC in both directions of the lattice. Here, $J$ and $\Delta$ represent hopping and pairing amplitude between nearest-neighbor sites in both directions. We choose them to be real for the COE class. Further, $\alpha$ denotes the range of particle-particle interactions between different sites. We noticed that the RPA works better for 2D models than 1D ones, even for larger $\alpha$ values (e.g., $\alpha>1.5$). This is due to the higher number of neighbors (coordination numbers) at a fixed distance for 2D rather than 1D, which improves the approximation of independent and uniformly distributed eigenphases $\theta_{\underline{n}}$ made of eigenvalues of $\hat{H}_0$. 
	
	We can again write the leading-order SFF for 2D model by Eq.~(\ref{SFF_COE_leading_order}). The Markov matrix $\mathcal{M}$ now can be mapped to a 2D XXZ-Heisenberg Hamiltonian in the Trotter regime of small parameters $(|J|,|\Delta| \ll 1)$:
	\begin{align}
		\mathcal{M}=(1-c_1 L_1L_2)\mathds{1}_{\mathcal{N}}+\sum_{\langle\vec{x},\vec{y}\rangle}\sum_\nu c_\nu s_{\vec{x}}^\nu s_{\vec{y}}^\nu+\mathcal{O}(J^4,\Delta^4),\label{mapped2D}
	\end{align}
	where $c_1=(|J|^2+|\Delta|^2)/2,c_2=c_3=(|J|^2-|\Delta|^2)/2$ and $s_{\vec{x}}^\nu,\nu=1,2,3$ are Pauli matrices at site $\vec{x}$. In the absence of pairing $(\Delta=0)$, we have a $SU(2)$ invariant mapped Hamiltonian in Eq.~\ref{mapped2D}, whose second-largest eigenvalue $\lambda_1$ is independent of total number of particles, $\hat{N}=\sum_{\vec{x}}\hat{n}_{\vec{x}}$. Thus, $\lambda_1$ can be calculated analytically in the single particle sector. We find $\lambda_1\approx 1-2\pi^2J^2/L^2$ in the thermodynamic limit, where $L={\rm max}(L_1,L_2)$. This leads to $t^* \propto L^2$, which is similar to 1D case. For $\Delta=J$, the mapped Hamiltonian in Eq.~\ref{mapped2D} becomes a 2D Ising Hamiltonian:
	\begin{align}
		\mathcal{M}&=(1-J^2 L_1L_2)\mathds{1}_\mathcal{N}+J^2\sum_{\langle\vec{x},\vec{y}\rangle} s^1_{\vec{x}}s^1_{\vec{y}}+\mathcal{O}(J^4,\Delta^4).
	\end{align}
	The second-largest eigenvalue of $\mathcal{M}$ for this case is $1-8J^2$ with a degeneracy of $L_1 L_2$. Therefore, we then have $t^*\propto \ln(L_1 L_2)$, indicating a logarithmic system-size dependence, which is again similar to the 1D case. While we demonstrate that our formalism works for 2D systems of the COE class, the above argument can be generalized for all three Wigner-Dyson classes in both 2D and three-dimensional systems. 
	\section{Conclusion}
	\label{Conclusion}
	This work extends many recent efforts to analytically calculate the spectral statistics in periodically kicked interacting many-body quantum systems in chaotic regimes to all three of Dyson's circular ensembles. Our work is the first study of such systems with a CSE class. For this, we have developed an ingenious scheme to include contributions from many diagrams of different permutations of basis states. We derived SFFs up to two leading orders in time for generic periodically kicked systems supporting the RPA. We showed that leading-order SFF is determined by identity permutation similar to the diagonal approximation in semiclassical proof \cite{Berry1985}. In contrast, the diagrams leading to second-order correction for $\mathcal{T}$-invariant systems with $\mathcal{T}^2=1$ are identical to the Sieber-Richter pairs \cite{Sieber2002}, even in pure quantum systems without classical limits. We provided general rules to calculate the contribution of different diagrams and discovered reduced diagrams that contain information on the contribution to SFF in the ergodic phase. Reduced diagrams allow more analytical control and can play a significant role in deriving the complete SFF in the ergodic phase. Our study also reveals an underlying stochastic mechanism determining the emergence of universal RMT behavior. More specifically, the general properties of the doubly stochastic matrix $\mathcal{M}$ determine the universal RMT form of SFF in all the three cases of $\mathcal{T}$-symmetry ($\mathcal{T}^2=1$, absence of $\mathcal{T}$, and $\mathcal{T}^2=-1$). Further, we studied chains of spinless and spinful fermions to derive system-size scaling of the Thouless time using the second-largest eigenvalue of $\mathcal{M}$, which can be obtained numerically for large system sizes and analytically calculated in the Trotter regime due to the $SU(2)$ symmetry of $\mathcal{M}$ for all the three classes in the presence of $U(1)$ symmetry. Without $U(1)$ symmetry, numerical study reveals that the second-largest eigenvalue of $\mathcal{M}$ is independent of system size. If the eigenvalue has degeneracy $\mathcal{O}(L^0)$, then $t^*$ is also $\mathcal{O}(L^0)$; whereas if the degeneracy is $\mathcal{O}(L^\zeta)$, then $t^*$ is $\mathcal{O}(\ln(L))$. In the end, we showed that similar system-size scalings of the Thouless time can also be found in higher dimensions.
	
	In recent years, many proposals have been made using cold atoms \cite{Dag23} and quantum simulators \cite{Joshi2022} to detect spectral statistics in various many-body quantum chaotic systems \cite{das2024proposalmanybodyquantumchaos}. There have also been recent efforts to measure SFF and various correlations in different many-body quantum chaotic systems using quantum processors \cite{SFF_QuantumProcessor} or computers \cite{Laurin2024}. We thus hope our findings, particularly system-size scaling of $t^*$ for the emergence of RMT SFF form with different unitary and anti-unitary symmetries (Tab.~\ref{Thouless_time_table})  can be tested in coming years using these above methods. The techniques developed in this article can be applied to derive other physically relevant quantities like correlation functions and entanglement entropies in periodically kicked interacting many-body quantum systems supporting the RPA \cite{Bertini2019PRL}. It would be interesting to check if the formalism can be extended to account for other symmetries leading to a ten-fold classification of Altland and Zirnbauer \cite{Altland_1997}. Some of the challenges in this direction in the near future are improving the RPA beyond the present form and deriving spectral statistics in autonomous systems. 	
	%The techniques developed in this article can be used to calculate other quantities like correlation function and entanglement entropy for periodically kicked systems supporting the RPA. Also, it would be interesting to see if the formalism can be extended to account for other symmetries leading to ten-fold classification of Altland and Zirnbauer \cite{Altland_1997}.
	
	\section*{Acknowledgment}
	This research was supported in part by the International Centre for Theoretical Sciences (ICTS) for participating in the program -  Stability of Quantum Matter in and out of Equilibrium at Various Scales    (code:ICTS/SQMVS2024/01). TP acknowledges support by European Research Council (ERC) through Advanced grant QUEST (Grant Agreement No. 101096208), and Slovenian Research and Innovation agency (ARIS) through the Program P1-0402 and Grants N1-0219, N1-0368. We would like to thank Ajay Sharma for useful discussions.
	\appendix
	\section{Derivation of the rules to evaluate $\mathcal{X}_\pi$ and $\mathcal{X}_\pi^{\{\underline{n},\underline{n}\}}$}
	\label{COE_rules_der}
We consider a permutation with the following diagram in Fig.~\ref{path_for_rule_der}.
	\begin{figure}[H]
		\centering
		\begin{tikzpicture}
			\def\rad{1.5}
			\draw[blue,dashed] (0,0) circle(\rad);
			\draw[red,thick,->] (60:\rad) arc(60:90:\rad);\draw[red,thick] (90:\rad) arc(90:150:\rad)--(30:\rad);\draw[red,thick,->] (30:\rad) arc(30:0:\rad);\draw[red,thick] (0:\rad) arc(0:-60:\rad)--(180:\rad);\draw[red,thick,->] (180:\rad) arc(180:210:\rad);\draw[red,thick] (210:\rad) arc(210:270:\rad)--(60:\rad);
			\draw (150:\rad) node[left]{$\underline{n}_{1}$};
			\draw (180:\rad) node[left]{$\underline{n}_{2}$};
			\draw (270:\rad) node[below]{$\underline{n}_{\tau_1}$};
			\draw (300:\rad) node[below right]{$\underline{n}_{\tau_1+1}$};
			\draw (30:\rad) node[right]{$\underline{n}_{\tau_2}$};
			\draw (60:\rad) node[above right]{$\underline{n}_{\tau_2+1}$};
		\end{tikzpicture}
		\caption{}
		\label{path_for_rule_der}
	\end{figure}
	\noindent
        The initial configuration of the states can be obtained from the blue circle by reading them in a counterclockwise direction, i.e., $\{\underline{n}_{1},\underline{n}_{2},...,\underline{n}_{\tau_1},\underline{n}_{\tau_1+1},...,\underline{n}_{\tau_2},\underline{n}_{\tau_2+1},...,\underline{n}_{t}\}$. Similarly, the configuration of the states after a permutation can be obtained from the red curve, i.e., $\{\underline{n}_{1},\underline{n}_{\tau_2},...,\underline{n}_{\tau_1+1},\underline{n}_{2},...,\underline{n}_{\tau_1},\underline{n}_{\tau_2+1},...,\underline{n}_{t}\}$. According to Eq.~(\ref{X_pi}), the contribution to the SFF is
	\begin{widetext}
		\begin{align}
			\mathcal{X}_\pi&= \sum_{\underline{n}_1,...,\underline{n}_t}V_{\underline{n}_1,\underline{n}_2}\left(\prod_{\tau=2}^{\tau_1-1}V_{\underline{n}_\tau,\underline{n}_{\tau+1}}\right)V_{\underline{n}_{\tau_1},\underline{n}_{\tau_1+1}}\left(\prod_{\tau=\tau_1+1}^{\tau_2-1}V_{\underline{n}_\tau,\underline{n}_{\tau+1}}\right)V_{\underline{n}_{\tau_2},\underline{n}_{\tau_2+1}}\left(\prod_{\tau=\tau_2+1}^{t-1}V_{\underline{n}_\tau,\underline{n}_{\tau+1}}\right)V_{\underline{n}_t,\underline{n}_1}\notag\\
			&\times V^*_{\underline{n}_1,\underline{n}_{\tau_2}}\left(\prod_{\tau=\tau_1+1}^{\tau_2-1}V^*_{\underline{n}_{\tau+1},\underline{n}_{\tau}}\right)V^*_{\underline{n}_{\tau_1+1},\underline{n}_{2}}\left(\prod_{\tau=2}^{\tau_1-1}V^*_{\underline{n}_\tau,\underline{n}_{\tau+1}}\right)V^*_{\underline{n}_{\tau_1},\underline{n}_{\tau_2+1}}\left(\prod_{\tau=\tau_2+1}^{t-1}V^*_{\underline{n}_\tau,\underline{n}_{\tau+1}}\right)V^*_{\underline{n}_t,\underline{n}_1}.
		\end{align}
		As the matrix $V$ is symmetric for $\mathcal{T}$-invaraint systems when $\mathcal{T}^2=1$, and  $\mathcal{M}_{\underline{n},\underline{n}'}=|V_{\underline{n},\underline{n}'}|^2$, we can rewrite the above expression as
		\begin{align}
			\mathcal{X}_\pi&= \sum_{\underline{n}_1,...,\underline{n}_t}\left(\prod_{\tau=2}^{\tau_1-1}\mathcal{M}_{\underline{n}_\tau,\underline{n}_{\tau+1}}\right)\left(\prod_{\tau=\tau_1+1}^{\tau_2-1}\mathcal{M}_{\underline{n}_\tau,\underline{n}_{\tau+1}}\right)\left(\prod_{\tau=\tau_2+1}^{t-1}\mathcal{M}_{\underline{n}_\tau,\underline{n}_{\tau+1}}\right)\mathcal{M}_{\underline{n}_t,\underline{n}_1}\notag\\
			&\qquad\times V_{\underline{n}_1,\underline{n}_2}V_{\underline{n}_{\tau_1},\underline{n}_{\tau_1+1}}V_{\underline{n}_{\tau_2},\underline{n}_{\tau_2+1}}V^*_{\underline{n}_1,\underline{n}_{\tau_2}}V^*_{\underline{n}_{\tau_1+1},\underline{n}_{2}}V^*_{\underline{n}_{\tau_1},\underline{n}_{\tau_2+1}}\notag\\
			&=\sum_{\substack{\underline{n}_1,\underline{n}_2\\ \underline{n}_{\tau_1},\underline{n}_{\tau_1+1}\\ \underline{n}_{\tau_2},\underline{n}_{\tau_2+1}}}\left(\mathcal{M}^{\tau_1-2}\right)_{\underline{n}_2,\underline{n}_{\tau_1}}\left(\mathcal{M}^{\tau_2-\tau_1-1}\right)_{\underline{n}_{\tau_1+1},\underline{n}_{\tau_2}}\left(\mathcal{M}^{t-\tau_2}\right)_{\underline{n}_{\tau_2+1},\underline{n}_1}\notag\\
			&\qquad\times V_{\underline{n}_1,\underline{n}_2}V_{\underline{n}_{\tau_1},\underline{n}_{\tau_1+1}}V_{\underline{n}_{\tau_2},\underline{n}_{\tau_2+1}}V^*_{\underline{n}_1,\underline{n}_{\tau_2}}V^*_{\underline{n}_{\tau_1+1},\underline{n}_{2}}V^*_{\underline{n}_{\tau_1},\underline{n}_{\tau_2+1}}.
		\end{align}
We insert the eigendecomposition of $\mathcal{M}^n=\sum_{i}\lambda_i^n\mathcal{M}^{(i)}$, where $\mathcal{M}^{(i)}=|\lambda_i\rangle\langle\lambda_i|$ to find
		\begin{align}
			\mathcal{X}_\pi&=\sum_{\substack{\underline{n}_1,\underline{n}_2\\ \underline{n}_{\tau_1},\underline{n}_{\tau_1+1}\\ \underline{n}_{\tau_2},\underline{n}_{\tau_2+1}}}\left(\sum_i\lambda_i^{\tau_1-2}\mathcal{M}^{(i)}_{\underline{n}_2,\underline{n}_{\tau_1}}\right)\left(\sum_j\lambda_j^{\tau_2-\tau_1-1}\mathcal{M}^{(j)}_{\underline{n}_{\tau_1+1},\underline{n}_{\tau_2}}\right)\left(\sum_k\lambda_k^{t-\tau_2}\mathcal{M}^{(k)}_{\underline{n}_{\tau_2+1},\underline{n}_1}\right)\notag\\
			&\qquad\times V_{\underline{n}_1,\underline{n}_2}V_{\underline{n}_{\tau_1},\underline{n}_{\tau_1+1}}V_{\underline{n}_{\tau_2},\underline{n}_{\tau_2+1}}V^*_{\underline{n}_1,\underline{n}_{\tau_2}}V^*_{\underline{n}_{\tau_1+1},\underline{n}_{2}}V^*_{\underline{n}_{\tau_1},\underline{n}_{\tau_2+1}}.
            \label{X_pi_example}
		\end{align}
	\end{widetext}
	The right-hand side of Eq.~(\ref{X_pi_example}) naturally leads to the rules outlined in Sec.~\ref{COE_Rules}.\\
	\section{Proof that the spectral radius of $\tilde{\mathcal{M}}$ is less than one}
        \label{Gersgorin}
	{\it Ger$\check{\rm s}$gorin circle theorem: Let $M$ be a complex  $\mathcal{N}\times\mathcal{N}$ matrix, with entries $m_{i,j}$. For each row index $i\in \{1,...,\mathcal{N}\}$, define}
    \begin{align}
        R_i=\sum_{j\neq i}|m_{i,j}|
    \end{align}
    \textit{as the sum of the absolute values of the non-diagonal entries in the $i^{\text{th}}$ row. Let $D(m_{i,i},R_i)\subseteq \mathds{C}$ be a closed disc centered at $m_{i,i}$ with radius $R_i$. Such a disc is called a Ger$\check{\rm s}$gorin disc. The theorem states that every eigenvalue of $M$ lies within at least one of the Ger$\check{\rm s}$gorin discs $D(m_{i,i},R_i)$} \cite{gershgorin1931uber,Horn_Johnson_1985}. 
	
	We use this theorem to show that the eigenvalues of the $\mathcal{N}\times\mathcal{N}$ matrix $\tilde{\mathcal{M}}$ have magnitudes less than one. The matrix elements of $\tilde{\mathcal{M}}$ are
	\begin{align}
		\tilde{\mathcal{M}}_{a,b}=V_{a,b}V^*_{b,a},
	\end{align}
	where $a,b=1,...,\mathcal{N}$. Since $V$ is a unitary matrix, all the columns of $V$ form a complete set of orthonormal basis states. The same is true for all the rows as well. However, the row states are generally different from the column states for a non-symmetric $V$. Next, we consider the sum of the absolute values of the elements of $\tilde{\mathcal{M}}$ along $a^{\text{th}}$ row:
	\begin{align}
		\sum_{b}|\tilde{\mathcal{M}}_{a,b}|&=\sum_b |V_{a,b}||V_{b,a}|.
	\end{align}
	The right-hand side of the above expression can be interpreted as the Euclidean inner product between two real non-negative vectors normalized to unity, which are\\
	\begin{minipage}{0.2\textwidth}
		\begin{align*}
			\begin{pmatrix}
				|V_{a,1}|\\
				|V_{a,2}|\\
				.\\
				.\\
				.\\
				|V_{a,\mathcal{N}}|
			\end{pmatrix},
		\end{align*}
	\end{minipage}
	\hfill
	\begin{minipage}{0.2\textwidth}
		\begin{align*}
			\begin{pmatrix}
				|V_{1,a}|\\
				|V_{2,a}|\\
				.\\
				.\\
				.\\
				|V_{\mathcal{N},a}|
			\end{pmatrix}.
		\end{align*}
	\end{minipage}\\
	If the two vectors are distinct for each $a$, their inner product is always less than 1, i.e.,
	\begin{align}
		\sum_b |V_{a,b}||V_{b,a}|<1\implies \sum_{b}|\tilde{\mathcal{M}}_{a,b}|<1.
	\end{align}
	Thus,  the radius of the Ger$\check{\rm s}$gorin disc, $R_a=\sum_{b\neq a}|\tilde{\mathcal{M}}_{a,b}|<1-|\tilde{\mathcal{M}}_{a,a}|$.
	The center of the disc is $\tilde{\mathcal{M}}_{a,a}$, and the farthest point of $a^{th}$ Ger$\check{\rm s}$gorin disc from the origin of the complex plane has a radial distance from the origin given by $R_a+|\tilde{\mathcal{M}}_{a,a}|<1$. Therefore, all the eigenvalues will have magnitude less than 1. If the two vectors are identical for some $a$, then the largest eigenvalue can be 1. Our numerical study of the spectra of $\tilde{\mathcal{M}}$ for Hamiltonian $\hat{H}_1$ in Eq.~(\ref{H01/2}) with arbitrary complex hopping parameters reveals that the largest eigenvalue of $\tilde{\mathcal{M}}$ is one when $\hat{H}_1$ has a non-conventional time reversal symmetry \cite{Haake2001}.
	
    \section{Derivation of the rules to evaluate $\mathcal{Y}_\pi$ and $\mathcal{Y}_\pi^{\{\underline{n},\underline{n}\}}$}
	\label{T_absent_rul_der}
	We again consider the permutation diagram in Fig.~\ref{path_for_rule_der}. According to the Eq.~(\ref{Y_pi}), the contribution to the SFF is
	\vspace{2cm}
	\begin{widetext}
		\begin{align}
			\mathcal{Y}_\pi&= \sum_{\underline{n}_1,...,\underline{n}_t}V_{\underline{n}_1,\underline{n}_2}\left(\prod_{\tau=2}^{\tau_1-1}V_{\underline{n}_\tau,\underline{n}_{\tau+1}}\right)V_{\underline{n}_{\tau_1},\underline{n}_{\tau_1+1}}\left(\prod_{\tau=\tau_1+1}^{\tau_2-1}V_{\underline{n}_\tau,\underline{n}_{\tau+1}}\right)V_{\underline{n}_{\tau_2},\underline{n}_{\tau_2+1}}\left(\prod_{\tau=\tau_2+1}^{t-1}V_{\underline{n}_\tau,\underline{n}_{\tau+1}}\right)V_{\underline{n}_t,\underline{n}_1}\notag\\
			&\times V^*_{\underline{n}_1,\underline{n}_{\tau_2}}\left(\prod_{\tau=\tau_1+1}^{\tau_2-1}V^*_{\underline{n}_{\tau+1},\underline{n}_{\tau}}\right)V^*_{\underline{n}_{\tau_1+1},\underline{n}_{2}}\left(\prod_{\tau=2}^{\tau_1-1}V^*_{\underline{n}_\tau,\underline{n}_{\tau+1}}\right)V^*_{\underline{n}_{\tau_1},\underline{n}_{\tau_2+1}}\left(\prod_{\tau=\tau_2+1}^{t-1}V^*_{\underline{n}_\tau,\underline{n}_{\tau+1}}\right)V^*_{\underline{n}_t,\underline{n}_1}.
		\end{align}
		Since the matrix $V$ is not symmetric in the absence of $\mathcal{T}$-symmetry, we define $\tilde{\mathcal{M}}_{\underline{n},\underline{n}'}=V_{\underline{n},\underline{n}'}V^*_{\underline{n}',\underline{n}}$ and $\mathcal{M}_{\underline{n},\underline{n}'}=V_{\underline{n},\underline{n}'}V^*_{\underline{n},\underline{n}'}$. Thus, the above expression for $\mathcal{Y}_\pi$ reduces to
		\begin{align}
			\mathcal{Y}_\pi&= \sum_{\underline{n}_1,...,\underline{n}_t}\left(\prod_{\tau=2}^{\tau_1-1}\mathcal{M}_{\underline{n}_\tau,\underline{n}_{\tau+1}}\right)\left(\prod_{\tau=\tau_1+1}^{\tau_2-1}\tilde{\mathcal{M}}_{\underline{n}_\tau,\underline{n}_{\tau+1}}\right)\left(\prod_{\tau=\tau_2+1}^{t-1}\mathcal{M}_{\underline{n}_\tau,\underline{n}_{\tau+1}}\right)\mathcal{M}_{\underline{n}_t,\underline{n}_1}\notag\\
			&\qquad\times V_{\underline{n}_1,\underline{n}_2}V_{\underline{n}_{\tau_1},\underline{n}_{\tau_1+1}}V_{\underline{n}_{\tau_2},\underline{n}_{\tau_2+1}}V^*_{\underline{n}_1,\underline{n}_{\tau_2}}V^*_{\underline{n}_{\tau_1+1},\underline{n}_{2}}V^*_{\underline{n}_{\tau_1},\underline{n}_{\tau_2+1}}\notag\\
			&=\sum_{\substack{\underline{n}_1,\underline{n}_2\\ \underline{n}_{\tau_1},\underline{n}_{\tau_1+1}\\ \underline{n}_{\tau_2},\underline{n}_{\tau_2+1}}}\left(\mathcal{M}^{\tau_1-2}\right)_{\underline{n}_2,\underline{n}_{\tau_1}}\left(\tilde{\mathcal{M}}^{\tau_2-\tau_1-1}\right)_{\underline{n}_{\tau_1+1},\underline{n}_{\tau_2}}\left(\mathcal{M}^{t-\tau_2}\right)_{\underline{n}_{\tau_2+1},\underline{n}_1}\notag\\
			&\qquad\times V_{\underline{n}_1,\underline{n}_2}V_{\underline{n}_{\tau_1},\underline{n}_{\tau_1+1}}V_{\underline{n}_{\tau_2},\underline{n}_{\tau_2+1}}V^*_{\underline{n}_1,\underline{n}_{\tau_2}}V^*_{\underline{n}_{\tau_1+1},\underline{n}_{2}}V^*_{\underline{n}_{\tau_1},\underline{n}_{\tau_2+1}}\notag\\
			&=\sum_{\substack{\underline{n}_1,\underline{n}_2\\ \underline{n}_{\tau_1},\underline{n}_{\tau_1+1}\\ \underline{n}_{\tau_2},\underline{n}_{\tau_2+1}}}\left(\sum_i\lambda_i^{\tau_1-2}\mathcal{M}^{(i)}_{\underline{n}_2,\underline{n}_{\tau_1}}\right)\left(\sum_j\chi_j^{\tau_2-\tau_1-1}\tilde{\mathcal{M}}^{(j)}_{\underline{n}_{\tau_1+1},\underline{n}_{\tau_2}}\right)\left(\sum_k\lambda_k^{t-\tau_2}\mathcal{M}^{(k)}_{\underline{n}_{\tau_2+1},\underline{n}_1}\right)\notag\\
			&\qquad\times V_{\underline{n}_1,\underline{n}_2}V_{\underline{n}_{\tau_1},\underline{n}_{\tau_1+1}}V_{\underline{n}_{\tau_2},\underline{n}_{\tau_2+1}}V^*_{\underline{n}_1,\underline{n}_{\tau_2}}V^*_{\underline{n}_{\tau_1+1},\underline{n}_{2}}V^*_{\underline{n}_{\tau_1},\underline{n}_{\tau_2+1}},
            \label{Y_pi_example}
		\end{align}
		where we use the eigendecomposition of $\mathcal{M}^n=\sum_{i}\lambda_i^n\mathcal{M}^{(i)}$ and $\tilde{\mathcal{M}}^n=\sum_i\chi_i^n\tilde{\mathcal{M}}^{(i)}$ in the last line, and $\mathcal{M}^{(i)}=\prescript{}{R}{|\lambda_i\rangle}\langle\lambda_i|_L$ and $\tilde{\mathcal{M}}^{(i)}=|\chi_i\rangle\langle\chi_i|$. The right-hand side in Eq.~(\ref{Y_pi_example}) naturally leads to the rules in Sec.~\ref{CUE_rules}.
	\end{widetext}
    
	\section{Derivation of the second-order correction for $\mathcal{T}^2=1$ using only the rules in Sec.~\ref{COE_Rules}}
	\label{COE_2nd_full_der}
	\subsubsection{Transposition ($T$)}
	A single transposition leads to three kinds of diagrams: (a) a transposition between nearest-neighbor states ($T^{(1)}$), (b) a transposition between next-nearest-neighbor states ($T^{(2)}$), and (c) a transposition of all other pairs of states ($T'$).\\
	\begin{figure}[h]
		\centering
		\begin{subfigure}[t]{0.3\linewidth}
			\begin{tikzpicture}
				\def\rad{0.75}
				\draw[blue,dashed] (-60:\rad) arc(-60:240:\rad);
				\draw[red,thick,->] (0:\rad) arc(0:90:\rad);\draw[red,thick] (90:\rad) arc(90:180:\rad)--(300:\rad);\draw[red,thick] (300:\rad)--(270:{\rad*cos(30)});\draw[red,thick](270:{\rad*cos(30)})--(240:\rad)--(0:\rad);
				\draw (180:\rad) node[left]{$a$};
				\draw (240:\rad) node[left]{$b$};
				\draw (300:\rad) node[right]{$c$};
				\draw (0:\rad) node[right]{$d$};
			\end{tikzpicture}
			\caption*{$T^{(1)}$}
		\end{subfigure}
        \hfill
		\begin{subfigure}[t]{0.3\linewidth}
			\begin{tikzpicture}
				\def\rad{0.75}
				\draw[blue,dashed] (-30:\rad) arc(-30:210:\rad);
				\draw[red,thick,->] (30:\rad) arc(30:90:\rad);\draw[red,thick] (90:\rad) arc(90:150:\rad)--(330:\rad);\draw[red,thick] (330:\rad)--(270:\rad);\draw[red,thick](270:\rad)--(210:\rad)--(30:\rad);
				\draw (150:\rad) node[above left]{$a$};
				\draw (210:\rad) node[left]{$b$};
				\draw (270:\rad) node[below]{$c$};
				\draw (330:\rad) node[right]{$d$};
				\draw (30:\rad) node[above right]{$e$};
			\end{tikzpicture}
			\caption*{$T^{(2)}$}
		\end{subfigure}
        \hfill
        \begin{subfigure}[t]{0.3\linewidth}
			\begin{tikzpicture}
				\def\rad{0.75}
				\draw[blue,dashed] (0,0) circle(\rad);
				\draw[red, thick,->] (30:\rad) arc(30:90:\rad);\draw[red,thick,->](90:\rad) arc(90:150:\rad)--(0:\rad)--(-150:\rad) arc(-150:-90:\rad);\draw[red,thick] (-90:\rad) arc(-90:-30:\rad)--(180:\rad)--(30:\rad);
				\draw (150:\rad) node[above left]{$a$};
				\draw (180:\rad) node[left]{$b$};
				\draw (210:\rad) node[below left]{$c$};
				\draw (30:\rad) node[above right]{$f$};
				\draw (0:\rad) node[right]{$e$};
				\draw (330:\rad) node[below right]{$d$};
			\end{tikzpicture}
			\caption*{$T'$}
		\end{subfigure}
		\caption{\justifying\small The state $b$ at time $\tau_1$ and the state $c$ at time $\tau_1+1$ are interchanged in transposition $T^{(1)}$. The state $b$ at time $\tau_1$ and the state $d$ at time $\tau_1+2$ are interchanged in transposition $T^{(2)}$. The state $b$ at time $\tau_1$ and the state $e$ at time $\tau_2$ are interchanged in transposition $T'$.}
	\end{figure}
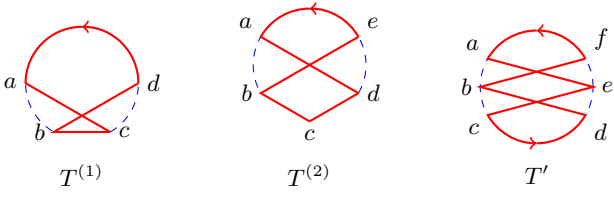
	\noindent Following the rules in Sec.~\ref{COE_Rules}, we can find the contribution of  the transpositions as
	\begin{align}
		\mathcal{X}_{T'}&=\left(\sum_i\lambda_i^{\tau_2-\tau_1-2}\mathcal{M}^{(i)}_{c,d}\right)\left(\sum_j\lambda_j^{t-\tau_2+\tau_1-2}\mathcal{M}^{(j)}_{f,a}\right)\notag\\
		&\times V_{a,b}V_{b,c}V_{d,e}V_{e,f} V^*_{a,e}V^*_{e,c}V^*_{d,b}V^*_{b,f},
		\label{X_T'}\\
		\mathcal{X}_{T^{(1)}}&=\left(\sum_i\lambda_i^{t-3}\mathcal{M}^{(i)}_{d,a}\right)V_{a,b}\mathcal{M}_{b,c}V_{c,d}V^*_{a,c}V^*_{b,d},
		\label{X_T1}\\
		\mathcal{X}_{T^{(2)}}&=\left(\sum_i\lambda_i^{t-4}\mathcal{M}^{(i)}_{e,a}\right)V_{a,b}\left(\mathcal{M}^2\right)_{b,d}V_{d,e}V^*_{a,d}V^*_{b,e}.
		\label{X_T2}
	\end{align}
	We separately calculate the contribution of different single transpositions and their cyclic and anticyclic variants by summing over allowed values of $\tau_1,\tau_2$ and multiplying by a factor of $2t$, respectively. For each case of transposition $1\leq\tau_1\leq t$. Since, each arc has minimum size 1 and we have PBC in time, $\tau_1+3\leq\tau_2\leq t+\tau_1-3$ for $T'$. Therefore,
	\begin{align}
		\bar{\mathcal{X}}_{T'}&=(2t)\frac{1}{2}\sum_{\tau_1=1}^t\sum_{\tau_2=\tau_1+3}^{t+\tau_1-3}\sum_{ij}\lambda_i^{\tau_2-\tau_1-2}\lambda_j^{t-\tau_2+\tau_1-2}Q^{ij}_{T'},
		\label{X_bar_T'}\\
		\bar{\mathcal{X}}_{T^{(1)}}&=(2t)\sum_{\tau_1=1}^t\sum_i\lambda_i^{t-3}Q^{i}_{T^{(1)}},
		\label{X_bar_T1}\\
		\bar{\mathcal{X}}_{T^{(2)}}&=(2t)\sum_{\tau_1=1}^t\sum_i\lambda_i^{t-4}Q^{i}_{T^{(2)}},
		\label{X_bar_T2}
	\end{align}
	where
	\begin{align}
		Q^{ij}_{T'}&=\mathcal{M}^{(i)}_{c,d}\mathcal{M}^{(j)}_{d,a}\notag\\
						&\times V_{a,b}V_{b,c}V_{d,e}V_{e,f} V^*_{a,e}V^*_{e,c}V^*_{d,b}V^*_{b,f},
		\label{Q_ij_T'}\\
		Q^i_{T^{(1)}}&=\mathcal{M}^{(i)}_{d,a}V_{a,b}\mathcal{M}_{b,c}V_{c,d}V^*_{a,c}V^*_{b,d},
		\label{Q_i_T1}\\
		Q^i_{T^{(2)}}&=\mathcal{M}^{(i)}_{e,a}V_{a,b}\left(\mathcal{M}^2\right)_{b,d}V_{d,e}V^*_{a,d}V^*_{b,e},
		\label{Q_i+T2}
	\end{align}
	and a factor of $1/2$ is inserted in (\ref{X_bar_T'}) to avoid double counting as changing the order of $\tau_1,\tau_2$ does not lead to a new transposition. We need the following relations for our next step
	\begin{align}
		\mathcal{M}^{(0)}_{a,b}&=\frac{1}{\mathcal{N}},
		\label{M0}\\
		\mathcal{M}^{(i)}\mathcal{M}^{(j)}&=\mathcal{M}^{(i)}\delta_{ij},
		\label{MiMj}\\
		\text{tr} \mathcal{M}^{(i)}&=1.
		\label{trMi}
	\end{align}
	Applying these relations along with the unitary property of the matrix $V$, we obtain
	\begin{align}
		Q^{00}_{T'}&=\frac{1}{\mathcal{N}},
		\label{Q00_T'}\\
		Q^{i0}_{T'}&=Q^{0i}_{T'}=\frac{\lambda_i^2}{\mathcal{N}},\; i=1,2,...,\mathcal{N}-1
		\label{Qi0_T'}\\
		Q^{(0)}_{T^{(1)}}&=\frac{1}{\mathcal{N}}\sum_i\lambda_i,
		\label{Q0_T1}\\
		Q^{(0)}_{T^{(2)}}&=\frac{1}{\mathcal{N}}\sum_i\lambda_i^2.
		\label{Q0_T2}
	\end{align}
	We next substitute Eqs.~(\ref{Q00_T'}),(\ref{Qi0_T'}) in Eq.~(\ref{X_bar_T'}), Eq.~(\ref{Q0_T1}) in Eq.~(\ref{X_bar_T1}), and Eq.~(\ref{Q0_T2}) in Eq.~(\ref{X_bar_T2}) to find
	\begin{align}
		\bar{\mathcal{X}}_{T'}&=\frac{t^2(t-5)}{\mathcal{N}}+\sum_{i\neq 0}\frac{2t^2(\lambda_i^3-\lambda_i^{t-2})}{\mathcal{N}(1-\lambda_i)}\notag\\
		&+t^2\sum_{\substack{i\neq 0\\j\neq 0}} \frac{\lambda_i^{t-4}\lambda_j-\lambda_i\lambda_j^{t-4}}{\lambda_i-\lambda_j}Q^{ij}_{T'},
		\label{X_bar_T'_result}\\
		\bar{\mathcal{X}}_{T^{(1)}}&=\frac{2t^2}{\mathcal{N}}+\frac{2t^2}{\mathcal{N}}\sum_{i\neq 0}\lambda_i+2t^2\sum_{i\neq 0}\lambda_i^{t-3}Q^{i}_{T^{(1)}},
		\label{X_bar_T1_result}\\
		\bar{\mathcal{X}}_{T^{(2)}}&=\frac{2t^2}{\mathcal{N}}+\frac{2t^2}{\mathcal{N}}\sum_{i\neq 0}\lambda_i^2+2t^2\sum_{i\neq 0}\lambda_i^{t-4}Q^{i}_{T^{(2)}}.
		\label{X_bar_T2_result}
	\end{align}
	We find the total contribution of all the single transpositions ($T$) by adding the contributions in Eqs.~(\ref{X_bar_T'_result}), (\ref{X_bar_T1_result}), and (\ref{X_bar_T2_result}):
	\begin{align}
		\bar{\mathcal{X}}_{T}&=\frac{t^2(t-1)}{\mathcal{N}}+\frac{2t^2}{\mathcal{N}}\sum_{i\neq 0}\frac{\lambda_i}{1-\lambda_i}\notag\\
		&+t^2\sum_{i\neq 0}(-\frac{2\lambda_i^{t-2}}{\mathcal{N}(1-\lambda_i)}+2\lambda_i^{t-3}Q^{i}_{T^{(1)}}+2\lambda_i^{t-4}Q^{i}_{T^{(2)}})\notag\\
		&+t^2\sum_{\substack{i\neq 0\\j\neq 0}} \frac{\lambda_i^{t-4}\lambda_j-\lambda_i\lambda_j^{t-4}}{\lambda_i-\lambda_j}Q^{ij}_{T'}.
		\label{X_bar_T_COE}
	\end{align}
	
	\subsubsection{Sub-sequence reversal ($S$)}
\begin{figure}[h]
		\centering
		\begin{tikzpicture}
			\def\rad{0.75}
			\draw[blue,dashed] (0,0) circle(\rad);
			\draw[red,thick,->] (30:\rad) arc(30:90:\rad);\draw[red,thick,->] (90:\rad) arc(90:150:\rad)--(-30:\rad) arc(-30:-90:\rad);\draw[red,thick] (-90:\rad) arc(-90:-150:\rad)--(30:\rad);
			\draw (150:\rad) node[above left]{$a$};
			\draw (210:\rad) node[below left]{$b$};
			\draw (330:\rad) node[below right]{$c$};
			\draw (30:\rad) node[above right]{$d$};
		\end{tikzpicture}
		\caption{\justifying\small A diagram representing sub-sequence reversal, where the order of states from time step $\tau_1$ to $\tau_2$ is reversed and the states at time steps $\tau_1$ and $\tau_2$ are denoted by $b$ and $c$, respectively.}
		\label{Sub_seq_Revrsal}
	\end{figure}
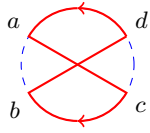

	A sub-sequence reversal permutation reverses the order of some consecutive states. Let us Consider a configuration of states on the blue circle for $t=9$ as $\{\underline{n}_1,\underline{n}_2,\underline{n}_3,\underline{n}_4,\underline{n}_5,\underline{n}_6,\underline{n}_7,\underline{n}_8,\underline{n}_9\}$. The configuration $\{\underline{n}_1,\underline{n}_2,\underline{n}_6,\underline{n}_5,\underline{n}_4,\underline{n}_3,\underline{n}_7,\underline{n}_8,\underline{n}_9\}$ can be obtained by performing a sub-sequence reversal from time steps 3 to 6. A sub-sequence reversal is shown diagrammatically in Fig. \ref{Sub_seq_Revrsal}.
	We apply the rules in Sec.~\ref{COE_Rules} to find
	\begin{align}
		\mathcal{X}_{S}&=\sum_{ij}\lambda_i^{\tau_2-\tau_1}\lambda_j^{t-\tau_2+\tau_1-2}Q^{ij}_{S},
	\end{align}
	where
	\begin{align}
		Q^{ij}_{S}&=\mathcal{M}^{(i)}_{b,c}\mathcal{M}^{(j)}_{d,a}V_{a,b}V_{c,d}V^*_{a,c}V^*_{b,d}
	\end{align}
	To determine the contribution of all the $S$ diagrams, we sum over all the allowed values of $\tau_1$ and $\tau_2$. Naturally, $1\leq \tau_1\leq t$, however, we observe that the $S$ diagrams resemble those of nearest-neighbor and next-nearest-neighbor transpositions. Therefore, the allowed values of $\tau_2$ should be such that the $S$ diagrams do not overlap with nearest-neighbor and next-nearest-neighbor transposition diagrams. This leads to $\tau_1+3\leq \tau_2\leq t+\tau_1-5$. Additionally, we observe that changing the order of $\tau_1,\tau_2$ leads to a different sub-sequence reversal shown in Fig.~\ref{Reversal_b}. However, the new sub-sequence reversal is just an anticyclic variant, shown in Fig.~\ref{Reversal_d}, of a sub-sequence reversal from $\tau_1+1$ to $\tau_2-1$ in Fig.~\ref{Reversal_c}. Since we sum over all allowed values of $\tau_1,\tau_2$ and explicitly include contribution of cyclic and anticyclic variants by including a factor of $2t$, we must include a factor of $1/2$ while summing the contribution of all the $S$ diagrams:
    \begin{figure}[h]
    \centering
    \begin{subfigure}{0.48\columnwidth} % 48% of the column width
        \centering
        \begin{tikzpicture}
				\def\rad{0.75}
                \draw[blue,dashed] (0,0) circle(\rad);
				\draw[red,thick,->] (30:\rad) arc(30:90:\rad);\draw[red,thick,->] (90:\rad) arc(90:150:\rad)--(0:\rad) arc(0:-90:\rad);\draw[red,thick] (-90:\rad) arc(-90:-180:\rad)--(30:\rad);
				\draw (180:\rad) node[left]{$\tau_1$};
				\draw (0:\rad) node[right]{$\tau_2$};
		\end{tikzpicture}
        \subcaption{}
        \label{Reversal_a}
    \end{subfigure}
    \hfill
    \begin{subfigure}{0.48\columnwidth}
        \centering
        \begin{tikzpicture}
				\def\rad{0.75}
                \draw[blue,dashed] (0,0) circle(\rad);
				\draw[red,thick,->] (180:\rad) arc(180:90:\rad);\draw[red,thick,->] (90:\rad) arc(90:0:\rad)--(-150:\rad) arc(-150:-90:\rad);\draw[red,thick] (-90:\rad) arc(-90:-30:\rad)--(180:\rad);
				\draw (180:\rad) node[left]{$\tau_1$};
				\draw (0:\rad) node[right]{$\tau_2$};
		\end{tikzpicture}
        \subcaption{}
        \label{Reversal_b}
    \end{subfigure}

    \vspace{0.3cm} % Vertical space between rows

    \begin{subfigure}{0.48\columnwidth}
        \centering
        \begin{tikzpicture}
				\def\rad{0.75}
                \draw[blue,dashed] (0,0) circle(\rad);
				\draw[red,thick,->] (0:\rad) arc(0:90:\rad);\draw[red,thick,->] (90:\rad) arc(90:180:\rad)--(-30:\rad) arc(-30:-90:\rad);\draw[red,thick] (-90:\rad) arc(-90:-150:\rad)--(0:\rad);
				\draw (-150:\rad) node[left]{$\tau_1+1$};
				\draw (-30:\rad) node[right]{$\tau_2-1$};
				\draw (180:\rad) node[left]{$\tau_1$};
				\draw (0:\rad) node[right]{$\tau_2$};
		\end{tikzpicture}
        \subcaption{}
        \label{Reversal_c}
    \end{subfigure}
    \hfill
    \begin{subfigure}{0.48\columnwidth}
        \centering
        \begin{tikzpicture}
				\def\rad{0.75}
                \draw[blue,dashed] (0,0) circle(\rad);
				\draw[red,thick,->] (180:\rad) arc(180:90:\rad);\draw[red,thick,->] (90:\rad) arc(90:0:\rad)--(-150:\rad) arc(-150:-90:\rad);\draw[red,thick] (-90:\rad) arc(-90:-30:\rad)--(180:\rad);
				\draw (180:\rad) node[left]{$\tau_1$};
				\draw (0:\rad) node[right]{$\tau_2$};
				\draw (-150:\rad) node[left]{$\tau_1+1$};
				\draw (-30:\rad) node[right]{$\tau_2-1$};
		\end{tikzpicture}
        \subcaption{}
        \label{Reversal_d}
    \end{subfigure}

    \caption{}
    \label{S_overcounting}
    \end{figure}
	
	\begin{align}
		\bar{\mathcal{X}}_{S}&=(2t)\frac{1}{2}\sum_{\tau_1=1}^t\sum_{\tau_2=\tau_1+3}^{t+\tau_1-5}\mathcal{X}_{S}\notag\\
		&=t^2\sum_{ij}\frac{\lambda_i^{t-4}\lambda_j^3-\lambda_i^3\lambda_j^{t-4}}{\lambda_i-\lambda_j}Q^{ij}_{S}\notag\\
		&=\frac{t^2(t-7)}{\mathcal{N}}+2t^2\sum_{i\neq 0}\frac{\lambda_i^3}{\mathcal{N}(1-\lambda_i)}\notag\\
		&-2t^2\sum_{i\neq 0}\frac{\lambda_i^{t-3}}{\mathcal{N}(1-\lambda_i)}+t^2\sum_{\substack{i\neq 0\\j\neq 0}}\frac{\lambda_i^{t-4}\lambda_j^3-\lambda_i^3\lambda_j^{t-4}}{\lambda_i-\lambda_j}Q^{ij}_{S},
		\label{X_bar_S_COE}
	\end{align}
	where we have used $Q^{00}_{S}=1/\mathcal{N},Q^{i0}_{S}=Q^{0i}_{S}=1/\mathcal{N}$.
	
	\subsubsection{Identity permutation with repetition ($R$)}
	
	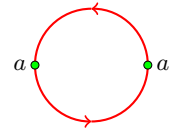
\begin{figure}[h]
		\centering
		\begin{tikzpicture}
			\def\radis{0.75}
			\draw[red,thick,->] (-90:\radis) arc(-90:90:{\radis});\draw[red,thick,->] (90:\radis) arc(90:270:\radis);
			\draw[fill=green] (180:{\radis}) circle(1.5pt) node[left]{$a$};
			\draw[fill=green] (0:{\radis}) circle(1.5pt) node[right]{$a$};
		\end{tikzpicture}
		\caption{\justifying\small A diagram representing identity permutation with repetition, where the states at time step $\tau_1$ and $\tau_2$ are the same and referred to as $a$.}
		\label{one_repetition}
	\end{figure}
	The contribution of the $R$ diagram in Fig.~\ref{one_repetition} can be found using the rules in Sec.~\ref{COE_Rules}. It is
	\begin{align}
		\mathcal{X}_{I}^{\{\underline{n},\underline{n}\}}&=\left(\sum_i\lambda_i^{\tau_2-\tau_1}\mathcal{M}^{(i)}_{a,a}\right)\left(\sum_j\lambda_j^{\tau_2-\tau_1}\mathcal{M}^{(j)}_{a,a}\right)\notag\\
		&=\sum_{ij}\lambda_i^{\tau_2-\tau_1}\lambda_j^{t-\tau_2+\tau_1}Q^{ij}_R,
	\end{align}
	where
	\begin{align}
		Q^{ij}_R&=\mathcal{M}^{(i)}_{a,a}\mathcal{M}^{(j)}_{a,a}.
	\end{align}
	For this case, all possible combinations of $\tau_1,\tau_2$ are allowed. Thus, $1\leq\tau_1\leq t$ and $\tau_1+1\leq\tau_2\leq t+\tau_1-1$. We also include a factor of $1/2$ to avoid double counting of pairs of $\tau_1,\tau_2$:
	\begin{align}
		\bar{\mathcal{X}}_{I}^{\{\underline{n},\underline{n}\}}&=(2t)\frac{1}{2}\sum_{\tau_1=1}^t\sum_{\tau_2=\tau_1+1}^{t+\tau_1-1}\sum_{ij}\lambda_i^{\tau_2-\tau_1}\lambda_j^{t-\tau_2+\tau_1} Q^{ij}_R\notag\\
		&=t^2\sum_{ij}\frac{\lambda_i^t\lambda_j-\lambda_i\lambda_j^t}{\lambda_i-\lambda_j}Q^{ij}_R.
	\end{align}
	We have $Q^{00}_R=1/\mathcal{N}$ and $Q^{0i}_R=Q^{i0}_R=1/\mathcal{N}$ for $i=1,...,\mathcal{N}-1$. Thus,
	\begin{align}
		\bar{\mathcal{X}}_{I}^{\{\underline{n},\underline{n}\}}&=\frac{t^2(t-1)}{\mathcal{N}}+2t^2\sum_{i\neq 0}\frac{\lambda_i}{\mathcal{N}(1-\lambda_i)}\notag\\
		&-2t^2\sum_{i\neq 0}\frac{\lambda_i^t}{\mathcal{N}(1-\lambda_i)}+t^2\sum_{\substack{i\neq 0\\j\neq 0}}\frac{\lambda_i^t\lambda_j-\lambda_i\lambda_j^t}{\lambda_i-\lambda_j}Q^{ij}_R.
		\label{X_bar_R_COE}
	\end{align}
	\subsubsection{Sub-sequence reversal with repetition ($SR$)}
	\begin{figure}[h]
		\centering
		\begin{tikzpicture}
			\def\rad{0.75}
			\draw[blue,dashed] (0,0) circle(\rad);
			\draw[red,thick,->] (30:\rad) arc(30:90:\rad);\draw[red,thick,->] (90:\rad) arc(90:150:\rad)--(-30:\rad) arc(-30:-90:\rad);\draw[red,thick] (-90:\rad) arc(-90:-150:\rad)--(30:\rad);	
			\draw[fill=green] (-150:{\rad}) circle(1.5pt);
			\draw[fill=green] (-30:{\rad}) circle(1.5pt);
			\draw (150:{\rad}) node[above left]{$a$};
			\draw (-150:{\rad}) node[below left]{$b$};
			\draw (-30:{\rad}) node[below right]{$b$};
			\draw (30:{\rad}) node[above right]{$c$};
		\end{tikzpicture}
		\caption{\justifying\small A diagram representing sub-sequence reversal with repetition, where the order of states from time step $\tau_1$ to $\tau_2$ is reversed and the states at time steps $\tau_1$ and $\tau_2$ are the same and denoted by $b$.}
		\label{one_rep_sub_reversal}
	\end{figure}
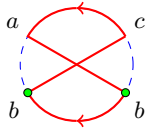
	The contribution of the $SR$ diagram in Fig.~\ref{one_rep_sub_reversal} can be calculated using the rules in Sec.~\ref{COE_Rules}. We obtain
	\begin{align}
		\mathcal{X}_{S}^{\{\underline{n},\underline{n}\}}&=\sum_{ij}\lambda_i^{\tau_2-\tau_1-2}\lambda_j^{t-\tau_2+\tau_1}Q^{ij}_{SR},
	\end{align}
	where
	\begin{align}
		Q^{ij}_{SR}&=\sum_{a,b,c}\mathcal{M}^{(i)}_{c,b}\mathcal{M}^{(j)}_{a,a}V_{a,b}V_{c,a}V^*_{a,c}V^*_{b,a}.
	\end{align}
	The allowed values of $\tau_1$ and $\tau_2$ are $1\leq\tau_1\leq t,\tau_1+3\leq\tau_2\leq t+\tau_1-3$. Therefore, the total contribution is
	\begin{align}
		\bar{\mathcal{X}}_{S}^{\{\underline{n},\underline{n}\}}&=(2t)\frac{1}{2}\sum_{\tau_1=1}^t\sum_{\tau_2=\tau_1+3}^{t+\tau_1-3}\sum_{ij}\lambda_i^{\tau_2-\tau_1-2}\lambda_j^{t-\tau_2+\tau_1}Q^{ij}_{SR}\notag\\
		&=t^2\sum_{ij}\frac{\lambda_i^{t-4}\lambda_j^3-\lambda_i\lambda_j^{t-2}}{\lambda_i-\lambda_j}Q^{ij}_{SR}.
	\end{align}
	Using $Q^{00}_{SR}=1/\mathcal{N},Q^{0i}_{SR}=Q^{i0}_{SR}=1/\mathcal{N}$, we find,
	\begin{align}
		\bar{\mathcal{X}}_{S}^{\{\underline{n},\underline{n}\}}&=\frac{t^2(t-5)}{\mathcal{N}}+2t^2\sum_{i\neq 0}\frac{\lambda_i^3}{\mathcal{N}(1-\lambda_i)}\notag\\
		&-2t^2\sum_{i\neq 0}\frac{\lambda_i^{t-2}}{\mathcal{N}(1-\lambda_i)}\notag\\
		&+t^2\sum_{\substack{i\neq 0\\j\neq 0}}\frac{\lambda_i^{t-4}\lambda_j^3-\lambda_i\lambda_j^{t-2}}{\lambda_i-\lambda_j}Q^{ij}_{SR}.
		\label{X_bar_SR_COE}
	\end{align}
	The second-order correction to the  SFF is
	\begin{align}
		K_{1}^{(2)}(t)&=\bar{\mathcal{X}}_{T}+\bar{\mathcal{X}}_{S}-\bar{\mathcal{X}}_{I}^{\{\underline{n},\underline{n}\}}-\bar{\mathcal{X}}_{S}^{\{\underline{n},\underline{n}\}}.
        \label{K2_COE_1}
	\end{align}
	Substituting Eqs.~(\ref{X_bar_T_COE}), (\ref{X_bar_S_COE}), (\ref{X_bar_R_COE}), and (\ref{X_bar_SR_COE}) in Eq.~(\ref{K2_COE_1}), we obtain
	\begin{widetext}
		\begin{align}
			K_1^{(2)}(t)&=-\frac{2t^2}{\mathcal{N}}+2t^2\sum_{i\neq 0}\left(\lambda_i^{t-3}Q^{i}_{T^{(1)}}+\lambda_i^{t-4}Q^{i}_{T^{(2)}}-\frac{\lambda_i^{t-3}}{\mathcal{N}(1-\lambda_i)}+\frac{\lambda_i^t}{\mathcal{N}(1-\lambda_i)}\right)\notag\\
			&+t^2\sum_{\substack{i\neq 0\\j\neq 0}} \left(\frac{\lambda_i^{t-4}\lambda_j-\lambda_i\lambda_j^{t-4}}{\lambda_i-\lambda_j}Q^{ij}_{T'}+\frac{\lambda_i^{t-4}\lambda_j^3-\lambda_i^3\lambda_j^{t-4}}{\lambda_i-\lambda_j}Q^{ij}_{S}-\frac{\lambda_i^t\lambda_j-\lambda_i\lambda_j^t}{\lambda_i-\lambda_j}Q^{ij}_R-\frac{\lambda_i^{t-4}\lambda_j^3-\lambda_i\lambda_j^{t-2}}{\lambda_i-\lambda_j}Q^{ij}_{SR}\right)\notag\\
			&=-\frac{2t^2}{\mathcal{N}}+\mathcal{O}(\lambda_1^t)
			\label{K2_COE}
		\end{align}
	\end{widetext}
	We include the contribution of all the Type \Rom{3} terms in Eq.~\ref{K2_COE}, which we ignore in our calculation with the reduced diagrams in Eq.~\ref{SFF_COE_2nd_order}.
	\section{Derivation of the second-order correction in the absence of $\mathcal{T}$-symmetry using only the rules in Sec.~\ref{CUE_rules}}
	\label{CUE_2nd_full_der}
	\subsubsection{Transposition}
	Following the rules in Sec.~\ref{CUE_rules}, the calculation for the cyclic variants of transposition is carried out in a similar way to the $\mathcal{T}^2=1$ case.
	\begin{align}
		\sum_{l=0}^{t-1}\sum_{T}\mathcal{Y}_{\mathcal{C}^lT}&=\frac{t^2(t-1)}{2\mathcal{N}}+\frac{t^2}{\mathcal{N}}\sum_{i\neq 0}\frac{\lambda_i-\lambda_i^{t-2}}{1-\lambda_i}\notag\\
		&+t^2\sum_{i\neq 0}\left(\lambda_i^{t-3}Q^{i}_{T^{(1)}}+\lambda_i^{t-4}Q^{i}_{T^{(2)}}\right)\notag\\
		&+\frac{t^2}{2}\sum_{\substack{i\neq 0\\j\neq 0}} \frac{\lambda_i^{t-4}\lambda_j-\lambda_i\lambda_j^{t-4}}{\lambda_i-\lambda_j}Q^{ij}_{T'}.
		\label{Y_T}
	\end{align}
The contribution from the anticyclic variants is
	\begin{align}
		\sum_{l=0}^{t-1}\sum_{T}\mathcal{Y}_{\mathcal{R}\mathcal{C}^lT}&=\frac{t^2}{2}\sum_{ij}\frac{\chi_i^{t-4}\chi_j-\chi_i\chi_j^{t-4}}{\chi_i-\chi_j}\tilde{Q}^{ij}_{T'}\notag\\
		&+t^2\sum_i\chi_i^{t-3}\tilde{Q}^{i}_{T^{(1)}}+t^2\sum_i\chi_i^{t-4}\tilde{Q}^{i}_{T^{(2)}},
	\end{align}
	where
	\begin{align}
		\tilde{Q}^{ij}_{T'}&=\tilde{\mathcal{M}}^{(i)}_{f,a}\tilde{\mathcal{M}}^{(j)}_{c,d}V_{a,b}V_{b,c}V_{d,e}V_{e,f}V^*_{e,a}V^*_{c,e}V^*_{b,d}V^*_{f,b},\\
		\tilde{Q}^{i}_{T^{(1)}}&=\tilde{\mathcal{M}}^{(i)}_{d,a}V_{a,b}V_{b,c}V_{c,d}V^*_{c,a}V^*_{b,c}V^*_{d,b},\\
		\tilde{Q}^{i}_{T^{(2)}}&=\tilde{\mathcal{M}}^{(i)}_{e,a}V_{a,b}V_{b,c}V_{c,d}V_{d,e}V^*_{d,a}V^*_{c,d}V^*_{b,c}V^*_{e,b}.
	\end{align}
        The total contribution from all single transposition diagrams is then: 
	\begin{widetext}
		\begin{align}
			\bar{\mathcal{Y}}_{T}&=\sum_{l=0}^{t-1}\sum_{T}\mathcal{Y}_{\mathcal{C}^lT}+\sum_{l=0}^{t-1}\sum_{T}\mathcal{Y}_{\mathcal{R}\mathcal{C}^lT}\notag\\
			&=\frac{t^2(t-1)}{2\mathcal{N}}+\frac{t^2}{\mathcal{N}}\sum_{i\neq 0}\frac{\lambda_i}{1-\lambda_i}+t^2\sum_{i\neq 0}\left(-\frac{\lambda_i^{t-2}}{\mathcal{N}(1-\lambda_i)}+\lambda_i^{t-3}Q^{i}_{T^{(1)}}+\lambda_i^{t-4}Q^{i}_{T^{(2)}}\right)+\frac{t^2}{2}\sum_{\substack{i\neq 0\\j\neq 0}} \frac{\lambda_i^{t-4}\lambda_j-\lambda_i\lambda_j^{t-4}}{\lambda_i-\lambda_j}Q^{ij}_{T'}\notag\\
			&+\frac{t^2}{2}\sum_{ij}\frac{\chi_i^{t-4}\chi_j-\chi_i\chi_j^{t-4}}{\chi_i-\chi_j}\tilde{Q}^{ij}_{T'}+t^2\sum_i\chi_i^{t-3}\tilde{Q}^{i}_{T^{(1)}}+t^2\sum_i\chi_i^{t-4}\tilde{Q}^{i}_{T^{(2)}}.
			\label{Y_bar_T}
		\end{align}
	\end{widetext}
	\subsubsection{Sub-sequence reversal}
The contribution of the sub-sequence reversal diagrams can be found again following the rules in Sec.~ \ref{CUE_rules} as       
	\begin{align}
		\bar{\mathcal{Y}}_{S}&=\frac{t}{2}\sum_{\tau_1=1}^t\sum_{\tau_2=\tau_1+3}^{t+\tau_1-5}\sum_{ij}\chi_i^{\tau_2-\tau_1}\lambda_j^{t-\tau_2+\tau_1-2}Q^{ij}_{S}\notag\\
		&+\frac{t}{2}\sum_{\tau_1=1}^t\sum_{\tau_2=\tau_1+3}^{t+\tau_1-5}\sum_{ij}\lambda_i^{\tau_2-\tau_1}\chi_j^{t-\tau_2+\tau_1-2}\tilde{Q}^{ij}_{S}\notag\\
%	\end{align}
%	\begin{align}
		&=\frac{t^2}{2}\sum_{ij}\frac{\chi_i^{t-4}\lambda_j^3-\chi_i^3\lambda_j^{t-4}}{\chi_i-\lambda_j}Q^{ij}_{S}\notag\\
		&+\frac{t^2}{2}\sum_{ij}\frac{\lambda_i^{t-4}\chi_j^3-\lambda_i^3\chi_j^{t-4}}{\lambda_i-\chi_j}\tilde{Q}^{ij}_{S},
		\label{Y_bar_S}
	\end{align}
	where 
	\begin{align}
		Q^{ij}_S&=\tilde{\mathcal{M}}^{(i)}_{b,c}\mathcal{M}^{(j)}_{d,a}V_{a,b}V_{c,d}V^*_{a,c}V^*_{b,d},\\
		\tilde{Q}^{ij}_S&=\mathcal{M}^{(i)}_{b,c}\tilde{\mathcal{M}}^{(j)}_{d,a}V_{a,b}V_{c,d}V^*_{c,a}V^*_{d,b}.
	\end{align}
	It can be shown that $Q^{i0}_{SR}=\tilde{Q}^{0i}_{SR}=\text{tr}\tilde{\mathcal{M}}^{(i)}/\mathcal{N}=1/\mathcal{N}$. Thus, we get
	\begin{align}
		\bar{\mathcal{Y}}_{S}&=\frac{t^2}{\mathcal{N}}\sum_i\frac{\chi_i^{t-2}-\chi_i^3}{\chi_i-1}\notag\\
		&+\frac{t^2}{2}\sum_{i}\sum_{j\neq 0}\frac{\chi_i^{t-4}\lambda_j^3-\chi_i^3\lambda_j^{t-4}}{\chi_i-\lambda_j}Q^{ij}_{S}\notag\\
		&+\frac{t^2}{2}\sum_{i\neq 0}\sum_{j}\frac{\lambda_i^{t-4}\chi_j^3-\lambda_i^3\chi_j^{t-4}}{\lambda_i-\chi_j}\tilde{Q}^{ij}_{S}.
		\label{Y_bar_S_final}
	\end{align}
	\subsubsection{Identity permutation with repetition}
The calculation of the cyclic variants of repetitions is also similar to $\mathcal{T}^2=1$ case.
	\begin{align}
		\sum_{l=0}^{t-1}\sum_{R}\mathcal{Y}^{\{\underline{n},\underline{n}\}}_{\mathcal{C}^lI}&=\frac{t^2(t-1)}{2\mathcal{N}}+\frac{t^2}{\mathcal{N}}\sum_{i\neq 0}\frac{\lambda_i}{1-\lambda_i}\notag\\
		&-\frac{t^2}{\mathcal{N}}\sum_{i\neq 0}\frac{\lambda_i^t}{1-\lambda_i}+\frac{t^2}{2}\sum_{\substack{i\neq 0\\j\neq 0}}\frac{\lambda_i^t\lambda_j-\lambda_i\lambda_j^t}{\lambda_i-\lambda_j}Q^{ij}_R,
	\end{align}
    Following the rules in Sec.~\ref{CUE_rules}, total contribution of all the anticyclic variants take the following form:
	\begin{align}
		\sum_{l=0}^{t-1}\sum_{R}\mathcal{Y}^{\{\underline{n},\underline{n}\}}_{\mathcal{R}\mathcal{C}^lI}&=\frac{t^2}{2}\sum_{ij}\frac{\chi_i^t\chi_j-\chi_i\chi_j^t}{\chi_i-\chi_j}\tilde{Q}^{ij}_R,
	\end{align}
	where
	\begin{align}
		\tilde{Q}^{ij}_R&=\sum_a\tilde{\mathcal{M}}^{(i)}_{a,a}\tilde{\mathcal{M}}^{(j)}_{a,a}.
	\end{align}
	Thus, we get the total contribution from the cyclic and anticyclic variants as 
	\begin{align}
		\bar{\mathcal{Y}}_{R}^{\{\underline{n},\underline{n}\}}&=\sum_{l=0}^{t-1}\sum_{R}\mathcal{Y}^{\{\underline{n},\underline{n}\}}_{\mathcal{C}^lI}+\sum_{l=0}^{t-1}\sum_{R}\mathcal{Y}^{\{\underline{n},\underline{n}\}}_{\mathcal{R}\mathcal{C}^lI}\notag\\
		&=\frac{t^2(t-1)}{2\mathcal{N}}+\frac{t^2}{\mathcal{N}}\sum_{i\neq 0}\frac{\lambda_i}{1-\lambda_i}\notag\\
		&-\frac{t^2}{\mathcal{N}}\sum_{i\neq 0}\frac{\lambda_i^t}{1-\lambda_i}+\frac{t^2}{2}\sum_{\substack{i\neq 0\\j\neq 0}}\frac{\lambda_i^t\lambda_j-\lambda_i\lambda_j^t}{\lambda_i-\lambda_j}Q^{ij}_R\notag\\
		&+\frac{t^2}{2}\sum_{ij}\frac{\chi_i^t\chi_j-\chi_i\chi_j^t}{\chi_i-\chi_j}\tilde{Q}^{ij}_R.
		\label{Y_bar_R}
	\end{align}
	\subsubsection{Sub-sequence reversal with repetition}
	Following    the rules in Sec.~\ref{CUE_rules}, we find the contribution from the sub-sequence reversal with a repetition diagrams as
          \begin{align}   
		\bar{\mathcal{Y}}_{SR}^{\{\underline{n},\underline{n}\}}&=\frac{t}{2}\sum_{\tau_1=1}^t\sum_{\tau_2=\tau_1+3}^{t+\tau_1-3}\sum_{ij}\chi_i^{\tau_2-\tau_1-2}\lambda_j^{t-\tau_2+\tau_1}Q^{ij}_{SR}\notag\\
		&+\frac{t}{2}\sum_{\tau_1=1}^t\sum_{\tau_2=\tau_1+3}^{t+\tau_1-3}\sum_{ij}\lambda_i^{\tau_2-\tau_1-2}\chi_j^{t-\tau_2+\tau_1}\tilde{Q}^{ij}_{SR}\notag\\
		&=\frac{t^2}{2}\sum_{ij}\frac{\chi_i^{t-4}\lambda_j^3-\chi_i\lambda_j^{t-2}}{\chi_i-\lambda_j}Q^{ij}_{SR}\notag\\
		&+\frac{t^2}{2}\sum_{ij}\frac{\lambda_i^{t-4}\chi_j^3-\lambda_i\chi_j^{t-2}}{\lambda_i-\chi_j}\tilde{Q}^{ij}_{SR},
		\label{Y_bar_SR_T_abs}
	\end{align}
	where
	\begin{align}
		Q^{ij}_{SR}&=\tilde{\mathcal{M}}^{(i)}_{b,c}\mathcal{M}^{(j)}_{a,a}V_{a,b}V_{c,a}V^*_{a,c}V^*_{b,a}\notag\\
		&=\left(\tilde{\mathcal{M}}\tilde{\mathcal{M}}^{(i)}\tilde{\mathcal{M}}\right)_{a,a}\mathcal{M}^{(j)}_{a,a}=\chi_i^2\tilde{\mathcal{M}}^{(i)}_{a,a}\mathcal{M}^{(j)}_{a,a},\\
		\tilde{Q}^{ij}_{SR}&=\mathcal{M}^{(i)}_{b,c}\tilde{\mathcal{M}}^{(j)}_{a,a}V_{a,b}V_{c,a}V^*_{c,a}V^*_{a,b}=\lambda_i^2\mathcal{M}^{(i)}_{a,a}\tilde{\mathcal{M}}^{(j)}_{a,a}.
	\end{align}
Thus, we can find 
	\begin{align}
		Q^{i0}_{SR}&=\frac{\chi_i^2}{\mathcal{N}}\text{tr}\tilde{\mathcal{M}}^{(i)}=\frac{\chi_i^2}{\mathcal{N}},
		\label{Q_i0_SR}\\
		\tilde{Q}^{0i}_{SR}&=\frac{1}{\mathcal{N}}\text{tr}\tilde{\mathcal{M}}^{(i)}=\frac{1}{\mathcal{N}}.
		\label{tQ_i0_SR}
	\end{align}
We substitute Eqs.~(\ref{Q_i0_SR}) and (\ref{tQ_i0_SR}) in Eq.~(\ref{Y_bar_SR_T_abs}) to find
	\begin{align}
		\bar{\mathcal{Y}}_{SR}^{\{\underline{n},\underline{n}\}}&=\frac{t^2}{\mathcal{N}}\sum_i\frac{\chi_i^3}{1-\chi_i}-\frac{t^2}{\mathcal{N}}\sum_i\frac{\chi_i^{t-2}}{1-\chi_i}\notag\\
		&+\frac{t^2}{2}\sum_{i}\sum_{j\neq 0}\frac{\chi_i^{t-4}\lambda_j^3-\chi_i\lambda_j^{t-2}}{\chi_i-\lambda_j}Q^{ij}_{SR}\notag\\
		&+\frac{t^2}{2}\sum_{i\neq 0}\sum_j\frac{\lambda_i^{t-4}\chi_j^3-\lambda_i\chi_j^{t-2}}{\lambda_i-\chi_j}\tilde{Q}^{ij}_{SR}.
		\label{Y_bar_SR}
	\end{align}
	Therefore, we get the total second-order correction to the SFF for systems in the absence of $\mathcal{T}$-symmetry as
	\begin{widetext}
		\begin{align}
			K^{(2)}_0(t)&=\bar{\mathcal{Y}}_{T}+\bar{\mathcal{Y}}_{S}-\bar{\mathcal{Y}}_{R}^{\{\underline{n},\underline{n}\}}-\bar{\mathcal{Y}}_{SR}^{\{\underline{n},\underline{n}\}}\notag\\
			&=t^2\sum_{i\neq 0}\left(-\frac{\lambda_i^{t-2}}{\mathcal{N}(1-\lambda_i)}+\lambda_i^{t-3}Q^{i}_{T^{(1)}}+\lambda_i^{t-4}Q^{i}_{T^{(2)}}\right)+\frac{t^2}{2}\sum_{\substack{i\neq 0\\j\neq 0}} \frac{\lambda_i^{t-4}\lambda_j-\lambda_i\lambda_j^{t-4}}{\lambda_i-\lambda_j}Q^{ij}_{T'}\notag\\
			&+t^2\sum_{ij}\frac{\chi_i^{t-4}\chi_j-\chi_i\chi_j^{t-4}}{\chi_i-\chi_j}\tilde{Q}^{ij}_{T'}+t^2\sum_i\chi_i^{t-3}\tilde{Q}^{i}_{T^{(1)}}+t^2\sum_i\chi_i^{t-4}\tilde{Q}^{i}_{T^{(2)}}\notag\\
			&+\frac{t^2}{2}\sum_{i}\sum_{j\neq 0}\frac{\chi_i^{t-4}\lambda_j^3-\chi_i^3\lambda_j^{t-4}}{\chi_i-\lambda_j}Q^{ij}_{S}+\frac{t^2}{2}\sum_{ij}\frac{\lambda_i^{t-4}\chi_j^3-\lambda_i^3\chi_j^{t-4}}{\lambda_i-\chi_j}\tilde{Q}^{ij}_{S}\notag\\
			&+\frac{t^2}{\mathcal{N}}\sum_{i\neq 0}\frac{\lambda_i^t}{1-\lambda_i}-\frac{t^2}{2}\sum_{\substack{i\neq 0\\j\neq 0}}\frac{\lambda_i^t\lambda_j-\lambda_i\lambda_j^t}{\lambda_i-\lambda_j}Q^{ij}_R-\frac{t^2}{2}\sum_{ij}\frac{\chi_i^t\chi_j-\chi_i\chi_j^t}{\chi_i-\chi_j}\tilde{Q}^{ij}_R\notag\\
			&+\frac{t^2}{\mathcal{N}}\sum_i\frac{\chi_i^{t-2}}{1-\chi_i}-\frac{t^2}{2}\sum_{ij}\frac{\chi_i^{t-4}\lambda_j^3-\chi_i\lambda_j^{t-2}}{\chi_i-\lambda_j}Q^{ij}_{SR}-\frac{t^2}{2}\sum_{ij}\frac{\lambda_i^{t-4}\chi_j^3-\lambda_i\chi_j^{t-2}}{\lambda_i-\chi_j}\tilde{Q}^{ij}_{SR}\notag\\
			&=\mathcal{O}(\lambda_1^t,\chi_0^t).\label{K2_CUE}
		\end{align}
	\end{widetext}
Equation~(\ref{K2_CUE}) shows the absence of universal term at second order in time, which explains the emergence of CUE SFF. Equation~\ref{K2_CUE} includes the contribution of all Type III terms, which are ignored in our calculation with the reduced diagrams in Eq.~\ref{K20_CUE}.   
	\bibliography{bibliographyRMT}
	
	\end{document}